\documentclass[twocolumn,twocolappendix]{aastex631}

\usepackage{amsmath}

\newcommand{\cstpl}[1]{\left\{\rule{0 cm}{#1}\right.}

\begin{document}

\title{Deuteration of Organic Molecules as a Probe of Starless Core and Filament Evolution in Barnard 10}


\correspondingauthor{Hanga Andras-Letanovszky, Yancy Shirley}
\email{hangaa@arizona.edu, yshirley@arizona.edu}

\author{Hanga Andras-Letanovszky}
\affiliation{Steward Observatory, The University of Arizona \\
933 N. Cherry Ave \\
Tucson, AZ 85719, USA}

\author{Yancy L. Shirley}
\affiliation{Steward Observatory, The University of Arizona \\
933 N. Cherry Ave \\
Tucson, AZ 85719, USA}

\author{Lucille J. Steffes}
\affiliation{Steward Observatory, The University of Arizona \\
933 N. Cherry Ave \\
Tucson, AZ 85719, USA}

\author{Brian Svoboda}
\affiliation{National Radio Astronomy Observatory\\
1101 Lopezville Road \\
Socorro, NM 87801, USA}

\author{Hannah Gruber}
\affiliation{Steward Observatory, The University of Arizona \\
933 N. Cherry Ave \\
Tucson, AZ 85719, USA}

\author{Samantha Scibelli}
\altaffiliation{Jansky Fellow of the National Radio Astronomy Observatory}
\affiliation{National Radio Astronomy Observatory\\
520 Edgemont Rd. \\
Charlottesville, VA 22903, USA}

\author{Emma Vertachnik}
\affiliation{Steward Observatory, The University of Arizona \\
933 N. Cherry Ave \\
Tucson, AZ 85719, USA}


\begin{abstract}

The deuterium fractionation of molecules in starless cores is sensitive to their dynamical histories, which recent magnetohydrodynamical simulations have shown to be extremely varied.
The deuterated isotopologues of formaldehyde (H$_2$CO) and methanol (CH$_3$OH) probe deuterium fractionation in simple organic molecules.
This complete survey targets 11 low-mass starless cores in the small, quiescent Barnard 10 (B10) region of the Taurus Molecular Cloud.
The cores were observed using the 12m Arizona Radio Observatory telescope with a 100\% detection rate in transitions of o/pH$_2$CO, HDCO, pD$_2$CO, A/E-CH$_3$OH, and CH$_2$DOH and the dense gas tracer N$_2$H$^+$. 
The HDCO and pD$_2$CO column densities and deuterium fractions are not correlated with those of the grain-surface deuteration tracer CH$_2$DOH, indicating significant gas-phase formation of deuterated formaldehyde. 
The observed deuterium fractions do not correlate with evolutionary indicators (e.g. core central density) or physical conditions (e.g. core mass).
We also find that cores within the northwestern filament have lower deuterium fractions than cores of similar densities in the other two filaments. 
These could indicate differential core evolution both between and within entire filaments.
Comparing the deuterium fractions of the B10 sample to published sources across different evolutionary stages suggests the inheritance of D$_2$CO from starless cores, although more surveys of organic deuteration are needed to conclusively determine inheritance. 

\end{abstract}




\section{Introduction} \label{sec:intro}

Starless cores are local over-densities of gas and dust on the scale of 0.1 pc that form in molecular clouds \citep{Bergin2007,DiFrancesco2007}.
These structures can be transient and may eventually disperse, but some fraction eventually collapse to form protostars. 
Starless cores that are bound by external pressure and self-gravity and that have kinematic evidence of infall are referred to as prestellar cores \citep{Keto2008,Andre2014,Caselli2025}.
Starless and prestellar cores represent the earliest stages of star formation, and set the initial physical and chemical conditions for the nascent star and its surrounding planetary system \citep{Ceccarelli2014}.
Their formation and boundedness are determined by the complex interplay of gravity, turbulence, and magnetic fields \citep{McKee1992, Krumholz2017}.
Like beads on a string, most starless and prestellar cores are found in parsec-scale gaseous filaments formed by large supersonic flows compressing gas in the parent molecular cloud \citep{Andre2014, Hacar2023}.
Starless and prestellar cores are cold ($\sim$ 10 K) and dense ($\geq 10^4$ cm$^{-3}$), making them readily identifiable in millimeter and submillimeter dust continuum emission observations of nearby star-forming regions \citep{Ward-Thompson1994,Motte1998A&A,Andre2010,Launhardt2013,Pattle2025}.
These conditions also mean cores are often easily detectable in rotational molecular line emission from dense gas tracers such as NH$_3$ \citep{Benson1989,Jijina1999,Seo2015,Friesen2017,Pineda2026} and N$_2$H$^+$ \citep{Caselli2002n2hp,Crapsi2005,Tafalla2015,Punanova2018}.

Magnetohydrodynamical simulations of star formation in molecular clouds, such as the STARFORGE simulations described in \cite{Offner2025}, point towards starless cores having a wide range of dynamical histories.
Not only do their lifetimes and evolution rates vary wildly, with the time from formation to collapse spanning $\sim$ 1--10 times the freefall timescale, but their evolution is often non-monotonic as well, with their central densities increasing and decreasing by up to an order of magnitude before either dissipating or rapidly increasing to form a protostar.
It is unknown what precisely controls starless core evolution rates; \cite{Offner2025} found core evolution to be stochastic, partially due to the turbulent flows that shape and impact star-forming regions.
Magnetic fields may play a large role in determining core evolution rates, as they could support cores, especially low-mass cores, against gravitational collapse.
However, magnetic field strengths are difficult to map accurately, especially on the scale of cores.
For example, magnetic field strength measurements towards the central Taurus Molecular Cloud span from 13 to 75 $\mu$G \citep{Chapman2011, Ward-Thompson2023}. 
Since magnetic energy is proportional to $B^2$, the actual importance of magnetic fields relative to gravity, external pressure, and turbulence for core stability is difficult to assess \citep{Galloway-Sprietsma2022, Scibelli2023}.
This makes a physical determination of core stability unfeasible for most starless cores. 
The rich chemistry of starless cores, however, may provide a probe of core evolution rates.
In particular, deuterium fractionation is sensitive to the dynamical histories of cores \citep{Galloway-Sprietsma2022}.

Deuterium fractionation is heavily favored in starless cores thanks to their low temperatures and high densities.
The main driver of deuteration in the gas phase is H$_2$D$^+$, which is produced by the exothermic reaction
$$\mathrm{H_3^+ + HD \rightarrow H_2D^+ + H_2 + 232 \  K} \ \ ,$$
which preferentially proceeds in the forward direction since the low ortho-to-para ratio of H$_2$ at $10$ K means that only a small fraction of H$_2$ molecules can overcome the activation energy barrier for the reverse direction \citep{Dalgarno1984,Brown1989,Roberts2003,Pagani2009}.
Additionally, at such low temperatures a common destructor of H$_2$D$^+$, CO, freezes out onto the ices on the surfaces of dust grains \citep{Bacmann2003,Walmsley2004,Jorgensen2005}.
This leads to an enhancement of H$_2$D$^+$, which can then pass on its deuterium to other molecules through ion-molecule reactions \citep{Ceccarelli2014}.
The deuteration of molecules on icy dust grain surfaces is also enhanced.
The H$_2$D$^+$ can be absorbed onto the ice and then lose its deuterium through dissociative electron recombination, leaving the deuterium atom free to attach to other molecules in the ice, such as CO \citep{Ceccarelli2014}.
Starless and prestellar cores thus build up deuterated molecules over time.
This has direct implications for interpreting core evolutionary history.
For instance, if two cores at the same density with similar initial conditions and environments have different deuterium fractions, the one with the higher deuterium fraction is likely evolving more slowly, as it has spent a longer time at its current density, allowing for a greater abundance of deuterated molecules \citep{Galloway-Sprietsma2022}.

In the past decade, it was discovered that complex organic molecules, which are molecules with six or more atoms including carbon \citep{Herbst2009}, are prevalent in the gas phase toward starless cores \citep{Bacmann2012,Jimenez-Serra2016,Scibelli2020,Scibelli2021,Jimenez-Serra2021,Megias2023,Scibelli2024,Scibelli2025}.
The precursors to these complex organics, H$_2$CO and CH$_3$OH, are abundant in starless and prestellar cores.
While there are numerous studies of deuterated organic molecules in protostellar cores (e.g., \citealt{Ceccarelli1998,Roberts2007,Bergman2011,Jorgensen2016,Bianchi2017,Koumpia2017,Persson2018, Manigand2019, Manigand2020,Riaz2022,Evans2023,Chahine2024,Okoda2024,Podio2024, Mercimek2025}), there are only a few surveys for deuterated organic molecules in starless and prestellar cores \citep{Bacmann2003,Chacon-Tanarro2019,Lin2023,Scibelli2025,Kulterer2026,Ferrer_Asensio2026} with only one survey of the entire starless core population of a region \citep{Ambrose2021}.

We thus perform a complete systematic survey of starless cores in the Barnard 10 (B10) region of the Taurus Molecular Cloud.
The B10 region is an ideal low-mass starless core evolution laboratory for a few reasons. First, the region is very quiescent and unevolved, with no nearby protostars or stars to generate feedback to disturb core evolution \citep{Hacar2013,Seo2015}.
Second, the region is relatively small with the central ``letter A-shaped" filamentary structure spanning $0.40$ pc in the plane of the sky (extending to $0.65$ pc including the southern isolated core Seo17; \citealt{Seo2015}). 
Thus it is possible that initial chemical conditions and external conditions may not vary much across the region (see Figure \ref{fig:B10_NH2_map}).
Third, B10 is relatively close by at a distance of only $\sim$ 135 pc \citep{Schlafly2014}.
We also have the benefit of multiple previously completed continuum maps and molecular maps and surveys of B10.
B10 has been mapped in the $1 \rightarrow 0$ transition of N$_2$H$^+$ \citep{Hacar2013, Tafalla2015} and the (1,1) and (2,2) inversion transitions of para-NH$_3$ \citep{Seo2015}, both of which are dense gas tracer molecules.
\cite{Scibelli2023} mapped the B10 region in 1.2 and 2.1 mm thermal dust continuum emission and, along with published \textit{Herschel Space Observatory} maps at $70 - 500$ $\mu$m \citep{Marsh2014,Marsh2016,Singh2022}, constructed 3D radiative transfer models of the starless cores within it.
The peak NH$_3$ positions of cores in B10 were also surveyed in CH$_3$OH \citep{Scibelli2020} and CH$_2$DOH, although not all sources were detected in CH$_2$DOH \citep{Ambrose2021}.

In this study, we surveyed 11 starless cores in the B10 region in both ortho-H$_2$CO and para-H$_2$CO and the two deuterated formaldehyde isotopologues, HDCO and para-D$_2$CO, achieving a 100\% detection rate.
The details of the observations are discussed in Section \ref{sec:obs}.
Section \ref{sec:data} covers our data analysis, namely our determination of molecular column densities and our method for quantifying correlations between observed quantities.
In Section \ref{sec:results}, we compare our formaldehyde and methanol isotopologue column densities and deuterium fractions to each other and to the column densities of the dense gas tracers para-NH$_3$ from \cite{Seo2015} and N$_2$H$^+$.
We also compare our organic deuterium fractions to the physical parameters determined by the 3D radiative transfer models in \cite{Scibelli2023}.
Through a spatial comparison of core central density and organic deuterium fractions in cores with respect to their parent filaments, we investigate the relative evolution rates of both the individual cores and the whole filaments.
We also place the B10 cores in the context of other starless cores and protostars from the literature to examine the possibility of deuterium inheritance from starless to protostellar phases.


\section{ARO 12m Observations} \label{sec:obs}

\begin{figure*}
    \centering
    \includegraphics[width=\linewidth]{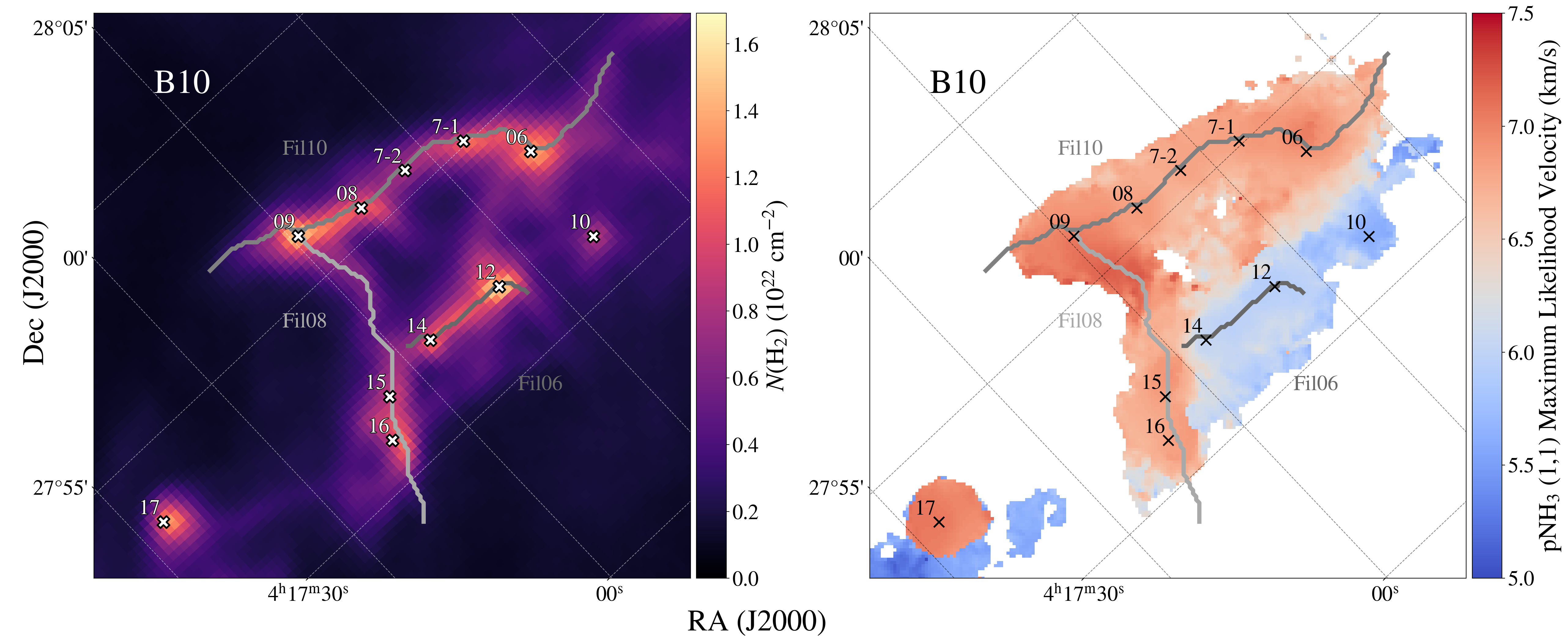}
    \caption{Left: 
    A map of the B10 region in the column density of H$_2$. 
    This map was converted from the optical depth map of the L1521 region of Taurus at 1 THz from \cite{Singh2022} using Equation \ref{eqn:NH2}, then reprojected to match the orientation of the map on the left.
    Right: A map of the B10 region in the maximum likelihood velocity of pNH$_3$ from observations of its (1,1) inversion transition (described in more detail in Section \ref{subsec:nestfit}). 
    Overplotted on both plots are velocity-coherent filaments in gray, numbered as in \cite{Hacar2013}, as well as crosses marking the core peak locations listed in Table \ref{tab:observedsources}. 
    }
    \label{fig:B10_NH2_map}
\end{figure*}

We observed 11 starless cores in the B10 region of the Taurus Molecular Cloud (Table \ref{tab:observedsources}), shown in an H$_2$ column density map in Figure \ref{fig:B10_NH2_map}.
Nine cores, with the prefix ``Seo'' in their names, are the pNH$_3$ (1,1) peak positions from \cite{Seo2015} (we will use the prefixes ``o" and ``p" to denote the ortho and para versions of molecules respectively).
A higher resolution 1.2 mm dust continuum map of B10 revealed that cores Seo07 and Seo13 split into distinct sources, Sci7-1 and Sci7-2 and Sci13-1 and Sci13-2 \citep{Scibelli2023}.
However, we do not include Sci13-1 and Sci13-2 because the sources are spatially confused with our ARO 12m beam.
We also omit the source Seo11 since there was a large discrepency between the 1.2 mm dust continuum position and the NH$_3$ position \citep{Scibelli2023}.
Seo11 is one of the weakest pNH$_3$ (1,1) intensity cores in the \cite{Seo2015} survey and the accuracy of its NH$_3$-derived position was likely affected by low signal-to-noise.

\begin{deluxetable*}{lccccl}
\tablewidth{0pt} 
\tablecaption{Observed Sources\label{tab:observedsources}}
\tablehead{
\colhead{Source} & \colhead{$\alpha$\tablenotemark{a}}& \colhead{$\delta$} & \colhead{$v_{\rm{LSR}}$\tablenotemark{b}} & \colhead{Filament\tablenotemark{c}} & \colhead{Other Core Names [Angular Separation]}\\
\colhead{} & \colhead{($^{h}:^{m}:^{s})$} & \colhead{($^{o}:^{\prime}:^{\prime\prime}$)}& \colhead{($\rm km~s^{-1}$)} & \colhead{} & \colhead{[($^{\prime\prime}$)]}
} 

\startdata 
Seo06 & 4:17:52.8 & +28:12:25.7 & 6.92 & Fil10 & HGBS J041751.9+281232\tablenotemark{d} [$11$], S14\tablenotemark{e} or WT23-8\tablenotemark{f} [$11$], H13-3\tablenotemark{c} [$42$], MC6\tablenotemark{g} [$50$]\\
Sci7-1 & 4:18:00.6 & +28:11:06.4 & 6.86 & Fil10 & HGBS J041800.2+281112\tablenotemark{d} [$8$], S19\tablenotemark{e} or WT23-7\tablenotemark{f} [$2$]\\
Sci7-2 & 4:18:03.3 & +28:09:06.8 & 6.82 & Fil10 & HGBS J041803.4+280905\tablenotemark{d} [$2$]\\
Seo08 & 4:18:03.7 & +28:07:16.9 & 6.88 & Fil10 & HGBS J041802.3+280736\tablenotemark{d} [$27$], S11\tablenotemark{e} or WT23-6\tablenotemark{f} [$20$]\\
Seo09 & 4:18:07.0 & +28:05:13.0 & 7.06 & Fil10 & HGBS J041807.8+280506\tablenotemark{d} [$19$], MC8\tablenotemark{g} [$4$], S5\tablenotemark{e} or WT23-5\tablenotemark{f} [$16$], H13-6\tablenotemark{c} [$31$] \\
Seo10 & 4:17:37.6 & +28:12:01.8 & 5.64 & Isolated & HGBS J041737.5+281205\tablenotemark{d} [$3$], S21\tablenotemark{e} [$5$] \\
Seo12 & 4:17:41.7 & +28:08:45.7 & 5.96 & Fil06 & HGBS J041741.9+280846\tablenotemark{d} [$2$], MC5-N\tablenotemark{h} [$4$], S2\tablenotemark{e} or WT23-1\tablenotemark{f} [$5$], H13-1\tablenotemark{c} [$46$]\\
Seo14 & 4:17:42.9 & +28:06:00.3 & 6.08 & Fil06 & HGBS J041743.1+280600\tablenotemark{d} [$4$], MC5-S\tablenotemark{h} [$4$], S7\tablenotemark{e} or WT23-2-S\tablenotemark{f} [$8$]\\
Seo15 & 4:17:41.0 & +28:03:49.9 & 6.86 & Fil08 & HGBS J041740.6+280413\tablenotemark{d} [$24$]\\
Seo16 & 4:17:36.1 & +28:02:56.7 & 6.84 & Fil08 & HGBS J041734.3+280303\tablenotemark{d} [$24$], S12\tablenotemark{e} or WT23-3\tablenotemark{f} [$21$]\\
Seo17 & 4:17:50.3 & +27:55:52.4 & 7.08 & Isolated & HGBS J041750.3+275604\tablenotemark{d} [$9$], S3\tablenotemark{e} [$13$], H13-2\tablenotemark{c} [$15$], Tau-C3\tablenotemark{i} [$23$]\\
\enddata
\tablecomments{ (a) Epoch J2000.0. (b) Maximum likelihood velocity at the core peak determined from pNH$_3$ (1,1) observations \citep{Seo2015}. (c) Filament numbers and core names are determined in \cite{Hacar2013}. Cores identified in N$_2$H$^+$ 1-0 emission are denoted with ``H13-\#". Note that the position quoted for H13-4 lies in between cores Sci7-2 [$54^{\prime\prime}$] and Seo08 [$57^{\prime\prime}$] and is therefore not listed. (d)  Closest Herschel Gould Belt Survey starless or prestellar core in \cite{Marsh2016} which are denoted ``HGBS J(coordinate name)". (e) Nearest SCUBA-2 observed starless core in \cite{Ward-Thompson2016} denoted by ``S\#". (f) SCUBA-2 observed starless core names in \cite{Ward-Thompson2023} which are a re-numbering of cores in \cite{Ward-Thompson2016} and are denoted by ``WT23-\#". (g) H$^{13}$CO$^+$ 1-0 core identified in \cite{Onishi2002} denoted by ``MC\#". (h) Core imaged at 1.3mm with ACA in \cite{Tokuda2020} modifying \cite{Onishi2002} name.  (i) Core identified at 100 $\mu$m with $A_{\rm{V}} > 4$ mag in \cite{Wood1994}.}
\end{deluxetable*}

\begin{figure}[h!]
\centering
\includegraphics[width = 1.08\linewidth]{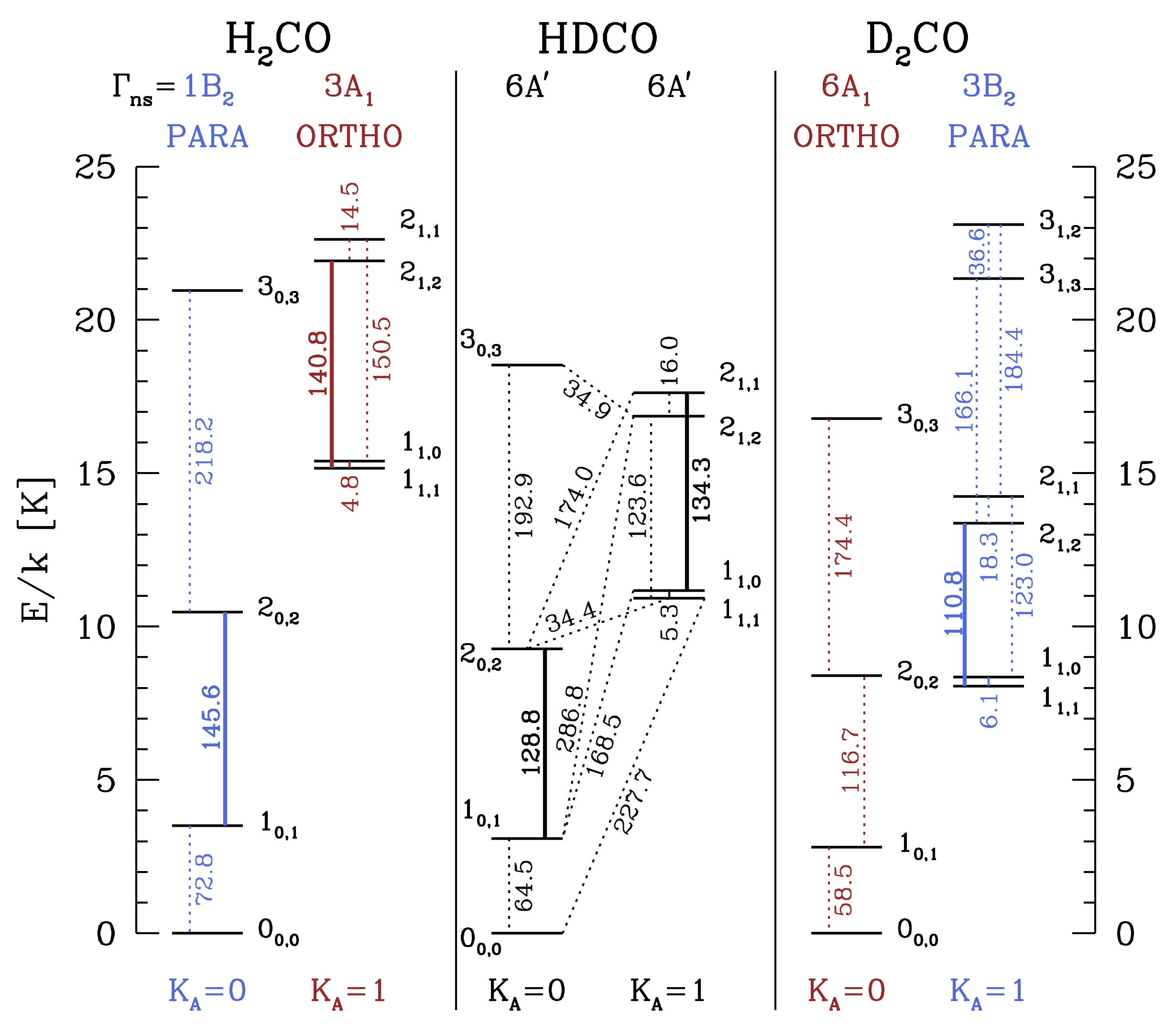}
\caption{All energy levels of H$_2$CO, HDCO, and D$_2$CO are shown with $E_u/k \leq 25$ K.  The symmetry and statistical weight of the nuclear spin states ($\Gamma_{ns}$) are given in C$_{2\rm{v}}$(M) for H$_2$CO and D$_2$CO and in C$_{\rm{s}}$(M) for HDCO (see Tables 9.1 and B.1 in \citealt{Bunker2005}).  Electric dipole allowed transitions are shown as dotted lines with the frequency given in GHz.  Ortho transitions are shown in red while para transitions are shown in blue.  The transitions observed in this paper are indicated as bold, solid lines.
\label{fig:EnergyLevels}}
\end{figure}

Observations of o/pH$_2$CO, HDCO, pD$_2$CO, CH$_2$DOH, A/E-CH$_3$OH, and N$_2$H$^+$  were conducted over multiple observing seasons using the Arizona Radio Observatory 12m telescope located at the Kitt Peak National Observatory (see Table \ref{tab:obs_transitions}).
We did not observe oD$_2$CO due to a lack of favorable low-energy rotational transitions within atmospheric windows (see Figure \ref{fig:EnergyLevels}). 
Observations were made with the facility MRI 3 mm and 2 mm sideband-separating, dual-polarization receivers.
The AROWS spectrometer was used for all observations with a spectral resolution of 39 kHz after Hanning smoothing corresponding to a velocity resolution of $\Delta v_{\rm{chan}} = 0.1 \, (116.9 \, \rm{GHz}/\nu_{\rm{GHz}})$ km s$^{-1}$.
Observations starting in 2023 made use of AROWS multi-window mode where four spectral windows could be steered within the 4 GHz IF bandpass in each sideband.
Unfortunately, both sidebands were only available for the Fall 2023 observing season and observations were limited to one sideband thereafter due to an AROWS board malfunction.
Prior spectral line observations exist for A/E-CH$_3$OH \citep{Scibelli2020} and CH$_2$DOH \citep{Ambrose2021} pointed toward the \cite{Seo2015} NH$_3$ positions.
Therefore, we observed A/E-CH$_3$OH and CH$_2$DOH toward the \cite{Scibelli2023} Sci7-1 and Sci7-2 positions along with additional follow-up observations towards Seo14, as it was originally a non-detection in the \cite{Ambrose2021} survey.
The details of each observing season are summarized in Table \ref{tab:obs_transitions}.
After observations were complete, a problem with the number of channels in AROWS \texttt{CLASS} headers was discovered that offset all velocities by one channel. This was corrected in post-processing for all spectra and reported $v_{LSR}$ values.

\begin{deluxetable*}{llDrCCLr}
\tablewidth{0pt} 
\tablecaption{Observed Transitions\label{tab:obs_transitions}}
\tablehead{
\colhead{Molecule} & \colhead{Transition} & \multicolumn2c{$\nu_\mathrm{tuning}$\tablenotemark{a}} & \colhead{$E_u/k$\tablenotemark{a}} & \colhead{$A_{ul}$\tablenotemark{a}} & \colhead{$\theta_{mb}$ on 12m} & \colhead{$\eta_{mb}$} &
\colhead{Season\tablenotemark{b}} \\
\colhead{} & \colhead{} & \multicolumn2c{(GHz)} & \colhead{(K)}& \colhead{($10^{-6} s^{-1}$)} & \colhead{$(^{\prime\prime})$} &\colhead{(\%)} & \colhead{} 
} 
\decimals
\startdata 
oH$_2$CO & J(K$_a$,K$_c$) = 2(1,2) $\rightarrow$ 1(1,1) & 140.839502 & 21.92 & 53.0 & 42.8 & 82.8\pm1.1 & 2023B \\
  &   & {}  &   &  &   & 78.7\pm1.4\tablenotemark{e} & 2025A \\
pH$_2$CO & J(K$_a$,K$_c$) = 2(0,2) $\rightarrow$ 1(0,1) & 145.602949 & 10.48 & 78.1 & 41.4 & 80.5\pm1.7 & 2025A \\
HDCO & J(K$_a$,K$_c$) = 2(0,2) $\rightarrow$ 1(0,1) & 128.812860 & 9.28 & 54.0 & 46.8 & 86.6\pm0.9 & 2021B \\
  &  & {} &  &  &  & 80.9\pm2.4\tablenotemark{f} & 2024A \\
    & J(K$_a$,K$_c$) = 2(1,1) $\rightarrow$ 1(1,0) & 134.284830 & 17.63 & 45.9 & 44.9 &86.6\pm0.9 & 2021B \\
  &  & {} &  &  &  & 80.9\pm2.4\tablenotemark{g} & 2023A/2024A \\
pD$_2$CO & J(K$_a$,K$_c$) = 2(1,2) $\rightarrow$ 1(1,1) & 110.837830 & 13.37 & 
25.8 & 54.4 & 86.9\pm0.9 & 2024A \\
A-CH$_3$OH\tablenotemark{c} & J(K$_a$,K$_c$) = 2(0,2) $\rightarrow$ 1(0,1) &  96.741371 & 6.96 & 3.4 & 62.3 & 89.2 \pm 2.8 & 2024B \\
E-CH$_3$OH\tablenotemark{c} & J(K$_a$,K$_c$) = 2(-1,2) $\rightarrow$ 1(-1,1) & 96.739358 & 12.54 & 2.6 & 62.3 & 89.2 \pm 2.8 & 2024B \\
CH$_2$DOH & J(K$_a$,K$_c$) = 2(0,2) e$_0$ $\rightarrow$ 1(0,1) e$_0$ &  89.407816 & 6.44 & 2.0 & 67.4 & 89.7\pm2.4\tablenotemark{h} & 2024B \\
  &  & {} & &  &  & 86.4\pm4.8\tablenotemark{i} & 2018A/2024B \\
N$_2$H$^+$\tablenotemark{d} & J = 1 $\rightarrow$ 0 & 93.173770 & 4.47 & 36.3 & 64.7 & 89.2\pm2.8 & 2024B \\
\enddata
\tablecomments{
(a) Spectroscopic data for each transition was taken from the Cologne Database for Molecular Spectroscopy \citep{CDMS} using the Splatalogue online database for astronomical spectroscopy (\url{https://splatalogue.online/}). 
The only exception is CH$_2$DOH, whose spectroscopic data was taken from the Jet Propulsion Laboratory Molecular Spectroscopy catalogue \citep{JPL}.
Detailed references for spectroscopic data for all molecules are listed in Appendix \ref{ap:spec_data_refs}.
(b) The observing seasons runs from February - June for the ``A" semester and from October - February for the ``B" semester.
(c) Both transitions of CH$_3$OH were observed in the same window, which was tuned to the frequency of the A-CH$_3$OH transition.
(d) The spectral window was tuned to the frequency for the strongest hyperfine transition $F_1 = 2 \rightarrow 1$ $F = 3 \rightarrow 2$.  The listed $E_u/k$ and $A_{ul}$ are for the unsplit transition at $\nu = 93.173398$ GHz.
(e) Seo17 only.
(f) Sci7-1 and Sci7-2 only.
(g) Sci7-1 and Sci7-2 (2024A) and Seo10, Seo15, and Seo16 (2023A) only.
(h) Sci7-1 and Sci7-2 only.
(i) Seo14 only; we averaged our observations (2024B) with those of \cite{Ambrose2021} in 2018A.
}
\end{deluxetable*}

All observations were conducted in absolute position switching mode with 5 minutes total integration time per scan made up of pairs of 30 seconds sampled on-source and off-source positions.
A common off position was used for all sources in B10 of $4^{h} 18^{m} 36^{s}.3$ $+28^{\rm{o}} 05^{\prime} 43^{\prime\prime}.6$ in the J2000.0 epoch.
The scans were calibrated using the chopper-wheel calibration method \citep{Kutner1981,Walker2016} and placed on the $T_A^*$ temperature scale.
Observations of the planets Venus, Mars, and Jupiter were observed to determine the main beam efficiency, $\eta_{mb}$, for each receiver during each observing season (Table \ref{tab:obs_transitions}).
The main beam full width at half-maximum (FWHM) is calculated from $\theta_{mb} = 60.9^{\prime\prime} (100 \, \rm{GHz}/\nu_{\rm{GHz}})$ for the ARO 12m receivers with 12 dB edge taper.
The final spectra were calibrated on the main-beam temperature scale, $T_{mb} = T_A^*/\eta_{mb}$.

The spectra were analyzed using the \texttt{CLASS} data analysis software \citep{Pety2005}.
A linear baseline was removed from each scan using a baseline window that extended from $-20$ km s$^{-1}$ to $+30$ km s$^{-1}$ excluding the velocity interval $\pm 3\sigma_v$ from the line velocity centroid, where $\sigma_v = \Delta v/\sqrt{8 \ln{2}}$ is the velocity dispersion determined from the full-width half-maximum linewidth, $\Delta v$, from a Gaussian fit to the sum of the raw scans.
For spectral lines with hyperfine structure, the excluded velocity intervals were determined from $-3\sigma_v$ from the highest frequency hyperfine line velocity centroid to $+3\sigma_v$ from the lowest frequency hyperfine line centroid in each blended clump of hyperfine lines.
A weighted-average using the baseline rms as the weight for each spectrum, $1/\sigma_{T_{mb}}^2$, was used to calculate the final spectrum.
The integrated intensity was then determined from the area under the spectral line within the $\pm 3\sigma_v$ interval.
The line profiles for oH$_2$CO and pH$_2$CO clearly show multiple Gaussian velocity components which were fit within \texttt{CLASS}. The integrated intensity of the component most closely matched to the $v_{LSR}$ of the HDCO lines was extracted by multiplying the total integrated intensity, calculated over all velocity components, by the fraction of the total area for the matched velocity component of the Gaussian fit.
$\Delta v$ and $v_{LSR}$ are reported in Tables \ref{tab:delV} and \ref{tab:vLSR} respectively.
The integrated intensities are reported in Table \ref{tab:intens_dv}.

\begin{longrotatetable}
\movetabledown=2cm
\begin{deluxetable*}{lCCCCCCCCCC}
\tablewidth{0pt}
\tablecaption{Beam Efficiency Corrected Integrated Intensities \label{tab:intens_dv}}
\tablehead{
\colhead{} & \multicolumn{1}{c}{oH$_2$CO} & \multicolumn{1}{c}{pH$_2$CO} & \multicolumn{2}{c}{HDCO} & \multicolumn{1}{c}{pD$_2$CO} & \multicolumn{1}{c}{A-CH$_3$OH\tablenotemark{a}} & \multicolumn{1}{c}{E-CH$_3$OH\tablenotemark{a}} & \multicolumn{1}{c}{CH$_2$DOH\tablenotemark{b}} & \multicolumn{1}{c}{N$_2$H$^+$} \\
\colhead{} & \multicolumn{1}{c}{$2(1,2) \rightarrow 1(1,1)$} & \multicolumn{1}{c}{$2(0,2) \rightarrow 1(0,1)$} & \multicolumn{1}{c}{$2(0,2) \rightarrow 1(0,1)$} & \multicolumn{1}{c}{$2(1,1) \rightarrow 1(1,0)$} & \multicolumn{1}{c}{$2(1,2) \rightarrow 1(1,1)$} & \multicolumn{1}{c}{$2(0,2) \rightarrow 1(0,1)$} & \multicolumn{1}{c}{$2(-1,2) \rightarrow 1(-1,1)$} & \multicolumn{1}{c}{$2(0,2) \mathrm{e}_0 \rightarrow 1(0,1) \mathrm{e}_0$} & \multicolumn{1}{c}{$1 \rightarrow 0$} \\
\colhead{Core} & \colhead{$I_{T_{mb}}$} & \colhead{$I_{T_{mb}}$} & \colhead{$I_{T_{mb}}$} & \colhead{$I_{T_{mb}}$} & \colhead{$I_{T_{mb}}$} & \colhead{$I_{T_{mb}}$} &  \colhead{$I_{T_{mb}}$} &  \colhead{$I_{T_{mb}}$} & \colhead{$I_{T_{mb}}$} \\
\colhead{} & \colhead{(K km s$^{-1}$)} &  \colhead{(K km s$^{-1}$)} & \colhead{(K km s$^{-1}$)} &  \colhead{(K km s$^{-1}$)} &  \colhead{(K km s$^{-1}$)} & \colhead{(K km s$^{-1}$)} & \colhead{(K km s$^{-1}$)} & \colhead{(K km s$^{-1}$)} &  \colhead{(K km s$^{-1}$)}  
} 
\startdata 
Seo06 & 0.351\pm0.013 & 0.343\pm0.009 & 0.178\pm0.005 & 0.040\pm0.004 & 0.010\pm0.002 & 0.566\pm0.021 & 0.424\pm0.004 & 0.012\pm0.002 & 1.764\pm0.056 \\
Sci7-1 & 0.360\pm0.007 & 0.302\pm0.008 & 0.186\pm0.006 & 0.034\pm0.003 & 0.013\pm0.002 & 0.317\pm0.010 & 0.237\pm0.008 & 0.025\pm0.005 & 1.650\pm0.052 \\
Sci7-2 & 0.328\pm0.007 & 0.314\pm0.009 & 0.153\pm0.006 & 0.026\pm0.002 & 0.011\pm0.002 & 0.239\pm0.008 & 0.174\pm0.006 & 0.017\pm0.004 & 1.331\pm0.042 \\
Seo08 & 0.264\pm0.009 & 0.292\pm0.008 & 0.208\pm0.006 & 0.042\pm0.003 & 0.029\pm0.003 & 0.299\pm0.011 & 0.224\pm0.003 & 0.023\pm0.004 & 2.306\pm0.073 \\
Seo09 & 0.269\pm0.014 & 0.293\pm0.008 & 0.281\pm0.007 & 0.058\pm0.004 & 0.031\pm0.003 & 0.483\pm0.018 & 0.366\pm0.003 & 0.034\pm0.004 & 2.761\pm0.087 \\
Seo10 & 0.671\pm0.016 & 0.519\pm0.014 & 0.108\pm0.006 & 0.025\pm0.003 & 0.006\pm0.002 & 0.466\pm0.019 & 0.352\pm0.005 & 0.029\pm0.006 & 0.597\pm0.020 \\
Seo12 & 0.880\pm0.017 & 0.626\pm0.018 & 0.253\pm0.006 & 0.077\pm0.005 & 0.027\pm0.003 & 0.621\pm0.025 & 0.475\pm0.005 & 0.019\pm0.002 & 2.403\pm0.076 \\
Seo14 & 0.792\pm0.033 & 0.732\pm0.018 & 0.235\pm0.006 & 0.065\pm0.004 & 0.013\pm0.002 & 0.755\pm0.028 & 0.558\pm0.005 & 0.011\pm0.002 & 1.224\pm0.039 \\
Seo15 & 0.160\pm0.005 & 0.118\pm0.004 & 0.083\pm0.006 & 0.018\pm0.003 & 0.008\pm0.002 & 0.328\pm0.013 & 0.233\pm0.005 & 0.018\pm0.003 & 1.376\pm0.044 \\
Seo16 & 0.103\pm0.003 & 0.093\pm0.004 & 0.095\pm0.005 & 0.022\pm0.004 & 0.007\pm0.002 & 0.360\pm0.015 & 0.267\pm0.004 & 0.015\pm0.003 & 1.342\pm0.043 \\
Seo17 & 0.261\pm0.008 & 0.196\pm0.006 & 0.147\pm0.005 & 0.034\pm0.002 & 0.024\pm0.002 & 0.200\pm0.008 & 0.137\pm0.004 & 0.022\pm0.004 & 2.203\pm0.070 \\
\enddata
\tablecomments{(a) From \cite{Scibelli2020}, except Sci7-1 and Sci7-2, which are from this work. (b) From \cite{Ambrose2021}, except Sci7-1 and Sci7-2, which are from this work, and Seo14, which was reobserved in this work. The uncertainties in \cite{Ambrose2021} were an order of magnitude too low due to an error in the placement of the decimal point, which has been corrected here.
}
\end{deluxetable*}
\end{longrotatetable}


\section{Data Analysis} \label{sec:data}

\subsection{pNH$_3$ Multi-component Fitting}\label{subsec:nestfit}

pNH$_3$ (1,1) and (2,2) observations that were originally presented in \cite{Seo2015} were re-analyzed using \texttt{NestFit} \citep{brian_svoboda_2021_4470028}, a Bayesian multi-component hyperfine fitting code developed by co-author B. Svoboda\footnote{\url{https://github.com/autocorr/nestfit}}.  
\texttt{NestFit} determines the number of velocity components and estimates their model parameters in a statistically rigorous and automated way using Bayes factors computed with nested sampling Monte Carlo \citep{Feroz08, Feroz09}.
The velocity-coherent filaments are identified from the maximum likelihood velocity from the posterior distributions (Steffes et al., in prep.; see Figure \ref{fig:B10_NH2_map}).
To calculate these values, we used posterior distributions assuming a two velocity-component model to find the maximum likelihood line-of-sight velocities for each pixel.
In the cases where the Bayes Factor comparing zero and one components suggests a single velocity-component model works best ($\ln(B_{0, 1}) \geq 11$) compared to the Bayes Factor comparing one and two components ($\ln(B_{1,2}) < 11$), we select the maximum likelihood value in the velocity posterior distribution for that pixel. 
In Figure \ref{fig:B10_NH2_map}, the single velocity component fits are most common near the central dust continuum defined spines of filaments Fil10 and Fil06, along with the southern region of Fil08. 
There are other areas of the B10 region where the two velocity-component model is a better fit, namely the region in between Fil10 and Fil06.
For example, there is a slight overlap at the southern edge of Fil06 and the redder gas associated with Fil08. 
However, these are very few pixels in these cases and the velocities shown in Figure \ref{fig:B10_NH2_map} correspond with the much higher signal-to-noise ratio components of pNH$_3$ gas compared to the secondary component.
In those cases there are two maximum likelihood velocities for each model, one at a redder velocity and one at a bluer velocity. 
The single velocity plotted in Figure \ref{fig:B10_NH2_map} was the better constrained velocity with the higher maximum likelihood.
However, in diffuse regions where the single-to-noise ratio was lower, this often led to the velocity posterior distributions being less well-constrained.
To select those velocities, we also compared the two-component posterior distributions with the single-component posterior distributions to find a dominant component and compared with the surrounding pixels to identify outliers likely caused by the relatively higher noise.
For the purposes of this analysis, we only use single velocities, as seen in Figure \ref{fig:B10_NH2_map}, which shows the velocity for the higher intensity, denser gas.
A more in-depth kinematic analysis of the multi-component gas can be found in Steffes et al. (in prep.).

\subsection{Calculating Beam-Averaged $N(\text{H}_2)$} \label{subsec:beam-avg_NH2}

We calculated the average column density of H$_2$ $\langle N(\text{H}_2) \rangle$ (Table \ref{tab:beam_avg_H2}) of each core within the FWHM beam size of each molecule's observed transition(s) on the ARO 12m telescope.
Maps of the B10 region of Taurus in optical depth and optical depth error determined at 1 THz were derived from \textit{Herschel Space Observatory} continuum images \cite{Singh2022}.
We converted these maps to H$_2$ column density (Figure \ref{fig:B10_NH2_map}) and column density error maps using (see Equation 73 in \citealt{Shirley2026})
\begin{equation}\label{eqn:NH2}
    N(\text{H}_2) = \frac{\tau_\nu}{\kappa_\nu r_{dg} \mu_{\mathrm{H}_2} m_\mathrm{H}} 
\end{equation}
where $\kappa_\nu = 10$ cm$^2$/g is the dust mass opacity at 1 THz  \citep{Singh2022}, $r_{dg} = 0.01$ is the dust+ice-to-gas mass ratio  \citep{Patra2025}, $\mu_{\mathrm{H}_2} = 2.809$ is the mean molecular weight for H$_2$ gas \citep{Evans2022, Shirley2026}, $m_\mathrm{H}$ is the mass of the Hydrogen atom, and $N$(H$_2$) is the column density of H$_2$.
We used \texttt{Photutils} \citep{photutils} to define apertures at the position of each core with diameters corresponding to the ARO 12m FWHM main beam size of the observed transition of each molecule (Table \ref{tab:obs_transitions}). 
For HDCO and CH$_3$OH, which had multiple transitions that were very close in beam size (differences of roughly $2^{\prime\prime}$ and $< 1^{\prime\prime}$ respectively), we used an average beam size for their transitions ($45.8^{\prime\prime}$ and $62.3^{\prime\prime}$ respectively).
We also used the average of the beam sizes of oH$_2$CO and pH$_2$CO ($42.1^{\prime\prime}$), as they were also close in size ($<2^{\prime\prime}$ apart).

We then determined the average H$_2$ column density in the aperture using 
\begin{equation}\label{eqn:6}
    \langle N(\text{H}_2) \rangle = \frac{4\theta_\mathrm{pix}^2}{\pi\theta_\mathrm{aper}^2}\sum_{\text{aper}}w_iN_i(\text{H}_2)
\end{equation}
where $\theta_{pix}$ is the pixel scale of the map, $\theta_\mathrm{aper}$ is the diameter of the circular aperture which is equal to the FWHM main beam size, the weight $w_i$ is the fractional overlap of the $i$-th pixel with the aperture, and  $N_i(\text{H}_2)$ is the column density of H$_2$ at that same pixel. 
The weighted sum of the column densities was computed using the \texttt{aperture\_photometry} function within \texttt{photutils}. 
 We then set the uncertainty on $\langle N(\text{H}_2) \rangle$ to be the weighted mean of the uncertainties of the pixels within the aperture on the $N$(H$_2$) error map, again weighted by the fractional overlaps $w_i$.

\subsection{Calculating Molecular Column Densities}\label{subsec:coldens_calcs}

We calculate molecular column densities using two methods: \texttt{RADEX} \citep{vanderTak2007} and the CTEX method as described in \cite{Mangum2015}. 
We briefly describe the methodology here and the details for each molecular calculation are given in Appendix \ref{ap:columndensityappexdix}.

\begin{figure}[ht!]
\centering
\includegraphics[scale=0.1,trim={0 20cm 0 15cm},clip]{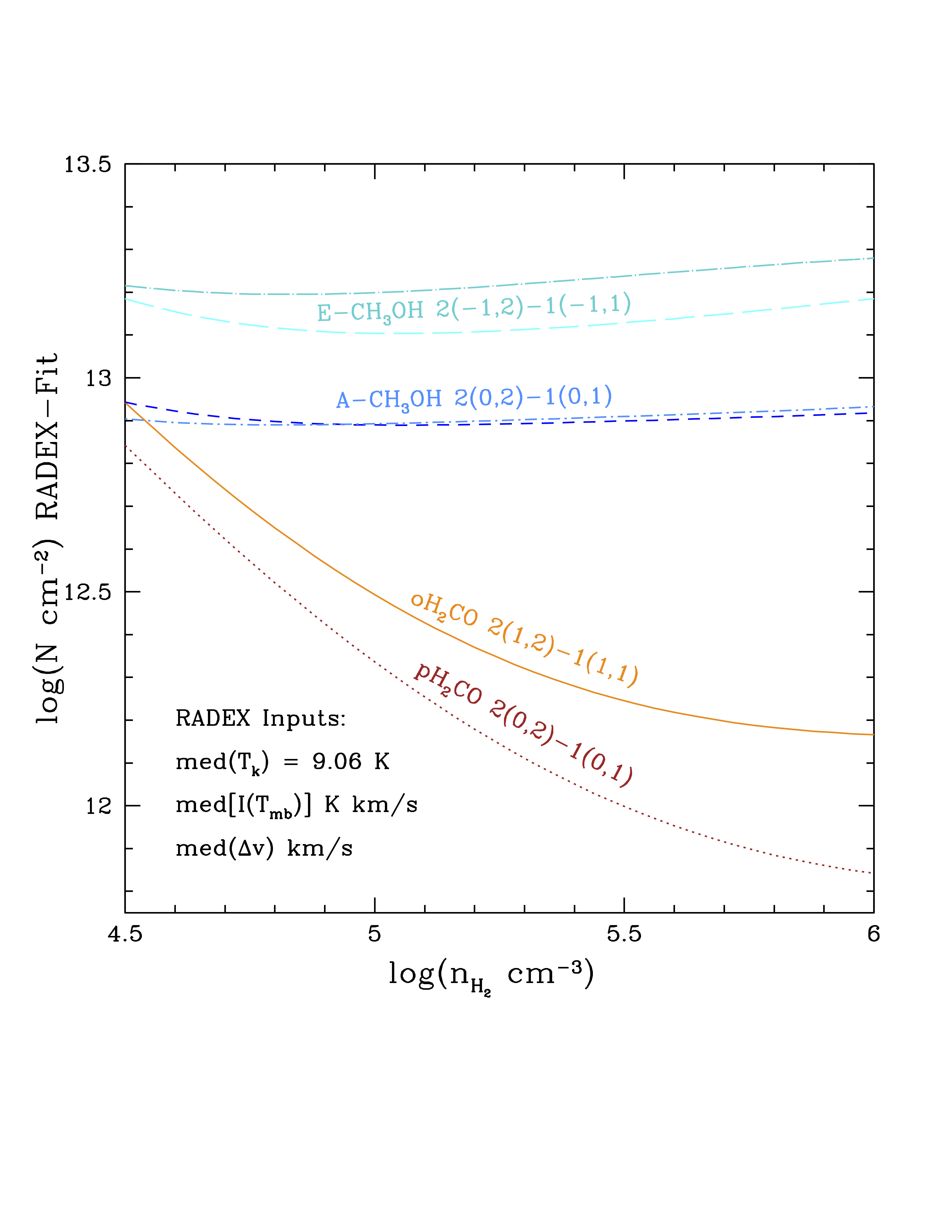}
\caption{The column density sensitivity is plotted for observed H$_2$CO and CH$_3$OH transitions to the input H$_2$ volume density.  For each input H$_2$ density, a \texttt{RADEX} model was run to calculate the column density that matches the median integrated intensity of the observed sources in each transition.  All models also assumed the median gas kinetic temperature of $9.06$ K and the median FWHM line width of each transition.  Collision rates with pH$_2$ for this plot were calculated by \cite{Wiesenfeld2013} for H$_2$CO (oH$_2$CO orange solid line, pH$_2$CO red dotted line) and by \cite{Rabli2010} (A-CH$_3$OH dark blue dashed line, E-CH$_3$OH cyan dashed line) and \cite{Dagdigian2024} (A-CH$_3$OH light blue dash-dotted line, E-CH$_3$OH turquoise dash-dotted line) for CH$_3$OH (see Appendix \ref{subsubsec:meth_coldens} for a comparison of the two collision rate calculations).
\label{fig:Radex_Nsensitivity}}
\end{figure}

\texttt{RADEX} in a non-LTE radiative transfer code that uses the escape probability formalism for a constant density, constant temperature sphere to perform a coupled statistical equilibrium and radiative transfer calculation \citep{vanderTak2007}.
The inputs for a single \texttt{RADEX} calculation are the H$_2$ density, gas kinetic temperature, the molecular collision rates, the observed FWHM line width, and the assumed column density of the molecule.
\texttt{RADEX} outputs the level populations, excitation temperature, and integrated intensity of the chosen spectral lines.
A grid of \texttt{RADEX} models, varying the inputs, may then be calculated to find the column densities for which the model integrated intensity matches the observed integrated intensity within the uncertainties.
Since the B10 sources have density and temperature gradients \citep{Scibelli2023}, the sensitivity of the \texttt{RADEX} calculation to these inputs must be assessed and suitable average densities and average temperatures need to be calculated.
For example, the column density determined from calculation of the 96.7 GHz A/E-CH$_3$OH transitions is a weak function of the core density input into the calculation \citep{Scibelli2020}.
If the input density is varied by more than an order of magnitude from $\log n_{\rm{H}_2} = 4.5 - 6.0$ with $ n_{\rm{H}_2}$ in units of cm$^{-3}$, the best-fitted A/E-CH$_3$OH column density varies by less than $10$\% for cores in B10 (Figure \ref{fig:Radex_Nsensitivity}).
This statement is not true for the o/pH$_2$CO 140 and 145 GHz transitions where the best-fitted column density can vary by factors of $6$ for input density in cm$^{-3}$ varying between $\log n_{\rm{H}_2} = 4.5 - 6.0$, resulting in a degeneracy between the input density and the column density. 
This makes using \texttt{RADEX} impractical for o/pH$_2$CO column density calculations since it is unclear which density is the correct one to pick for each core.
It is also only possible to perform \texttt{RADEX} calculations if molecular collision rates with pH$_2$ are published, which they are not for CH$_2$DOH, HDCO, and D$_2$CO at this time of submission of this paper.
 
A second approach to calculating column densities is the constant excitation temperature (CTEX) approximation, a statistical equilibrium approximation in which all transitions of a molecule are assumed to have the same excitation temperature populated by a Boltzmann distribution \citep{Mangum2015}.
There are different approaches to calculating the column density depending on whether the emission is optically thin ($\tau < 0.2$) or not ($\tau \geq 0.2$) and whether or not the optical depth $\tau$ is able to be accurately determined ($\tau/\sigma_{\tau} > 3$, i.e., from high signal-to-noise observations of spectrally-resolved hyperfine transitions such as our N$_2$H$^+$ 1-0 observations).
If the emission is optically thin or when the optical depth can be accurately determined, then the column density $N_{tot}$ is given by
\begin{equation}\label{eqn:CTEX_Ntot}
    \begin{split}
        \text{Case 1: } & \text{if $\tau_{tot} < 0.2$:} \\
        N_{tot} &= \frac{8\pi k\nu^2}{hg_uA_{ul}c^3}\frac{J_\nu(T_{ex}) Q(T_{ex})e^{E_u/kT_{ex}}}{f[J_\nu(T_{ex}) - J_\nu(T_{bg})]}\,I \\
        \text{Case 2: } &\text{if $\tau_{tot}$ well constrained:} \\
        N_{tot} &= \frac{4\pi^{3/2}k\nu^2}{\sqrt{\text{ln}(2)}\,hg_uA_{ul}c^3} J_{\nu}(T_{ex})Q(T_{ex})e^{E_u/kT_{ex}} \tau_{tot} \Delta v,
    \end{split}
\end{equation}
where $\nu$ is the frequency of the transition, $k$ is the Boltzmann constant, $h$ is Planck's constant, $c$ is the speed of light, $\Delta v$ is the linewidth, $I$ is the integrated intensity, $J_\nu(T) = (h\nu/k) [e^{h\nu/kT} - 1]^{-1}$ is the Planck function in temperature units calculated for either the excitation temperature ($T_{ex}$) or the background radiation temperature ($T_{bg} = 2.725$ K; \citealt{Fixsen2009}), $Q(T_{ex})$ is the partition function, $g_u$ is statistical weight of the upper level, $A_{ul}$ is the spontaneous emission coefficient of the given transition, $E_u$ is the rotational energy of the upper level, and $f$ is the source-beam filling fraction, and $\tau_{tot}$ is the total optical depth \citep{Caselli2022ion,Mangum2015}.
The criteria of $\tau_{tot} < 0.2$ is chosen such that the correction factor to the column density for optical depth, $C_{\tau} = \tau/[1 - \exp{(-\tau)}]$ \citep{Goldsmith1999}, is a less than $10$\% correction to the optically thin (Case 1) column density.  
If the optical depth is not well constrained and it is unclear if the emission is optically thin, then an alternative approach (Case 3) uses the radiative transfer equation to calculate a model integrated intensity, $I_{\rm{model}}$, to find combinations of $N_{tot}$ and $T_{ex}$ that match the observed intensity,
\begin{equation}\label{eqn:Imodel}
    \begin{split}
    I_{\rm{model}} = {} & f[J_\nu(T_{ex}) - J_\nu(T_{bg})] \\ 
    {} & \times \int 1 - \exp{\left[ -\frac{\sqrt{\text{ln}(2)}c^3}{4\pi^{3/2}\nu^3}\frac{A_{ul}g_uN_{tot}}{\Delta v}\right.}  \\ 
     & \times \left.\frac{e^{-E_u/kT_{ex}}(e^{h\nu/kT_{ex}}-1)}{ Q(T_{ex})}\phi(v, \Delta v)\right] d\nu \;\;, \\
    \end{split}
\end{equation}
where $\phi(v, \Delta v) = \exp{[-4\text{ln}(2)((v-v_{\rm{LSR}})/\Delta v)^2]}$ for a Gaussian profile\footnote{For a line profile with hyperfine structure, then we can write $\phi(v, \Delta v) = \sum_i R_i \exp{[-4\text{ln}(2)((v-v_{\rm{LSR}} - \delta v_i)/\Delta v)^2]}$ where $R_i$ is the relative intensity ($\sum_i R_i$ = 1) and $\delta v_i$ is the velocity offset of the i$^{\rm{th}}$ hyperfine transition.}.
The integral may be evaluated numerically using Simpson's Rule\footnote{The optical depth criteria of $\tau_{tot} < 0.2$ is strict for Case 1 because the derivation of this equation assumes that the equivalent width of the emission line is $W_v = \int 1 - e^{-\tau(v)} dv \approx \int \tau(v) dv$ by keeping only the linear order terms of the Maclaurin series expansion of the exponential \citep{Mangum2015}.  
Higher order terms in the Maclaurin expansion could be calculated to increase the $\tau_{tot}$ criteria (e.g., see Section 9.2 of \citealt{Draine2011}); however,  Equation \ref{eqn:Imodel} (Case 3) obviates higher order approximations by calculating $W_v$ numerically without the Maclaurin series approximation.}.
Our observations of o/pH$_2$CO, HDCO, pD$_2$CO, CH$_2$DOH, and N$_2$H$^+$ span these three cases.
Furthermore, since pNH$_3$ has multiple rotational energy level ladders that are not radiatively connected, a modified version of the CTEX approximation is used with two temperatures ($T_{ex}$ for the inversion transitions and $T_{rot}$ for the ratio of populations across different K ladders).
We show the calculation of the column densities of HDCO and o/pH$_2$CO below and the details for column density calculations of the other molecules may be found in Appendix \ref{ap:columndensityappexdix}.

\begin{figure*}[ht!]
\centering
\includegraphics[scale=0.55,trim={0 0 0 1cm}]{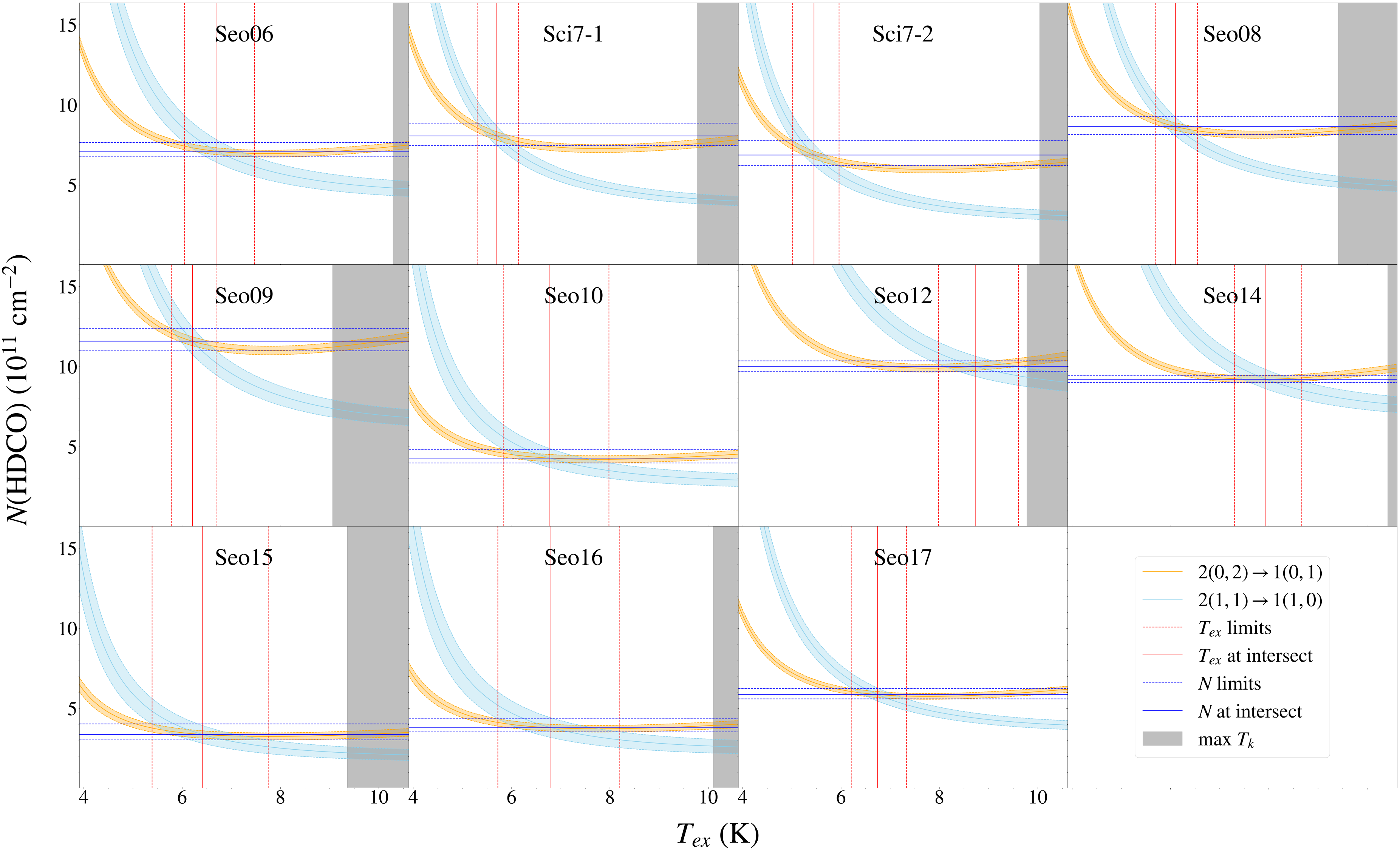}
\caption{The column density of HDCO is plotted as a function of the excitation temperature in the CTEX approximation.  The orange curves and blue curves represent the observations of the $2(0,2) \rightarrow 1(0,1)$ and $2(1,1) \rightarrow 1(1,0)$ transitions respectively.  The overlap of the curves determines the range of column density and $T_{ex}$ for each core (shown as red and blue lines respectively).  The shaded grey regions show $T_{ex} > T_{k}$.
\label{fig:HDCO_CTEX}}
\end{figure*}

The optically thin CTEX method (Case 1) may be applied to the HDCO $J (K_a, K_c) = 2(0,2) \rightarrow 1(0,1)$ (128.9 GHz) and $J (K_a, K_c) = 2(1,1) \rightarrow 1(0,1)$ (134.3 GHz) transitions.
This pair of transitions are separated by $\Delta E_u/k = 8.35$ K and therefore can provide simultaneous constraints on the excitation temperature and column density (see Figure \ref{fig:EnergyLevels}).
A curve of possible total column densities $N_{tot}$ at different excitation temperatures is generated from Case 1 of Equation \ref{eqn:CTEX_Ntot} for each observed transition. 
We assumed the filling fraction $f=1$, since the beam sizes of the 12m for each transition are similar enough 
that the difference in filling fraction is negligible.
From the intersections of the $N_{tot}$ vs. $T_{ex}$ curves with errors for both transitions, the central value and upper and lower limits for both column density and excitation temperature (Figure \ref{fig:HDCO_CTEX}) were determined (Tables \ref{tab:coldens} and \ref{tab:temps_tau_thetas}).
Case 2 of Equation \ref{eqn:CTEX_Ntot} was used to check that the optically thin assumption was valid by setting $\tau_{tot} = 0.2$ and finding that $N(\rm{HDCO})_{\rm{Case 1}} <$  $N(\tau_{tot} = 0.2)_{\rm{Case 2}}$ for all sources.
The upper errorbar value of $T_{ex}$ for each core was less than the core's average gas kinetic temperature reported in \cite{Scibelli2023} indicating that the HDCO transitions are sub-thermally populated.

\begin{deluxetable*}{lCCCCCCCCC}
\tablewidth{0pt} 
\tablecaption{Column Densities\label{tab:coldens}}
\tablehead{
\colhead{Core} & \colhead{$N(\mathrm{oH_2CO})$} & \colhead{$N(\mathrm{pH_2CO})$} & \colhead{$N(\mathrm{HDCO})$} & \colhead{$N(\mathrm{pD_2CO})$} & \colhead{$N$(A-CH$_3$OH)} & \colhead{$N$(E-CH$_3$OH)} & \colhead{$N(\mathrm{CH_2DOH})$\tablenotemark{a}} & \colhead{$N(\mathrm{N_2H^+})$} & \colhead{$N(\mathrm{pNH_3})$} \\
\colhead{} & \colhead{($10^{12}$ cm$^{-2}$)} & \colhead{($10^{12}$ cm$^{-2}$)} & \colhead{($10^{11}$ cm$^{-2}$)} & \colhead{($10^{10}$ cm$^{-2}$)} & \colhead{($10^{13}$ cm$^{-2}$)} &  \colhead{($10^{13}$ cm$^{-2}$)} &  \colhead{($10^{12}$ cm$^{-2}$)} & \colhead{($10^{12}$ cm$^{-2}$)} & \colhead{($10^{14}$ cm$^{-2}$)}
} 
\startdata
Seo06 & 1.90^{+0.27}_{-0.19} & 0.92^{+0.15}_{-0.11} & 7.11^{+0.54}_{-0.35} & 6.24^{+1.69}_{-1.45} & 1.41^{+0.10}_{-0.08} & 1.59^{+0.10}_{-0.09} & 1.14^{+0.29}_{-0.26} & 5.06^{+0.25}_{-0.24} & 1.11^{+0.22}_{-0.20} \\
Sci7-1 & 2.30^{+0.30}_{-0.23} & 1.03^{+0.18}_{-0.14} & 8.06^{+0.80}_{-0.60} & 8.93^{+2.25}_{-1.93} & 0.77^{+0.05}_{-0.04} & 0.86^{+0.05}_{-0.04} & 2.35^{+0.62}_{-0.56} & 4.56^{+0.22}_{-0.21} & 1.13^{+0.16}_{-0.14} \\
Sci7-2 & 2.21^{+0.38}_{-0.28} & 1.17^{+0.28}_{-0.20} & 6.87^{+0.90}_{-0.67} & 7.95^{+2.28}_{-1.88} & 0.56^{+0.04}_{-0.03} & 0.60^{+0.04}_{-0.03} & 1.55^{+0.52}_{-0.47} & 2.95^{+0.14}_{-0.14} & 0.60^{+0.16}_{-0.14} \\
Seo08 & 1.59^{+0.20}_{-0.16} & 0.90^{+0.13}_{-0.10} & 8.65^{+0.64}_{-0.49} & 18.73^{+2.70}_{-2.34} & 0.71^{+0.04}_{-0.04} & 0.80^{+0.05}_{-0.04} & 2.21^{+0.45}_{-0.41} & 5.83^{+0.14}_{-0.14} & 1.89^{+0.13}_{-0.13} \\
Seo09 & 1.65^{+0.26}_{-0.20} & 0.90^{+0.14}_{-0.10} & 11.59^{+0.79}_{-0.60} & 19.97^{+2.77}_{-2.40} & 1.22^{+0.08}_{-0.07} & 1.38^{+0.09}_{-0.08} & 3.11^{+0.51}_{-0.47} & 7.62^{+0.16}_{-0.16} & 2.59^{+0.19}_{-0.19} \\
Seo10 & 3.94^{+0.93}_{-0.53} & 1.45^{+0.40}_{-0.24} & 4.29^{+0.55}_{-0.30} & 3.47^{+1.46}_{-1.18} & 1.17^{+0.11}_{-0.09} & 1.28^{+0.12}_{-0.10} & 2.62^{+0.68}_{-0.60} & 1.85^{+0.60}_{-0.50} & 0.58^{+0.59}_{-0.35} \\
Seo12 & 4.33^{+0.26}_{-0.19} & 1.41^{+0.12}_{-0.10} & 10.02^{+0.34}_{-0.30} & 15.80^{+1.97}_{-1.80} & 1.50^{+0.11}_{-0.09} & 1.75^{+0.11}_{-0.09} & 1.85^{+0.27}_{-0.24} & 7.63^{+0.32}_{-0.31} & 2.11^{+0.19}_{-0.18} \\
Seo14 & 4.23^{+0.47}_{-0.37} & 1.79^{+0.19}_{-0.15} & 9.22^{+0.25}_{-0.21} & 8.03^{+1.39}_{-1.28} & 1.87^{+0.17}_{-0.13} & 2.05^{+0.16}_{-0.13} & 1.03^{+0.27}_{-0.24} & 3.39^{+0.46}_{-0.42} & 0.76^{+0.38}_{-0.29} \\
Seo15 & 0.87^{+0.20}_{-0.12} & 0.32^{+0.10}_{-0.06} & 3.38^{+0.66}_{-0.34} & 5.07^{+2.13}_{-1.55} & 0.77^{+0.06}_{-0.05} & 0.81^{+0.06}_{-0.05} & 1.71^{+0.40}_{-0.36} & 4.00^{+0.24}_{-0.23} & 0.92^{+0.24}_{-0.20} \\
Seo16 & 0.53^{+0.10}_{-0.05} & 0.24^{+0.06}_{-0.04} & 3.80^{+0.56}_{-0.26} & 4.55^{+1.94}_{-1.48} & 0.83^{+0.06}_{-0.05} & 0.93^{+0.06}_{-0.06} & 1.40^{+0.35}_{-0.31} & 4.11^{+0.42}_{-0.40} & 1.25^{+0.21}_{-0.18} \\
Seo17 & 1.39^{+0.14}_{-0.11} & 0.52^{+0.06}_{-0.05} & 5.87^{+0.37}_{-0.27} & 14.86^{+2.12}_{-1.84} & 0.46^{+0.03}_{-0.03} & 0.47^{+0.03}_{-0.03} & 2.10^{+0.47}_{-0.43} & 7.03^{+0.39}_{-0.37} & 2.61^{+0.20}_{-0.19} \\
\enddata
\tablecomments{(a) Column density values are calculated using the JPL catalog entry for CH$_2$DOH for which the parameters are listed in Table \ref{tab:obs_transitions}.  If the Lille Spectroscopic Database catalog entry from April 2026 is used instead, then all column densities should be multiplied by a factor of $0.87$ (see Appendix \ref{subsubsec:CTEX} for details).}
\end{deluxetable*}

\begin{deluxetable*}{lCCCCCCL}
\tablewidth{0pt} 
\tablecaption{Relevant Temperatures, Total Optical Depths, and Source Sizes\label{tab:temps_tau_thetas}}
\tablehead{
\colhead{} & \colhead{HDCO} & \multicolumn{2}{c}{A-CH$_3$OH} & \colhead{E-CH$_3$OH} & \multicolumn{3}{c}{N$_2$H$^+$}\\
\colhead{Core} & \colhead{$T_{ex}$\tablenotemark{a}} & \colhead{$T_{ex}$\tablenotemark{b}} & \colhead{$\langle T_k \rangle_\mathrm{beam}$\tablenotemark{c}} & \colhead{$T_{ex}$} & \colhead{$T_{ex}$} & \colhead{$\tau_\mathrm{tot}$} &
\colhead{$\theta_s$} \\
\colhead{} & \colhead{(K)} & \colhead{(K)} & \colhead{(K)} & \colhead{(K)} & \colhead{(K)} &  \colhead{} &  \colhead{($^{\prime\prime}$)} 
} 
\startdata
Seo06 & 6.71^{+0.76}_{-0.66} & 8.21^{+0.53}_{-0.54} & 9.22\pm0.66 & 7.72^{+0.50}_{-0.50} & 4.50\pm0.05 & 5.133\pm0.139 & 128^{+33}_{-20} \\
Sci7-1 & 5.70^{+0.44}_{-0.40} & 7.97^{+0.47}_{-0.47} & 9.07\pm0.58 & 7.35^{+0.42}_{-0.42} & 6.23\pm0.10 & 4.168\pm0.080 & 80^{+9}_{-7} \\
Sci7-2 & 5.45^{+0.51}_{-0.44} & 7.80^{+0.57}_{-0.57} & 9.09\pm0.72 & 7.01^{+0.47}_{-0.47} & 6.63\pm0.19 & 2.335\pm0.002 & 94^{+15}_{-11} \\
Seo08 & 6.10^{+0.45}_{-0.41} & 7.98^{+0.43}_{-0.42} & 9.07\pm0.52 & 7.38^{+0.38}_{-0.38} & 5.87\pm0.08 & 5.301\pm0.006 & 103^{+14}_{-11} \\
Seo09 & 6.21^{+0.48}_{-0.43} & 7.73^{+0.38}_{-0.38} & 8.60\pm0.46 & 7.32^{+0.36}_{-0.36} & 5.79\pm0.06 & 6.285\pm0.001 & 98^{+12}_{-9} \\
Seo10 & 6.78^{+1.20}_{-0.95} & 7.41^{+0.55}_{-0.55} & 8.59\pm0.70 & 6.69^{+0.45}_{-0.45} & 4.72\pm0.35 & 1.879\pm0.334 & 72^{+13}_{-9} \\
Seo12 & 8.74^{+0.87}_{-0.76} & 8.15^{+0.58}_{-0.58} & 9.13\pm0.72 & 7.71^{+0.56}_{-0.55} & 5.31\pm0.07 & 5.911\pm0.116 & 83^{+9}_{-7} \\
Seo14 & 7.94^{+0.72}_{-0.64} & 7.89^{+0.65}_{-0.65} & 8.96\pm0.81 & 7.28^{+0.58}_{-0.57} & 6.33\pm0.40 & 1.917\pm0.051 & 68^{+8}_{-6} \\
Seo15 & 6.41^{+1.34}_{-1.02} & 7.73^{+0.53}_{-0.53} & 8.88\pm0.65 & 7.06^{+0.45}_{-0.45} & 5.73\pm0.18 & 2.984\pm0.016 & 73^{+9}_{-7} \\
Seo16 & 6.80^{+1.40}_{-1.08} & 7.94^{+0.55}_{-0.55} & 9.06\pm0.68 & 7.31^{+0.49}_{-0.49} & 4.40\pm0.08 & 4.599\pm0.298 & 105^{+24}_{-16} \\
Seo17 & 6.74^{+0.59}_{-0.52} & 7.94^{+0.44}_{-0.44} & 9.06\pm0.54 & 7.32^{+0.41}_{-0.40} & 5.52\pm0.07 & 6.253\pm0.190 & 82^{+9}_{-7} \\
\enddata
\tablecomments{(a) Assumed to also be $T_{ex}$ for o/pH$_2$CO and pD$_2$CO. (b) Assumed to also be $T_{ex}$ for CH$_2$DOH. (c) Average $T_k$ was determined from the \texttt{NestFit} re-analysis of  pNH$_3$ (1,1) and (2,2) observations (Steffes et al., in prep.) within the average CH$_3$OH beam size and was used to calculate $N$(A-CH$_3$OH) and $N$(E-CH$_3$OH) with \texttt{RADEX}. Details are found in Appendix \ref{ap:columndensityappexdix}.}
\end{deluxetable*}

The column density calculations for o/pH$_2$CO required the model intensity approach (Equation \ref{eqn:Imodel}) because we only had observations of one transition of each spin species (Figure \ref{fig:EnergyLevels}), and they were potentially optically thick with no reliable way to determine the optical depth in the 140.8 GHz or 145.6 GHz transitions (Case 3).
The column density was determined by calculating $I_{\rm{model}}$ for each combination of $N_{tot}$ and $T_{ex}$ within a predefined grid and keeping only those combinations that produced an intensity within $1\sigma$ of the observed intensity.
For each core's grid, the limits on $T_{ex}$ were the lower and upper limits on the HDCO $T_{ex}$ of that core derived from its uncertainties and the lower and upper limits on $N_{tot}$ were $10^{11}$ cm$^{-2}$ and $10^{13}$ cm$^{-2}$.
We assumed that $T_{ex}$ was the same as for HDCO in each core to calculate the column density of o/pH$_2$CO with uncertainties (Table \ref{tab:coldens}).
We calculated the column density for oH$_2$CO and pH$_2$CO separately and added them together for a total H$_2$CO column density.
The mean and standard deviation of the ortho to para ratio for H$_2$CO across the B10 sample is $2.3\pm0.4$, which is significantly below the statistical ratio of 3. 
Seo12 has the closest ratio at 3.1; the remaining cores are all at or below 2.7.
The mean and standard deviation of published measurements toward low-mass starless cores is $2.2\pm0.6$ \citep{Minh1995,Dickens1999,Punanova2025}, which is similar to the B10 core mean.

\subsection{Quantifying Correlations} \label{subsec:correl}

We quantify 109 correlations among column densities, core physical parameters, deuteration ratios, and other molecular abundance ratios using the Spearman rank correlation coefficient $r_s$ (see Section 14.7 in \citealt{Zwillinger2000}).
This method takes two samples of the same size and independently ranks the observations within each from smallest to largest.
The Spearman rank correlation cofficient is then calculated as
\begin{equation} \label{eq:Spearmanr}
    r_s = \frac{
    n\sum_{i=1}^n u_iv_i - \left(\sum_{i=1}^n u_i  \right) \left(\sum_{i=1}^n v_i  \right)
    }{
    \sqrt{
    \left[ n\sum_{i=1}^n u_i^2 - \left(\sum_{i=1}^n u_i  \right)^2 \right]
    \left[ n\sum_{i=1}^n v_i^2 - \left(\sum_{i=1}^n v_i  \right)^2 \right]
    }
    }
\end{equation}
where $n$ is the sample size, $u_i$ is the rank of the $i$-th observation from the first sample, and $v_i$ is the rank of the $i$-th observation for the second sample (from Section 14.7 in \citealt{Zwillinger2000}).
The numerator is the covariance of $u_i$ and $v_i$, and the denominator is the product of the standard deviations of $u_i$ and $v_i$.
The value of $r_s$ ranges from -1 to 1, with 1 indicating a perfect positive correlation, -1 a perfect negative correlation (anti-correlation), and 0 a complete lack of correlation.
The primary advantage of using $r_s$ is that in comparing the ranks of the observations, rather than their raw values, it makes no assumptions about the linearity of the relationship between the observations.
Any monotonically increasing (or decreasing) relationship would produce the same value of $r_s$ as a linear relationship as long as the number and ranks of the observations are the same.
This is ideal for our case, as it requires no assumptions about the precise form of any potential relationship between quantities.

\begin{figure}[h!]\centering
\includegraphics[width=\linewidth]{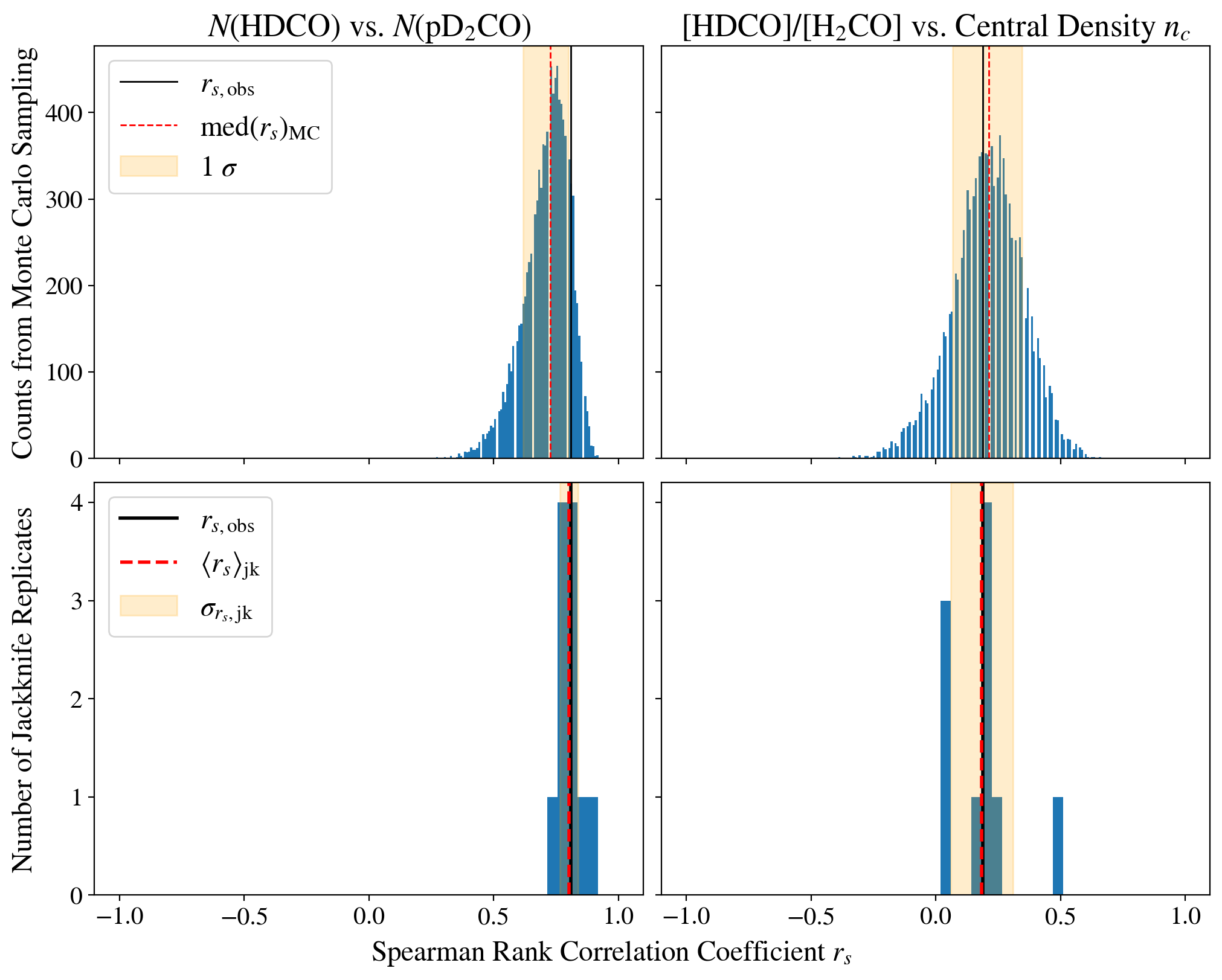}
\caption{Example histograms for Monte Carlo sampling (top row) and jackknife resampling (bottom row) for a good correlation, $N$(HDCO) vs. $N$(pD$_2$CO) with $r_s = 0.81$ (left column) and for a poor correlation, $n_c$ vs. [HDCO]/[H$_2$CO] with $r_s = 0.13$ (right column).
\label{fig:correl_histograms_ex}}
\end{figure}

One issue with interpreting $r_s$ is that it does not inherently account for uncertainties in data.
To estimate the distribution of potential $r_s$ values for each pair of quantities we compared, we used a Monte Carlo based method to create randomly sampled distributions based on the uncertainties in the observed data, for which we then calculated the $r_s$ values.
For each data point we created a probability density function (PDF) centered at the observed value with its width determined by the errorbars of the datapoint.
We tested different choices for the shape of the PDF, accounting for errorbars that are not symmetric, as is discussed in Appendix \ref{ap:MC_PDFS}.
In most cases, the choice of PDF shape did not significantly affect the resulting $r_s$ distribution.
Thus, for the best balance between accuracy and simplicity, we chose the dimidiated Gaussian \citep{Barlow2026}:
\begin{equation}\label{eq:dimid_gauss}
    p(x) = \cstpl{1.1 cm}
    \begin{array}{cc}
        \frac{1}{\sqrt{2\pi}\sigma^-}e^{-\frac{(x-\rm{med}(x))^2}{2(\sigma^-)^2}}, & x < med(x) \\
        \frac{1}{\sqrt{2\pi}\sigma^+}e^{-\frac{(x-\rm{med}(x))^2}{2(\sigma^+)^2}}, & x > med(x)
    \end{array} 
\end{equation}
where $\rm{med}(x)$ is the median of the distribution, which we set to be the observed data. 
This PDF essentially uses the left half of a Gaussian with $\sigma=\sigma^-$ and the right half of a Gaussian with $\sigma=\sigma^+$.
Using the PDFs, we then randomly drew a value for each data point $10^4$ times, creating $10^4$ pairs of randomly sampled datasets for the two quantities.
We then calculated $r_s$ for each pair of datasets to create a distribution of $10^4$ $r_s$ values for statistical analysis.
Example Monte Carlo sampling distributions of $r_s$ for good and bad correlations may be found in Figure \ref{fig:correl_histograms_ex}.
Note that the regular gaps in the Monte Carlo histograms are likely due to the fact that the ranks of the data are discrete, meaning $r_s$ can only take on certain values as a result.
We used the median (med($r_s)_\mathrm{MC}$) rather than the mean of the $r_s$ distribution due to its lower susceptibility to outliers.
To calculate the uncertainty in $r_s$, we used the 1-$\sigma$ confidence intervals around med($r_s)_\mathrm{MC}$.
Finally, we also calculated the skewness ($\gamma_{1,r_s}$), also known as the Fisher's moment coefficient of skewness, of the distribution (see Section 2.2.24.1 in \cite{Zwillinger2000})
with $\gamma_{1,r_s} = 0$ indicating a symmetric distribution, $\gamma_{1,r_s} > 0$ representing a distribution leaning to the left with a tail extending to the right, and vice versa for $\gamma_{1,r_s} < 0$. 
These statistics for all comparisons may be found in Tables \ref{tab:MC_jk_N}, \ref{tab:MC_jk_RD}, and \ref{tab:MC_jk_molratios}.

The other potential issue with our correlation analysis arises from our small sample size of 11 sources.
We estimate the possible variance in $r_s$ due to the influence of individual cores through jackknife resampling.
In doing so, for each pair of quantities, we calculated $r_s$ multiple times, each time removing one core from our sample beforehand.
This gave us a distribution of $r_s$ (the jackknife replicates), of which we then took the mean ($\langle r_s \rangle_\mathrm{jk}$) and standard deviation ($\sigma_{r_s,\mathrm{jk}}$).
Example jackknife distributions of $r_s$ for good and bad correlations may again be found in Figure \ref{fig:correl_histograms_ex}.
The jackknife resampling statistics for all comparisons may be found in Tables \ref{tab:MC_jk_N}, \ref{tab:MC_jk_RD}, and \ref{tab:MC_jk_molratios}.

\begin{figure*}
    \centering
\includegraphics[width=1\linewidth]{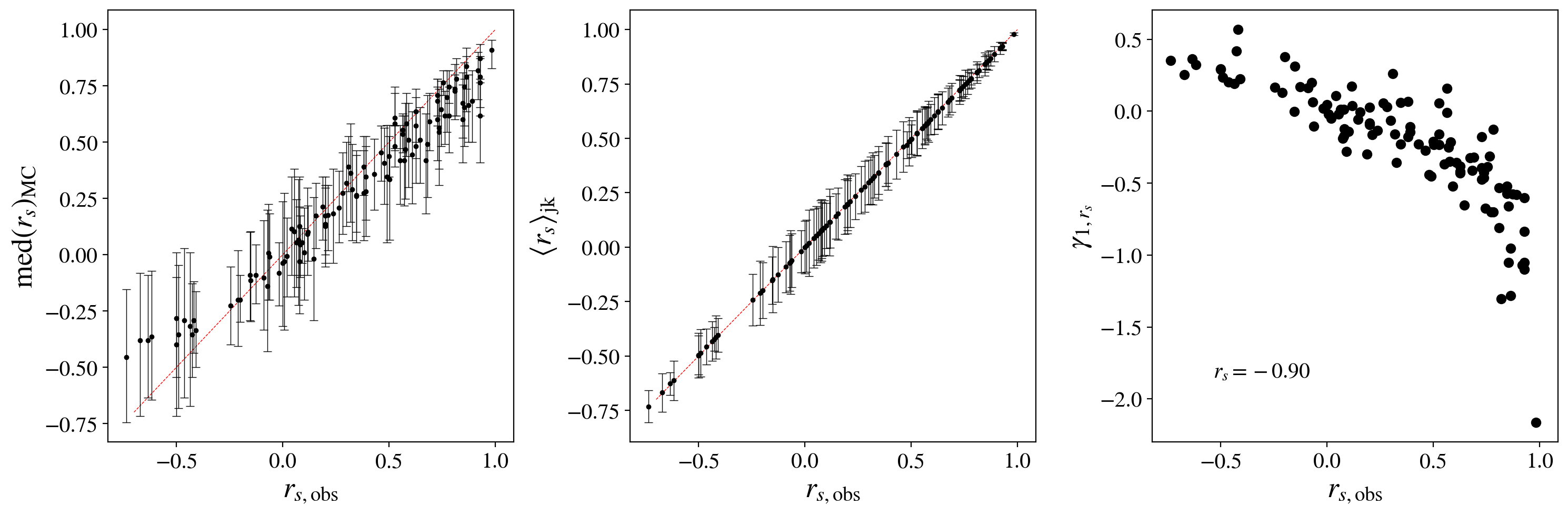}
\caption{The median $r_s$ from the Monte Carlo sampling method (med($r_s$)$_\mathrm{MC}$) (left), the mean $r_s$ from jackknife resampling ($\langle r_s \rangle_\mathrm{jk}$ (center), and the the skewness of Monte Carlo distribution ($\gamma_{1,r_s}$)(right) plotted against $r_s$ for every comparison.
The red dotted line is a one-to-one line.
Included on the right is $r_s$ for $r_{s,\mathrm{obs}}$ and $\gamma_{1,r_s}$.
\label{fig:rs_estimate_comp}}
\end{figure*}

The Spearman rank correlation coefficients estimated using our Monte Carlo method and jackknife resampling agree very well with the Spearman rank correlation coefficients of the observed data across the board.
This can be seen in Figure \ref{fig:rs_estimate_comp}, which shows a comparison of the Monte Carlo and jackknife estimated $r_s$ values against $r_{s,\mathrm{obs}}$ for all comparisons.
The jackknife means are almost exactly one-to-one with $r_{s,\mathrm{obs}}$.
The Monte Carlo median $r_s$ values are almost always within 1$\sigma$ of the observed $r_s$, with the discrepancy increasing the further $r_{s,\mathrm{obs}}$ is from zero.
It's worth noting that these departures from a one-to-one correspondance with $r_{s,\mathrm{obs}}$ in the Monte Carlo estimates for $r_{s,\mathrm{obs}}$ values far from zero is likely a systematic issue of the method.
In fact, plotting the skewness of the Monte Carlo distributions against $r_{s,\mathrm{obs}}$ for the same comparisons shows a strong anticorrelation with $r_{s,\mathrm{obs}}$ with $r_s=-0.90$ (Figure \ref{fig:rs_estimate_comp}).
This all points towards the hard cut-off of $r_s$ at 1 and -1, as well as the generally lower chances of getting a more extreme value of $r_s$, causing the distributions to skew towards lower values of $r_s$.
Regardless, if $r_{s,\mathrm{obs}}$ indicates a strong (anti)correlation ($|r_s| \geq 0.7$), the Monte Carlo estimate rarely indicates a poor correlation.
The only exceptions are comparisons involving [pD$_2$CO]/[HDCO], as that abundance ratio has the largest uncertainties (see Table \ref{tab:dfracs}).
As such, these comparisons have values of med$(r_s)_\mathrm{MC}$ that tend to lower values than other comparisons, differing from the corresponding $r_{s,\mathrm{obs}}$ by up to 0.31.
Thus, our Monte Carlo method and jackknife resampling indicate that the Spearman rank correlations derived from our observed data, with reasonable uncertainties, are robust.


\section{Results and Discussion} \label{sec:results}

For easy visual comparison of correlations between sets of quantities, we used the Python package \texttt{NetworkX} \citep{networkx} to represent them as graphs (see Figure \ref{fig:NCorrelationGraph}).
Each node represents a quantity such as a molecular column density or a physical parameter (i.e., core central density).
The edge between two nodes represents the Spearman rank correlation coefficient of the plot of those two quantities for all of the sources. 
They are placed using the Kamada-Kawai path-length cost-function \citep{Kamada1989}, as implemented in \texttt{NetworkX} \citep{networkx}, which optimizes node placement using a given set of weights that help determine the lengths of the edges connecting the nodes.
We calculated the weight of an edge connecting two nodes representing quantities $x$ and $y$ as $1 - |r_s(x,y)|$.
The closer together two nodes are, the more strongly correlated or anti-correlated they generally are.
It is important to note that there is not necessarily a direct correspondence between $r_s$ and the edge lengths, since preserving such a correspondence while ensuring all the nodes are connected by edges would be impossible in a two dimensional space.
Regardless, the graphs serve as a way to succinctly and clearly show which quantities are most and least correlated.
We define the correlation between two quantities as being ``well" correlated if they have a Spearman rank correlation coefficient of $r_s \geq 0.7$.

\subsection{Chemical Comparisons} \label{subsec:chemical}

\begin{figure*}[ht!]\centering
\includegraphics[width=.49\linewidth,trim={0 0 0 1cm},clip]{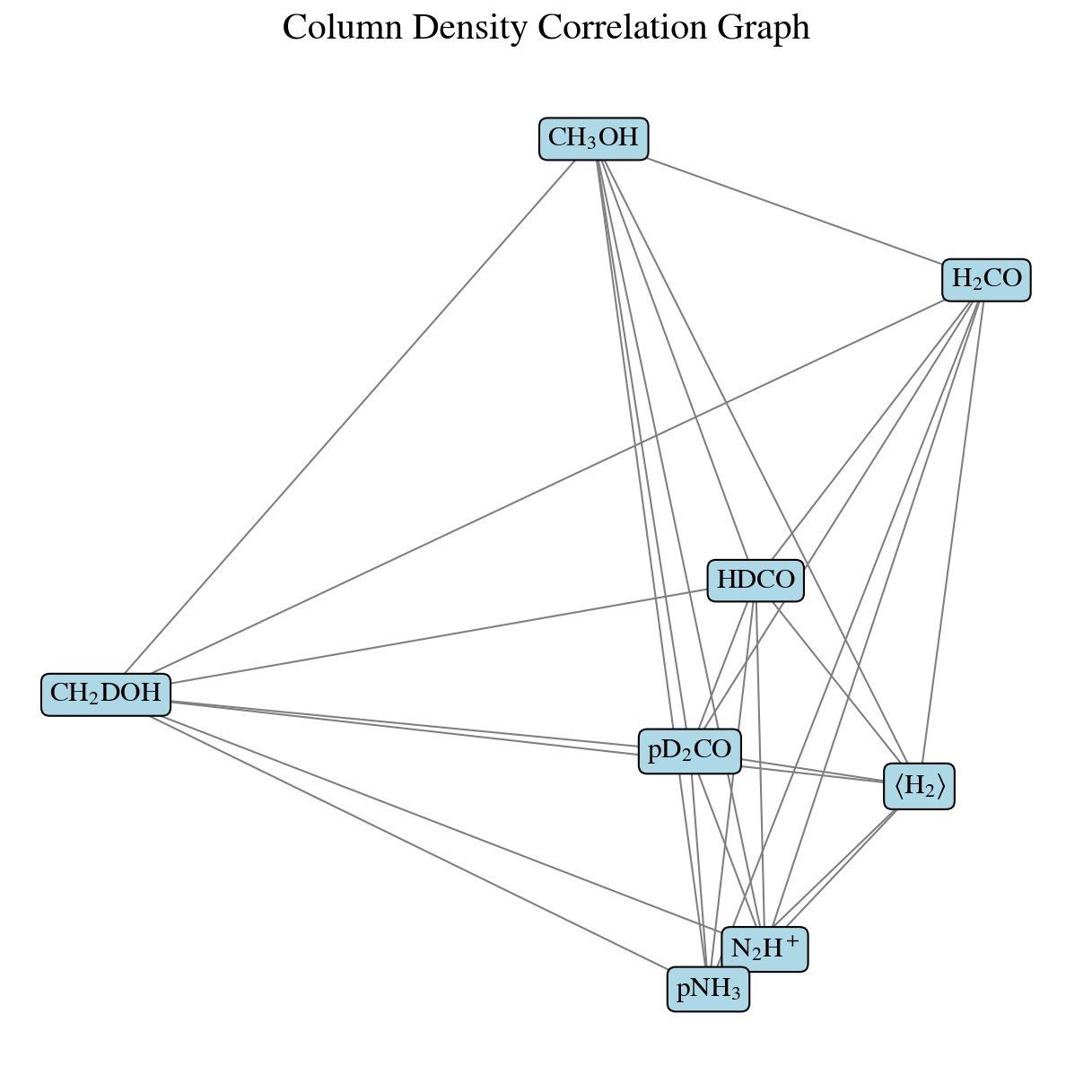}
\includegraphics[width=.49\linewidth]{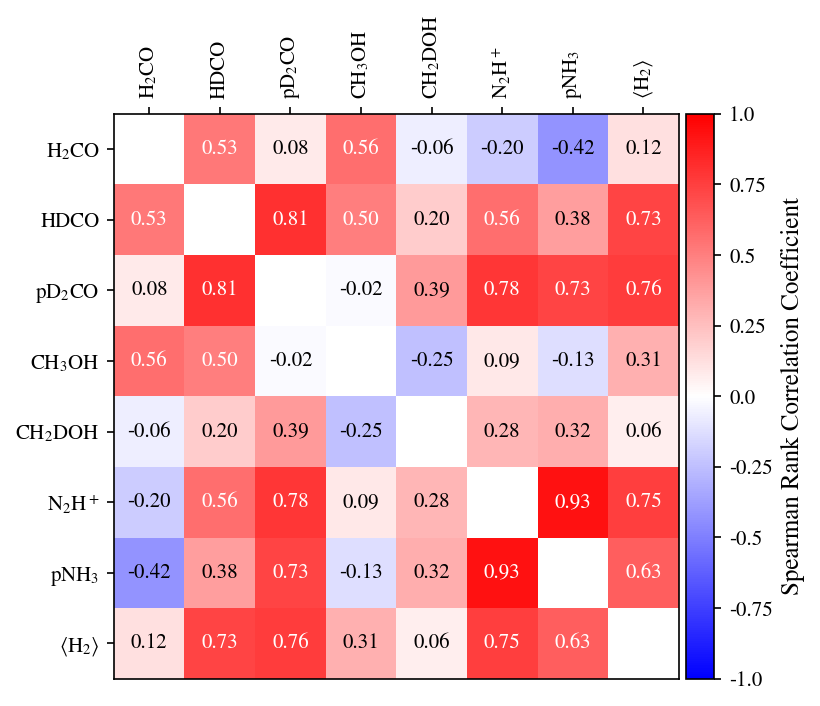}
\caption{Left: The correlations between different column densities are shown as a graph. Each node represents the column density of the given molecule. The distance between nodes was weighted with $1 - |r_s|$, where $r_s$ is the Spearman rank correlation coefficient, such that more correlated or anticorrelated nodes are more closely spaced in the graph.
Note that each column density was compared to the average H$_2$ column density within the beam size of the relevant transition on the ARO 12m with the exceptions of $N$(pNH$_3$) and $N$(N$_2$H$^+$), which were compared to $N$(H$_2$) at the core peak.
Right: The Spearman rank correlation values between different column densities are shown as a color-coded grid.
Each square corresponds to a different pair of quantities, and is color-coded by $r_s$, which is also shown within the square for reference.
\label{fig:NCorrelationGraph}}
\end{figure*}

Figure \ref{fig:NCorrelationGraph} shows a graph of the correlations between the column densities of the molecules analyzed in our survey (Table \ref{tab:MC_jk_N}).
A strong correlation is observed between the column densities of dense gas tracers N$_2$H$^+$ and pNH$_3$ ($r_s=0.93$; med$(r_s)_\mathrm{MC}=0.87^{+0.06}_{-0.11}$; $\langle r_s \rangle_\mathrm{jk}=0.92\pm0.02$).
N$_2$H$^+$ also has a strong correlation with N(H$_2$) at the core peak ($r_s=0.75$; med$(r_s)_\mathrm{MC}=0.76\pm0.05$; $\langle r_s \rangle_\mathrm{jk}=0.75\pm0.06$). 
These correlations are not surprising since NH$_3$ and N$_2$H$^+$ have traditionally been used as tracers of dense molecular gas in starless cores \citep{Benson1989, Caselli2002n2hp}.
In contrast, H$_2$CO and CH$_3$OH are not well correlated with pNH$_3$, N$_2$H$^+$, or beam-averaged H$_2$ ($|r_s| \leq 0.42$).
These lack of correlations are also not surprising as it is well known that H$_2$CO and CH$_3$OH freeze out of the gas phase in cold, dense starless cores resulting in spatial chemical differentiation with cold dense gas tracers \citep{Tafalla2006,Spezzano2017,Punanova2022,Punanova2025}.

Deuterated molecules are also thought to be tracers of the cold dense molecular gas.
The column densities of both pD$_2$CO and HDCO correlate well with each other ($r_s = 0.81$; med$(r_s)_\mathrm{MC}=0.73^{+0.07}_{-0.11}$; $\langle r_s \rangle_\mathrm{jk}=0.80\pm0.04$) as well with the beam-averaged H$_2$ column density (for HDCO: $r_s=0.73$; med$(r_s)_\mathrm{MC}=0.71^{+0.07}_{-0.09}$; $\langle r_s \rangle_\mathrm{jk}=0.72\pm0.05$ and for pD$_2$CO: $r_s=0.76$; med$(r_s)_\mathrm{MC}=0.62^{+0.11}_{-0.13}$; $\langle r_s \rangle_\mathrm{jk}=0.76\pm0.06$).
Dense gas tracers like N$_2$H$^+$ have been shown to trace the dust (or H$_2$) peaks well in cores like L1544 \citep{Spezzano2017} and L1521E \citep{Nagy2019}.
Maps of L1544 by \cite{Chacon-Tanarro2019} showed that HDCO and D$_2$CO also peak close to the dust peak, meaning they are likely in the same class of molecule as dense gas tracers.
By contrast, CH$_2$DOH is essentially uncorrelated with H$_2$ ($r_s=0.06$; med$(r_s)_\mathrm{MC}=0.05^{+0.19}_{-0.20}$; $\langle r_s \rangle_\mathrm{jk}=0.06\pm0.12$).
This could be because, as shown in the same \cite{Chacon-Tanarro2019} maps of L1544, CH$_2$DOH traces slightly outer layers of the core where gas-phase CO has depleted enough for deuteration to be favored, but freezes out in the center of the core.

The column densities of pD$_2$CO and HDCO don't correlate well with CH$_2$DOH (for pD$_2$CO: $r_s=0.39$; med$(r_s)_\mathrm{MC}=0.35^{+0.18}_{-0.19}$; $\langle r_s \rangle_\mathrm{jk}=0.39\pm0.12$ and for HDCO: $r_s=0.20$; med$(r_s)_\mathrm{MC}=0.17\pm{0.17}$; $\langle r_s \rangle_\mathrm{jk}=0.20\pm0.12$).
CH$_3$OH and CH$_2$DOH are thought to primarily form in cold starless cores on icy dust grain surfaces through the successive hydrogenation or deuteration of CO \citep{Tielens1982,Watanabi2002,Ceccarelli2014}.
A fraction of the molecules are then non-thermally desorbed into the gas phase by mechanisms which are still being studied and their relative importance debated \citep{Vasyunin2013,Vasyunin2017,Scibelli2021,Wakelam2021,Paulive2022,Furuya2022,Faure2025,Bariosco2025,Borshcheva2025,Riedel2025}.
CH$_3$OH and CH$_2$DOH cannot form primarily in the gas phase in cold starless cores due to a well-known bottleneck in ion-molecule chemistry (CH$_3$OH$_2^+$ + e$^-$ does not readily form CH$_3$OH; \citealt{Geppert2006}).
In contrast, H$_2$CO does not have a gas phase bottleneck and can form readily through both gas phase and grain surface reactions in cold starless cores \citep{Bacmann2016}.
The low average H$_2$CO ortho to para ratio (less than $3$) observed in the B10 cores (see Section \ref{subsec:coldens_calcs}) is consistent with recent laboratory experiments by \cite{Yocum2023} and protoplanetary disk observations by \cite{Terwisscha_van_Scheltinga2021} indicating that low H$_2$CO ortho to para ratios may indicate a significant, if not dominant, cold gas-phase formation route for the H$_2$CO in starless cores.
This difference in formation routes may explain the poor correlation between the column densities of H$_2$CO and CH$_3$OH ($r_s = $0.56; med$(r_s)_\mathrm{MC}=0.54^{+0.09}_{-0.08}$; $\langle r_s \rangle_\mathrm{jk}=0.56\pm0.09$).
Prior studies have found a range in the relative contributions of gas-phase vs. ice surface chemistry formation routes for HDCO and pD$_2$CO.
\cite{Bacmann2003} showed that gas-phase deuteration with freeze-out could not on its own explain observed formaldehyde deuterium fractionation in cores. 
An evolutionary analysis of high-mass star-forming regions by \cite{Zahorecz2017} also supported primarily grain surface deuteration of H$_2$CO in high-mass cores as well.
Recent observations of the protoplanetary disk of IRS 63 by \cite{Podio2024} found that D$_2$CO traces the shocked region where a streamer is impacting the disk, which they argue implies that the D$_2$CO is being liberated from grain surface ices and is thus primarily formed on grain surfaces. 
However, if grain-surface deuteration were the primary formation route, then we would expect to see a much stronger correlation between deuterated isotopolgoues of formaldehyde and CH$_2$DOH.
This lack of correlation observed in our survey thus indicates that the gas-phase component of formaldehyde deuteration is significant.
In the case of IRS 63, D$_2$CO deuterated in the gas phase may have frozen out onto grain surfaces to then be liberated later by the accretion shock onto the disk.

\begin{figure*}[ht!]\centering
\includegraphics[width=0.49\linewidth,trim={0 0 0 1cm},clip]{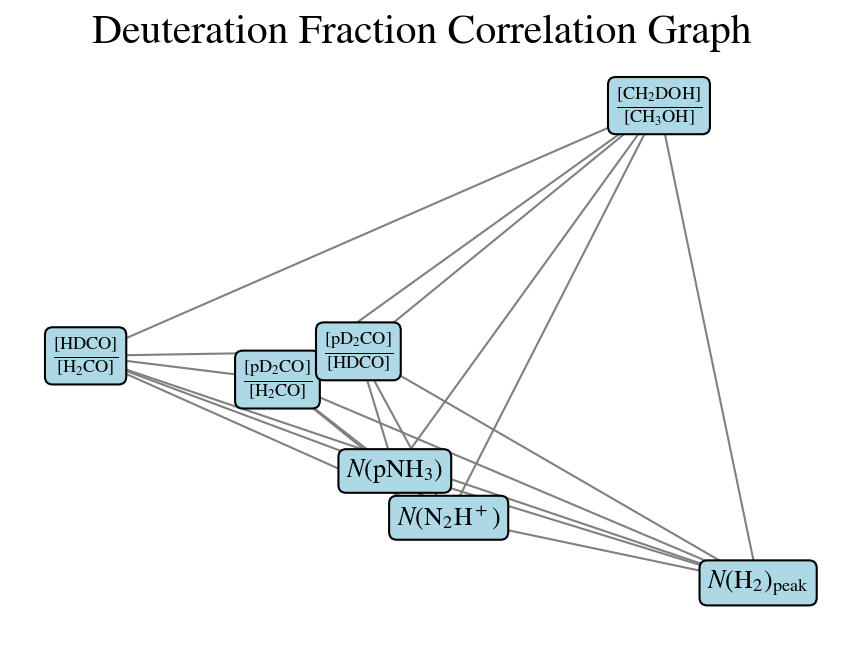}
\includegraphics[width=0.49\linewidth]{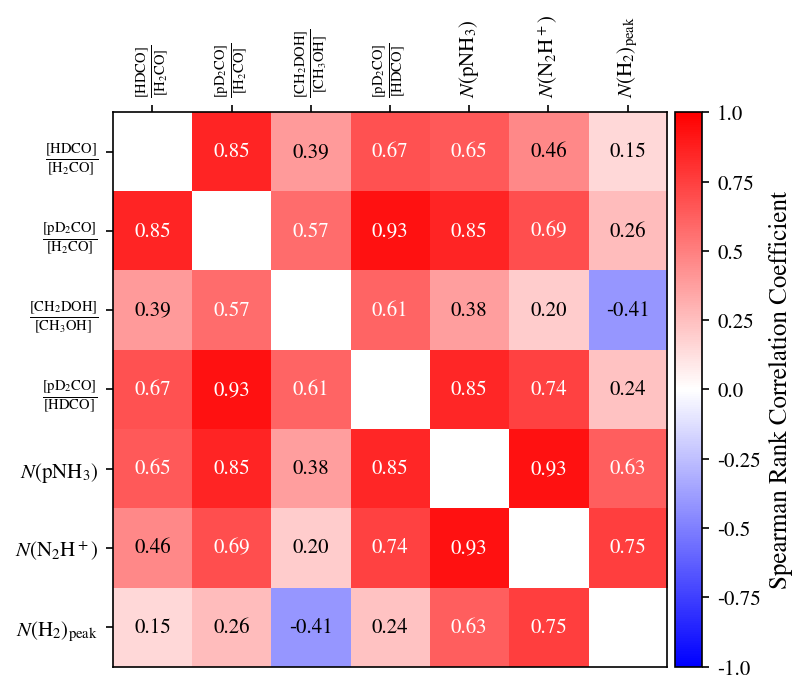}
\caption{Left: The correlations between different deuterium fractions and dense gas tracer column densities are shown as a graph. Each node represents the given quantity. 
The distance between nodes was weighted with $1 - |r_s|$, where $r_s$ is the Spearman rank correlation coefficient, such that more correlated or anticorrelated nodes are more closely spaced in the graph.
Right: The Spearman rank correlation values between different deuterium fractions and dense gas tracer column densities are shown as a color-coded grid.
Each square corresponds to a different pair of quantities, and is color-coded by $r_s$, which is also shown within the square for reference.
\label{fig:DFracCorrelationGraph}}
\end{figure*}

Given their lack of correlation with CH$_2$DOH, we can also explore the contribution of the gas-phase formation route to the HDCO and pD$_2$CO column densities. 
pD$_2$CO correlates very well with the dense gas tracers (for N$_2$H$^+$: $r_s=0.78$; med$(r_s)_\mathrm{MC}=0.75\pm{0.07}$; $\langle r_s \rangle_\mathrm{jk}=0.78\pm0.04$ and for pNH$_3$: $r_s=0.73$; med$(r_s)_\mathrm{MC}=0.68^{+0.10}_{-0.11}$; $\langle r_s \rangle_\mathrm{jk}=0.72\pm0.05$), whereas HDCO does not correlate well with either (for N$_2$H$^+$: $r_s=0.56$; med$(r_s)_\mathrm{MC}=0.58^{+0.06}_{-0.08}$; $\langle r_s \rangle_\mathrm{jk}=0.56\pm0.08$ and for pNH$_3$: $r_s=0.38$; med$(r_s)_\mathrm{MC}=0.39^{+0.14}_{-0.13}$; $\langle r_s \rangle_\mathrm{jk}=0.38\pm0.10$).
N$_2$H$^+$ can only form via gas-phase ion molecule chemistry.
This result may imply a primarly gas-phase formation route for D$_2$CO and a more mixed gas-phase and grain surface formation for HDCO. 
This could also be an indication that multiply-deuterated isotopologues are better tracers of dense molecular gas.
It is also possible that D$_2$CO traces a more central region than HDCO, with HDCO freezing out further from the center of the core.
However, both of the latter two results require resolved spatial maps of these molecules in the B10 cores to confirm.
Factors such as different formation timescales or desorption rates for deuterated formaldehyde and deuterated methanol may contribute to the lack of correlation of HDCO and pD$_2$CO with CH$_2$DOH.
However, this lack of correlation along with the strong correlation of pD$_2$CO with the dense gas tracers N$_2$H$^+$ and pNH$_3$ and even the strong correlation of both HDCO and pD$_2$CO with H$_2$ are together evidence of a significant gas-phase component to the single-deuteration and especially double-deuteration of formaldehyde.

Looking now the deuterium fractionation of the molecules, this can be quantified in two different ways: a straight ratio of the abundances of the deuterated and nondeuterated isotopologue of a molecule, and the D/H ratio.
As defined in \cite{Manigand2019}, the D/H ratio is the actual abundance ratio of deuterium atoms to hydrogen atoms betwen two given isotopologues, accounting for the different positions on the molecule in which the deuterium may be:
\begin{equation}\label{eq:DH}
    \frac{[\text{X}\text{H}_{n-i}\text{D}_i]}{[\text{X}\text{H}_n]} = \binom{n}{i}\left(\frac{[\text{D}]}{[\text{H}]}\right)^i \ \ ,
\end{equation}
where $n$ is the total number of hydrogen atoms in the nondeuterated isotopologue, $i$ is the total number of  deuterium atoms in the deuterated isotopologue, and $\binom{n}{i} = \frac{n!}{i!(n-1)!}$ is the number of combinations of $i$ deuterium atoms out of $n$ possible hydrogen atoms.
The expression for the D/H ratio resulting from the comparison of two deuterated isotopologues is then:
\begin{equation}\label{eq:DH_pt2}
    \frac{[\text{X}\text{H}_{n-i}\text{D}_i]}{[\text{X}\text{H}_{n-j}\text{D}_j]} = \frac{\binom{n}{i}}{\binom{n}{j}}\left(\frac{[\text{D}]}{[\text{H}]}\right)^{i-j} \ \ , \ i > j >0  \ .
\end{equation}
Thus for our deuterium fractions, we can see that
\begin{equation}\label{eq:DH_dfracs}
    \rm \frac{[D]}{[H]} =
    \begin{cases}
        \rm\frac{1}{2}\frac{[HDCO]}{[H_2CO]} \\
        \rm\sqrt{3\frac{[pD_2CO]}{[H_2CO]}} \\
        \rm\frac{1}{3}\frac{[CH_2DOH]}{[CH_3OH]} \\
        \rm6\frac{[pD_2CO]}{[HDCO]}
    \end{cases}
    \  ,
\end{equation}
where we accounted for the statistical D$_2$CO ortho to para ratio of 2 ([D$_2$CO] = [oD$_2$CO] + [pD$_2$CO] implies [D$_2$CO]/[pD$_2$CO] = 2 + 1 = 3).
The deuterium fractions and corresponding D/H ratios of all cores are reported in Table \ref{tab:dfracs}.
However, in all three cases the abundance ratio and D/H ratio are directly proportional to each other.
Thus for the purposes of determining correlations, using the D/H ratio would produce the exact same result as the abundance ratio, since the two quantities have the same ranks.
As such, we perform the remaining analysis with only the deuterated-nondeuterated abundance ratios, which we will refer to henceforth as the deuterium fractions.
The resulting Spearman rank correlation coefficients and associated statistics are shown in Table \ref{tab:MC_jk_RD}.

\begin{deluxetable*}{lCCCCCCCC}
\tablewidth{0pt} 
\tablecaption{Deuterium Fractions\label{tab:dfracs}}
\tablehead{
\colhead{} & \multicolumn{2}{c}{$\rm[HDCO]/[H_2CO]$} & \multicolumn{2}{c}{$\rm[pD_2CO]/[H_2CO]$} & \multicolumn{2}{c}{$\rm[pD_2CO]/[HDCO]$} &  \multicolumn{2}{c}{$\rm[CH_2DOH]/[CH_3OH]$} \\
\colhead{Core} & \colhead{Abundance Ratio} & \colhead{D/H} & \colhead{Abundance Ratio} & \colhead{D/H} & \colhead{Abundance Ratio} & \colhead{D/H} & \colhead{Abundance Ratio} & \colhead{D/H}
} 
\startdata 
Seo06 & 0.25^{+0.05}_{-0.04} & 0.126^{+0.026}_{-0.022} & 0.022^{+0.009}_{-0.007} & 0.26^{+0.05}_{-0.05} & 0.09^{+0.03}_{-0.03} & 0.53^{+0.18}_{-0.15} & 0.04^{+0.01}_{-0.01} & 0.013^{+0.004}_{-0.004} \\
Sci7-1 & 0.24^{+0.06}_{-0.05} & 0.121^{+0.028}_{-0.023} & 0.027^{+0.011}_{-0.008} & 0.28^{+0.05}_{-0.05} & 0.11^{+0.04}_{-0.03} & 0.66^{+0.23}_{-0.19} & 0.14^{+0.05}_{-0.04} & 0.048^{+0.016}_{-0.014} \\
Sci7-2 & 0.20^{+0.06}_{-0.05} & 0.102^{+0.032}_{-0.025} & 0.023^{+0.012}_{-0.008} & 0.27^{+0.06}_{-0.05} & 0.12^{+0.05}_{-0.04} & 0.69^{+0.30}_{-0.23} & 0.13^{+0.06}_{-0.05} & 0.045^{+0.019}_{-0.016} \\
Seo08 & 0.35^{+0.07}_{-0.06} & 0.174^{+0.034}_{-0.029} & 0.075^{+0.021}_{-0.017} & 0.48^{+0.06}_{-0.06} & 0.22^{+0.05}_{-0.04} & 1.30^{+0.28}_{-0.24} & 0.15^{+0.04}_{-0.03} & 0.049^{+0.013}_{-0.011} \\
Seo09 & 0.46^{+0.10}_{-0.08} & 0.228^{+0.049}_{-0.041} & 0.079^{+0.023}_{-0.019} & 0.49^{+0.07}_{-0.06} & 0.17^{+0.03}_{-0.03} & 1.03^{+0.21}_{-0.18} & 0.12^{+0.03}_{-0.02} & 0.040^{+0.009}_{-0.008} \\
Seo10 & 0.08^{+0.03}_{-0.02} & 0.040^{+0.013}_{-0.010} & 0.006^{+0.004}_{-0.003} & 0.14^{+0.04}_{-0.04} & 0.08^{+0.04}_{-0.03} & 0.48^{+0.26}_{-0.20} & 0.11^{+0.04}_{-0.03} & 0.036^{+0.013}_{-0.011} \\
Seo12 & 0.17^{+0.02}_{-0.02} & 0.087^{+0.008}_{-0.008} & 0.028^{+0.005}_{-0.005} & 0.29^{+0.03}_{-0.03} & 0.16^{+0.03}_{-0.02} & 0.95^{+0.15}_{-0.14} & 0.06^{+0.01}_{-0.01} & 0.019^{+0.004}_{-0.004} \\
Seo14 & 0.15^{+0.02}_{-0.02} & 0.077^{+0.009}_{-0.009} & 0.013^{+0.004}_{-0.003} & 0.20^{+0.03}_{-0.03} & 0.09^{+0.02}_{-0.02} & 0.52^{+0.10}_{-0.09} & 0.03^{+0.01}_{-0.01} & 0.009^{+0.003}_{-0.003} \\
Seo15 & 0.28^{+0.11}_{-0.08} & 0.141^{+0.057}_{-0.040} & 0.042^{+0.028}_{-0.019} & 0.36^{+0.10}_{-0.09} & 0.15^{+0.09}_{-0.06} & 0.90^{+0.52}_{-0.38} & 0.11^{+0.03}_{-0.03} & 0.036^{+0.012}_{-0.010} \\
Seo16 & 0.50^{+0.15}_{-0.12} & 0.249^{+0.077}_{-0.058} & 0.059^{+0.037}_{-0.026} & 0.42^{+0.12}_{-0.11} & 0.12^{+0.06}_{-0.05} & 0.72^{+0.38}_{-0.30} & 0.08^{+0.03}_{-0.02} & 0.026^{+0.009}_{-0.007} \\
Seo17 & 0.31^{+0.05}_{-0.04} & 0.154^{+0.025}_{-0.022} & 0.078^{+0.019}_{-0.016} & 0.48^{+0.06}_{-0.05} & 0.25^{+0.05}_{-0.04} & 1.52^{+0.30}_{-0.27} & 0.23^{+0.07}_{-0.06} & 0.075^{+0.023}_{-0.019} \\
\enddata
\tablecomments{The abundance ratios have been rounded. 
The D/H ratios were calculated using the unrounded abundance ratios.}
\end{deluxetable*}

In Figure \ref{fig:DFracCorrelationGraph}, [pD$_2$CO]/[H$_2$CO] and [HDCO]/[H$_2$CO] correlate well with each other ($r_s=0.85$; med$(r_s)_\mathrm{MC}=0.75^{+0.11}_{-0.17}$; $\langle r_s \rangle_\mathrm{jk}=0.85\pm0.03$), and neither correlates very well with [CH$_2$DOH]/[CH$_3$OH] (for [HDCO]/[H$_2$CO]: $r_s=0.39$; med$(r_s)_\mathrm{MC}=0.28\pm{0.20}$; $\langle r_s \rangle_\mathrm{jk}=0.39\pm0.09$ and for [pD$_2$CO]/[H$_2$CO]: $r_s=0.57$; med$(r_s)_\mathrm{MC}=0.42\pm{0.20}$; $\langle r_s \rangle_\mathrm{jk}=0.57\pm0.07$).
Interestingly, neither correlates very well with the column densities of the dense gas tracers or H$_2$ (at the dust peak), with the exception of [pD$_2$CO]/[H$_2$CO] which strongly correlates with pNH$_3$ (although the discrepancy from the Monte Carlo results is large: $r_s=0.85$; med$(r_s)_\mathrm{MC}=0.65^{+0.13}_{-0.17}$; $\langle r_s \rangle_\mathrm{jk}=0.85\pm0.03$).
On the surface, this seems to contradict the idea of deuteration being favored in dense environments; however, as will be discussed in Subsection \ref{subsec:physical}, this could instead be a sign that deuteration is revealing unique evolutionary information about the cores.

The ratio of doubly-to-singly deuterated formaldehyde, [pD$_2$CO]/[HDCO], mostly follows the same correlation pattern as [pD$_2$CO]/[H$_2$CO] since both of these ratios are well correlated with each other ($r_s=0.93$; $\langle r_s \rangle_\mathrm{jk}=0.92\pm0.01$).
A notable exception is a weaker correlation with [HDCO]/[H$_2$CO] ($r_s=0.67$).
However, these results should be interpreted cautiously since [pD$_2$CO]/[HDCO] has larger uncertainties with larger discrepancies between $r_s$ and the results of the Monte Carlo analysis than for other ratios (see Section \ref{subsec:correl}).
For example, even though $r_s=0.93$ for the correlation with [pD$_2$CO]/[H$_2$CO], the med$(r_s)_\mathrm{MC}=0.62^{+0.16}_{-0.21}$ and is significantly lower due to the larger errors.

\begin{figure*}[ht!]\centering
\includegraphics[width=0.49\linewidth]{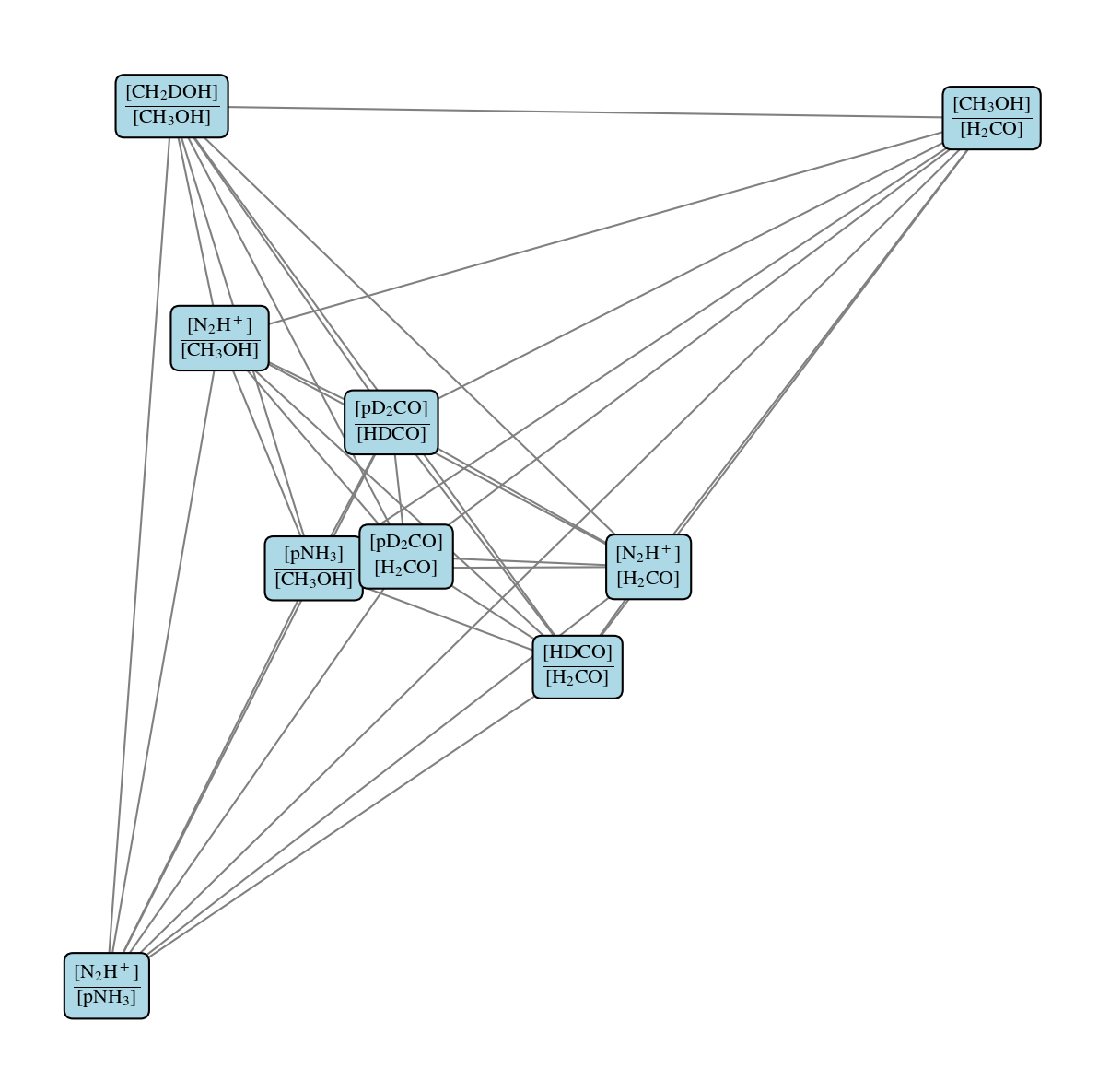}
\includegraphics[width=0.49\linewidth]{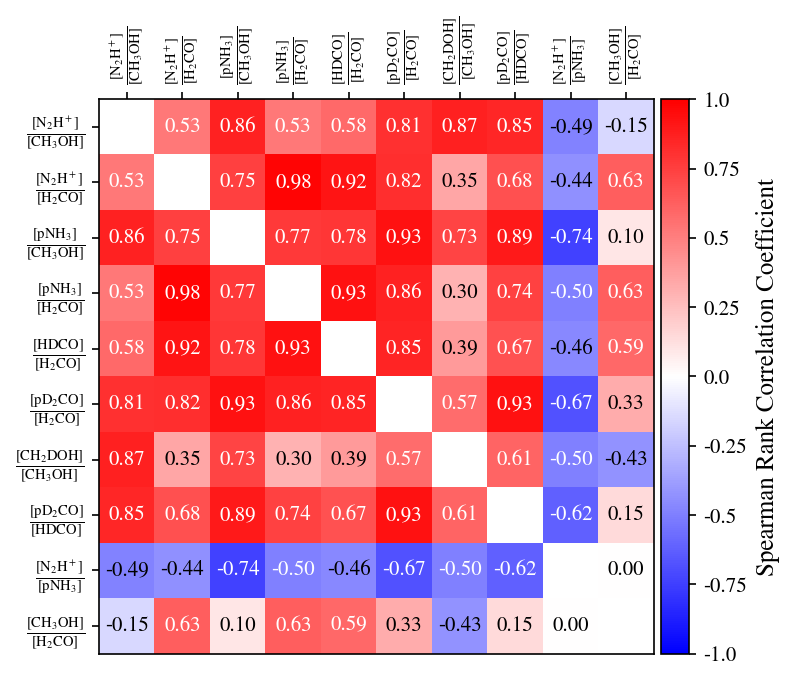}
\caption{Left: The correlations between deuterium fractions and other molecular abundance ratios are shown as a graph. Each node represents the given quantity. 
The distance between nodes was weighted with $1 - |r_s|$, where $r_{s}$ is the Spearman rank correlation coefficient, such that more correlated or anticorrelated nodes are more closely spaced in the graph.
Note that $\mathrm{[pNH_3]/[H_2CO]}$ was excluded from the graph for legibility because it is almost perfectly correlated with $\mathrm{[N_2H^+]/[H_2CO]}$, resulting in their nodes almost completely overlapping.
Right: The Spearman rank correlation values between deuterium fractions and other molecular abundance ratios are shown as a color-coded grid.
Each square corresponds to a different pair of quantities, and is color-coded by $r_s$, which is also shown within the square for reference.
\label{fig:MolRatioCorrelationGraph}}
\end{figure*}

Finally, we also compared abundance ratios of all of our observed molecules to the organic deuterium fractions and to each other (Table \ref{tab:MC_jk_molratios} and Figure \ref{fig:MolRatioCorrelationGraph}). 
We categorized the molecules into ``early-time" molecules (H$_2$CO, CH$_3$OH), whose abundance peaks earlier in the core's lifetime when the core is less dense and freeze out in the central regions at later times when the core is more dense, and ``late-time" molecules (deuterated organics, N$_2$H$^+$, and pNH$_3$), whose abundance peaks later when the core has higher density.
We focused our analysis on comparisons including ratios of ``late" to ``early" molecules and the deuterium fractions. 
We also included $\mathrm{[N_2H^+]/[pNH_3]}$, $\mathrm{[H_2CO]/[CH_3OH]}$, and $\mathrm{[pD_2CO]/[HDCO]}$ in our comparisons as well, although there is no clear chemical interpretation of the comparisons that include them since they are ratios of molecules of the same ``evolutionary" category.
Looking at comparisons between late/early ratios (excluding deuterium fractions), 4 out of the 6 comparisons (67\%) correlate well with $|r_s|\geq0.7$.
Looking instead at whether $\langle r_s \rangle_\mathrm{jk}$ is within a standard deviation of 0.7 gives the same result.
Instead accounting for whether med$(r_s)_\mathrm{MC}$ is within $1\sigma$ of 0.7 brings the percentage up to 100\%.
Including the deuterium fractions brings the fraction of good correlations up to 13 out of 18 (72\%). 
This percentage again remains the same looking at $\langle r_s \rangle_\mathrm{jk}$ and its standard deviation, and jumps to 89\% when considering med$(r_s)_\mathrm{MC} + 1\sigma$.
Interestingly, $\mathrm{[CH_2DOH]/[CH_3OH]}$ correlates the least often with the other abundance ratios. 
Excluding it from the sample brings the fraction of correlations up to 11/14 (79\%) for $r_{s,\mathrm{obs}}$ and $\langle r_s \rangle_\mathrm{jk} + \sigma_{r_s,\mathrm{jk}}$ and 100\% for  med$(r_s)_\mathrm{MC} + 1\sigma$.
Regardless, we can see that ratios of late/early time molecules and deuterium fractions tend to correlate with each other across board.
This confirms that molecules in the same evolutionary category tend to behave similarly across the B10 cores, and thus lends credence to the deuterium fractions as tracers of chemical maturity.

\subsection{Physical Comparisons} \label{subsec:physical}

We compare the deuterium fractions of formaldehyde and methanol to physical and evolutionary properties taken from the best-fit 3D radiative transfer models and dendrogram analysis presented in \cite{Scibelli2023} (Table \ref{tab:physparams}).
Seo17 was not included in the \cite{Scibelli2023} sample of cores, so we omitted it from this set of comparisons.
The three physical properties we used are the core mass $M(R_c)$, the strength of the interstellar radiation field (ISRF), and the ``plateau radius" $r_\mathrm{plat}$.
The plateau radius was calculated as 
\begin{equation}\label{eq:r_plat}
    r_\mathrm{plat} = \sqrt{r_xr_y} \ \ \ ,
\end{equation}
where $r_x$ and $r_y$ are the radii defining the size of the density plateau in the spheroidal Plummer profile (see Equation 5 of \citealt{Scibelli2023}) used to model the B10 cores.
Although the kinetic temperature was also determined from an analysis of the pNH$_3$ (1,1) and (2,2) inversion transitions in \cite{Seo2015}, we omitted it from our analysis due to the lack of dynamic range, with all cores in B10 being within 1 K of each other.

The three evolutionary properties we compare to are the central density of the core, $n_c$, the average H$_2$ density within the core dendrogram from \cite{Scibelli2023}, $\langle n(\mathrm{H_2})\rangle_\mathrm{dendro}$, and the virial ratio as determined in \cite{Scibelli2023}.
We estimated $\langle n(\mathrm{H_2})\rangle_\mathrm{dendro}$ as the number of H$_2$ molecules within the dendrogram area averaged over a sphere with a cross-sectional area equal to that of the dendrogram. 
The resulting equation is
\begin{equation}
    \langle n(\mathrm{H_2})\rangle_\mathrm{dendro} = \frac{3\sqrt{\pi}}{4D\theta_\mathrm{pix}}    \frac{\sum_\mathrm{dendro} w_iN_i(\mathrm{H_2}) \ \ }{(\sum_\mathrm{dendro} w_i)^{3/2}} \ \ ,
\end{equation}
with $D=135$ pc the approximate distance to B10 \citep{Schlafly2014}, $\theta_\mathrm{pix}$ the pixel scale of the map, $N_i(\mathrm{H_2})$ the H$_2$ column density of a given pixel, and $w_i$ the fractional overlap of said pixel with the core dendrogram.
The virial ratio is defined as
\begin{equation}\label{eq:virial_ratio}
    \alpha_{\rm{K,G,P}} = \frac{2\Omega_K}{|\Omega_G|+|\Omega_P|} \ \ \ ,
\end{equation}
including the internal kinetic energy $\Omega_K$ supporting the core against collapse, and the gravitational potential energy $\Omega_G$ and the external pressure $\Omega_P$ as compression terms.
In theory, a virial ratio of $1$ indicates virial equilibrium. 
A virial ratio $>$ 1 would indicate an unbound core and a virial ratio $<$ 1 indictes a bound core, by external pressure and gravity.
However, to truly reflect the virial state of the core, we would also need to include the magnetic energy $\Omega_B$ as an additional support term in the numerator of the virial ratio ($\alpha_{\rm{K,B,G,P}}$).
Unfortunately, accurate maps of magnetic field strengths in molecular clouds, especially at the spatial dimensions of cores, are difficult to measure.
Towards the central Taurus Molecular Cloud a range of magnetic field strengths from as high as 75 $\mu$G \citep{Chapman2011} to as low as 13 $\mu$G \citep{Ward-Thompson2023} in B10 has been estimated using different observing techniques and methods.
$\Omega_B$ is proportional to magnetic energy and therefore depends on $B^2$ amplifying any uncertainty in $B$ to be much larger for $\Omega_B$.
Considering that the ``pressure-bound" cores in B10 would require, on average, an effective difference between the external and internal magnetic fields of only 15 $\mu$G to be brought back to virial equilibrium \citep{Scibelli2023}, it is clear that the observational constraints on $\Omega_B$ are not strong enough to accurately judge the true virial state of the cores in B10. 
Although the virial ratio without $\Omega_B$ is not an accurate measure of the boundedness of the cores, the balance between the internal kinetic energy and bounding terms is still valuable information, so we include it in our analysis.

\begin{deluxetable*}{lCCCCCCCC}
\tablewidth{0pt} 
\tablecaption{Core Physical and Evolutionary Parameters \label{tab:physparams}}
\tablehead{
\colhead{Core} & \colhead{$n_c$\tablenotemark{a}\tablenotemark{b}} & \colhead{$\langle n\mathrm{(H_2)\rangle_{dendro}}$} & \colhead{$\alpha_\mathrm{K,G,P}$\tablenotemark{a}} & \colhead{$M(R_C)$\tablenotemark{a}} & \colhead{ISRF\tablenotemark{a}\tablenotemark{b}} & \colhead{$r_\mathrm{plat}$\tablenotemark{ab}} \\
\colhead{} & \colhead{($10^5$ cm$^{-3}$)} & \colhead{($10^5$ cm$^{-3}$)} & \colhead{} & \colhead{(M$_\odot$)} & \colhead{} & \colhead{($^{\prime\prime}$)}
} 
\startdata 
Seo06 & 0.63\pm0.22 & 2.04\pm0.22 & 0.17\pm0.13 & 0.17\pm0.06 & 1.75^{+0.43}_{-0.43} & 42.68\pm0.88 \\
Sci7-1 & 0.58\pm0.19 & 1.77\pm0.29 & 0.08\pm0.04 & 0.12\pm0.04 & 0.87^{+0.19}_{-0.19} & 30.76\pm2.33 \\
Sci7-2 & 1.00\pm0.50 & 3.19\pm0.20 & 0.78\pm0.21 & 0.04\pm0.00 & 0.68^{+0.16}_{-0.16} & 15.22\pm0.47 \\
Seo08 & 8.67\pm1.25 & 3.60\pm0.23 & 1.21\pm0.10 & 0.20\pm0.01 & 0.60^{+0.40}_{-0.30} & 13.31\pm0.49 \\
Seo09 & 3.00\pm0.50 & 2.14\pm0.35 & 0.20\pm0.02 & 0.62\pm0.02 & 1.00^{+1.00}_{-0.40} & 44.24\pm1.64 \\
Seo10 & 2.00\pm0.50 & 1.12\pm0.20 & 1.46\pm0.76 & 0.16\pm0.02 & 0.87^{+0.19}_{-0.19} & 30.61\pm0.44 \\
Seo12 & 9.80\pm0.40 & 2.70\pm0.48 & 0.40\pm0.04 & 0.36\pm0.05 & 2.00^{+1.00}_{-1.00} & 23.05\pm2.81 \\
Seo14 & 1.70\pm0.87 & 3.49\pm0.37 & 0.46\pm0.35 & 0.11\pm0.06 & 0.96^{+0.54}_{-0.54} & 21.60\pm1.91 \\
Seo15 & 1.50\pm0.50 & 4.31\pm0.28 & 1.13\pm0.45 & 0.07\pm0.01 & 0.30^{+0.30}_{-0.30} & 28.43\pm5.72 \\
Seo16 & 5.50\pm0.50 & 2.20\pm0.21 & 2.73\pm0.33 & 0.24\pm0.02 & 2.00^{+1.00}_{-1.00} & 28.77\pm0.44 \\
\enddata
\tablecomments{(a) Taken from \cite{Scibelli2023} (b) Values and uncertainties determined from the best-fit models in \cite{Scibelli2023} either using the mean and standard deviation of best-fit model values or, if all best-fit models gave the same value, estimated uncertainties using the model grid spacing.}
\end{deluxetable*}

\begin{figure*}[ht!]\centering
\includegraphics[width=0.59\linewidth,trim={0 0 0 1cm},clip]{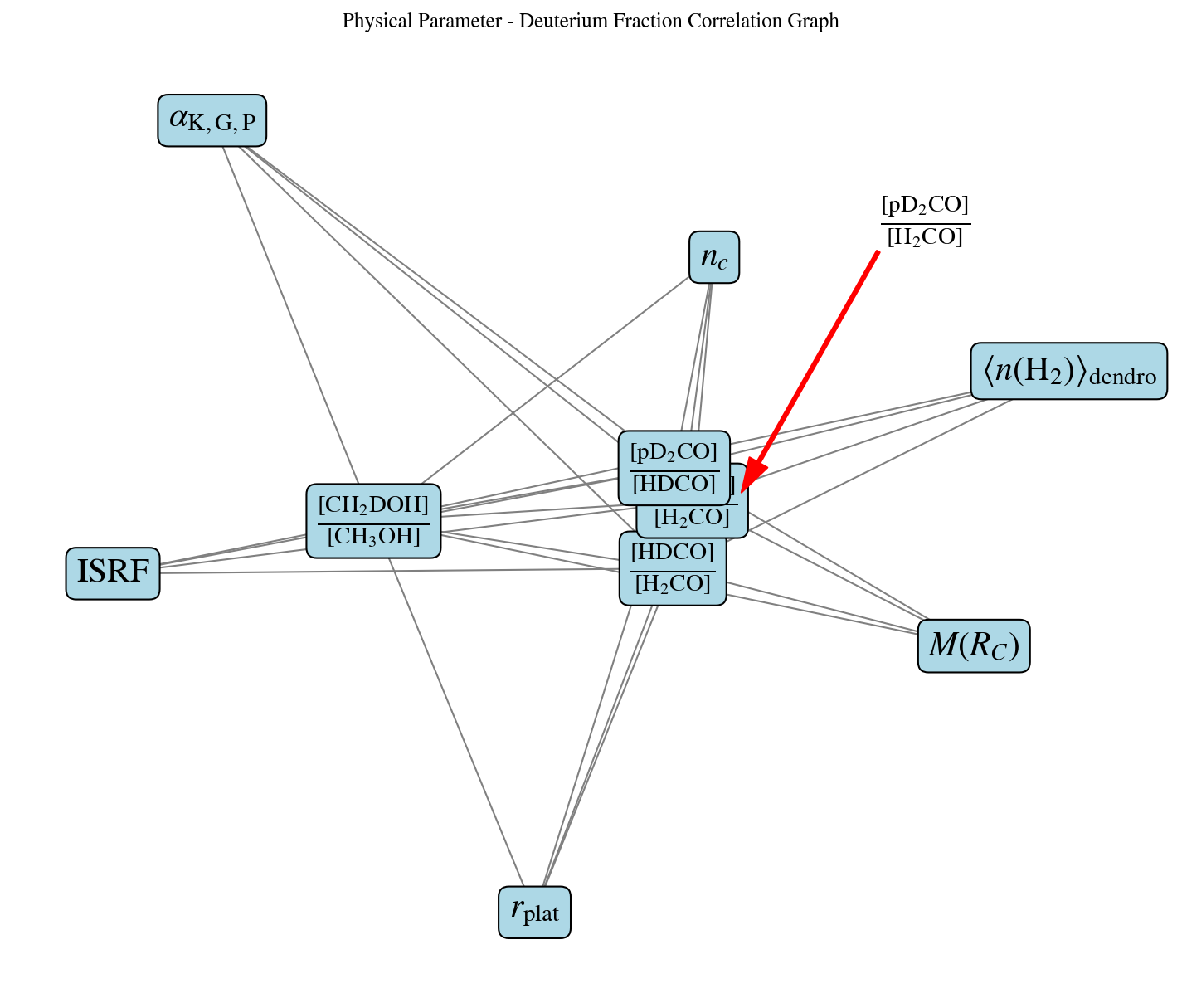}
\includegraphics[width=0.39\linewidth]{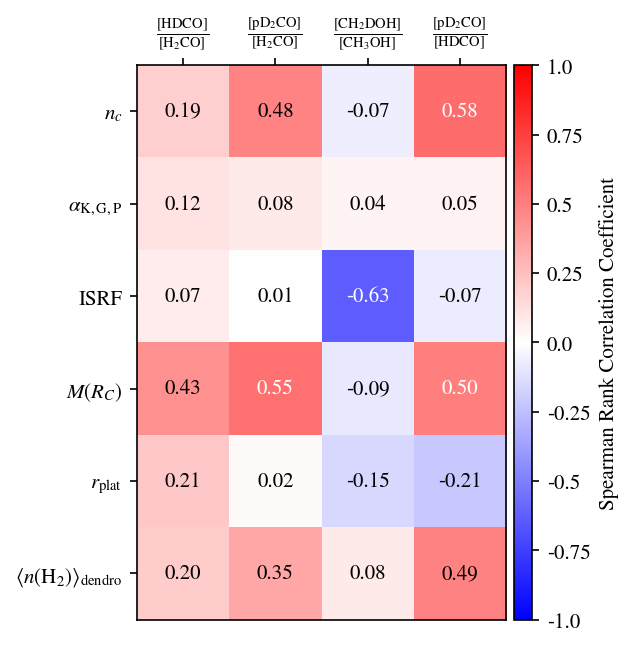}
\caption{Left: The correlations between molecular deuterium fractions and core physical parameters (from left to right: strength of the ISRF, virial ratio, density plateau radius, core mass, central density, and average core density) are shown as a graph.  The distance between nodes corresponds to $1 - |r_s|$ where $r_s$ is the Spearman rank correlation coefficient, such that nodes with higher correlation or anticorrelation are more closely spaced in the graph.  
Note that it is impossible to plot this graph without the labels for the [pD$_2$CO]/[HDCO] and the [pD$_2$CO]/[H$_2$CO] nodes overlapping. The hidden node is [pD$_2$CO]/[H$_2$CO], as indicated by the red arrow. 
Right: The Spearman rank correlation values between different deuterium fractions and core physical parameters are shown as a color-coded grid.
Each square corresponds to a different pair of quantities, and is color-coded by $r_s$, which is also shown within the square for reference.
\label{fig:PhysParamCorrelationGraph}}
\end{figure*}

We plot the Spearman rank correlation coefficients between
the deuteration fractions and core physical and evolutionary parameters (Table \ref{tab:MC_jk_RD}) in Figure \ref{fig:PhysParamCorrelationGraph}.
None of the physical or evolutionary properties are well correlated with any of the deuterium fractions.
The strongest correlation is between [CH$_2$DOH]/[CH$_3$OH] and ISRF ($r_s = -0.63$), but given that the ISRF is relatively poorly constrained, the original models having used a very coarse grid, concluding a strong correlation is dubious, which is reflected in the distribution of $r_s$ values (med($r_s$)$_\mathrm{MC}$ = $-0.38^{+0.29}_{-0.25}$; $\langle r_s \rangle_\mathrm{jk}=-0.63\pm0.05$).

The lack of correlation between $n_c$ and the deuterium fractions is particularly striking; for example, its correlation coefficient with [HDCO]/[H$_2$CO] is $r_s=0.19$ (med$(r_s)_\mathrm{MC}=0.21^{+0.13}_{-0.15}$; $\langle r_s \rangle_\mathrm{jk}=0.18\pm0.13$) (Figure \ref{fig:PhysParamCorrelationGraph}).
The deuterium fractions also correlate poorly with the average dendrogram H$_2$ density (e.g., $r_s = 0.35$ (med$(r_s)_\mathrm{MC}=0.26^{+0.18}_{-0.21}$; $\langle r_s \rangle_\mathrm{jk}=0.34\pm0.12$) for [pD$_2$CO]/[H$_2$CO]).
$\mathrm{[pD_2CO]/[HDCO]}$ has only a marginally better correlation coefficient with both central density ($r_s = 0.58$) and the core average density ($r_s = 0.49$), but neither is strongly correlated. 
The lack of correlation between core density and deuteration was also seen for a sample of starless cores in the L1251 region from observations of oNH$_2$D \citep{Galloway-Sprietsma2022}.
Due to the very sub-thermally populated energy levels of oNH$_2$D, a small uncertainty on $T_{ex}$ results in a large uncertainty in the oNH$_2$D column density due to the exponential sensitivity of upper energy level of the 85.9 GHz transition,  $N(\rm{oNH}_2\rm{D}) \propto e^{20.1/T_{ex}}$.
HDCO does not suffer from this problem and therefore the column densities and deuteration ratios are inherently less uncertain than they are for oNH$_2$D.
Our survey confirms this lack of correlation between density and organic molecule deuteration ratios with smaller errors.
Thus, the deuteration of organics does not simply trace any single physical or evolutionary parameter, but rather reveals unique information about the evolution of the core.

\subsection{Spatial Comparisons}\label{subsec:spatial_comp}

We also compare the deuterium fractions with the spatial positions of the cores in their velocity-coherent filaments to probe filamentary-scale evolution.
The filamentary structure of the central Taurus Molecular Cloud was first quantitatively described by \cite{Hacar2013}, who used the multiple velocity components in C$^{18}$O emission to identify velocity-coherent filaments (also called fibers in that paper).
The velocity-coherent filaments from a \texttt{NestFit} reanalysis of the \cite{Seo2015} pNH3 map (Steffes et al., in prep.) agree well with the \cite{Hacar2013} filaments, so we use the Hacar labels for clarity (Figure \ref{fig:B10_NH2_map}, Table \ref{tab:observedsources}). 
The filament labeled Fil06 contains cores Seo12 and Seo14.
The filament labeled Fil08 contains cores Seo15 and Seo16 while the filament labeled Fil10 contains cores Seo06, Sci 7-1, Sci 7-2, Seo08, and Seo09.
Seo10 and Seo17 are isolated cores not belonging to any filaments.

\begin{figure}[ht!]\centering
\includegraphics[width=\linewidth]{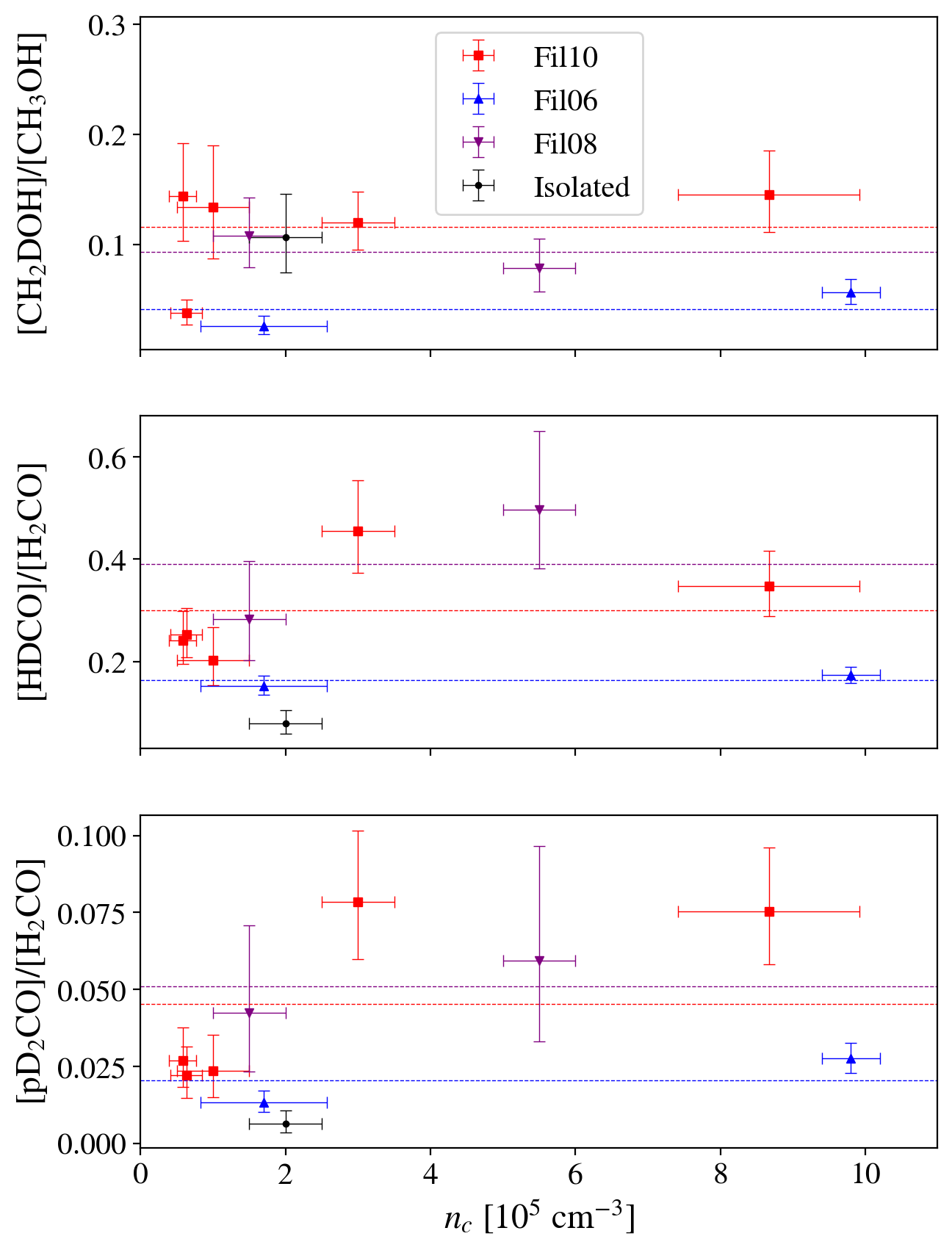}
\caption{The deuterium fractions are plotted as a function of central density.
The shape and color of each datapoint corresponds to the spatial location of the corresponding starless core.
Red squares are located in the eastern Fil10 filament, blue upwards-pointing triangles in the western Fil06 filament, and purple downwards-pointing triangles in the southern Fil08 filament.
Black circles represent isolated cores.
Dashed lines indicate the mean values for each filament. 
\label{fig:nc_dfrac_plot}}
\end{figure}

\begin{figure*}[ht!]\centering
\includegraphics[width=\linewidth]{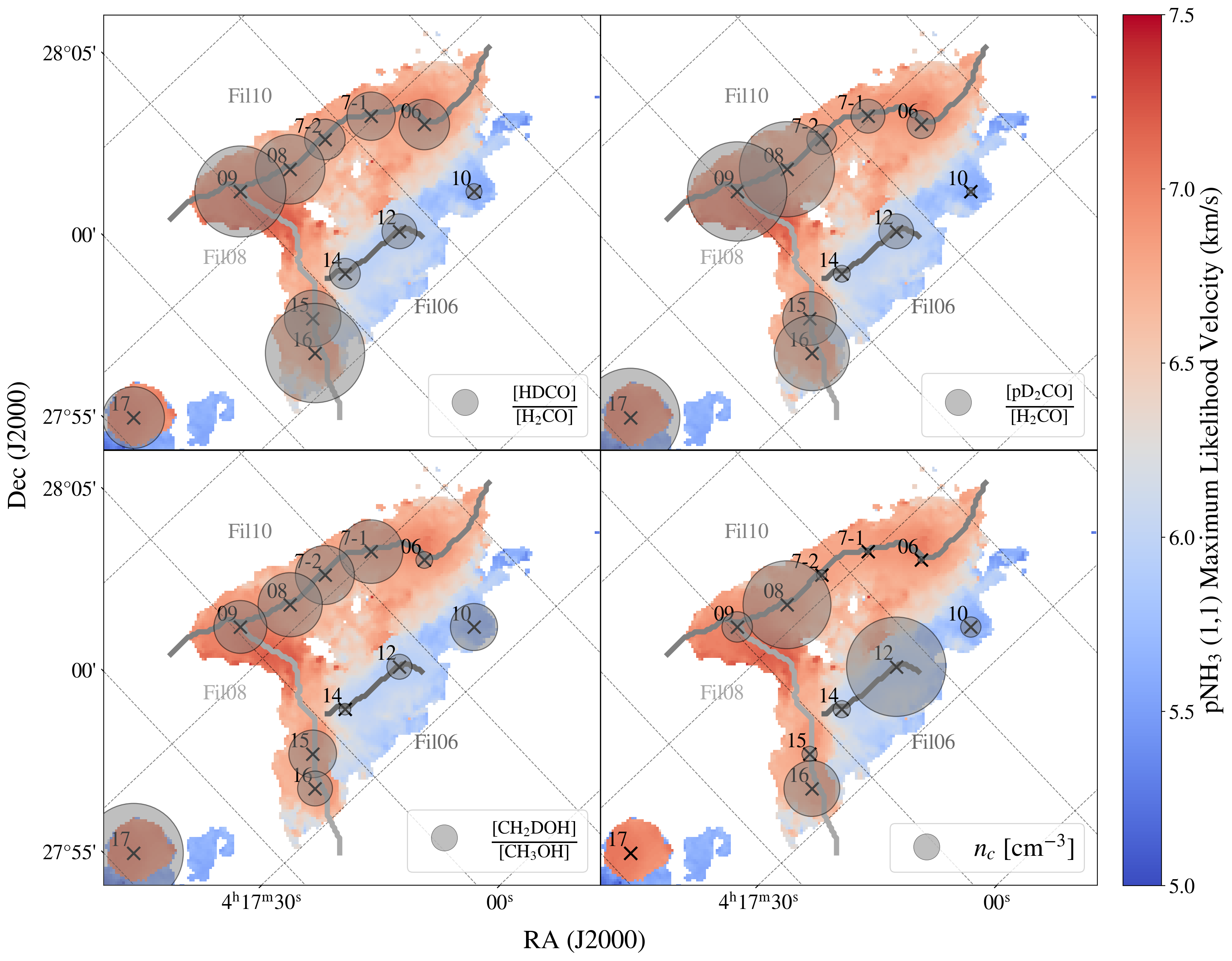}
\caption{Quantities are plotted as grey circles on a map of the maximum likelihood velocity of pNH$_3$ (1,1) in B10.
The circles are overlaid on the corresponding core's location, marked with a cross, and the size of the circle corresponds to the relative magnitude of the quantity.
The filaments are also overlaid in grey.
Going clockwise from the top left the plotted quantities are $\rm [HDCO]/[H_2CO]$, $\rm [pD_2CO]/[H_2CO]$, the core central density $n_c$, and $\rm [CH_2DOH]/[CH_3OH]$.  
\label{fig:spatial_correls}}
\end{figure*}

We find significant spatial correlations in the [HDCO]/[H$_2$CO], [pD$_2$CO]/[H$_2$CO], and [CH$_2$DOH/CH$_3$OH] deuterium fractions and the filaments in which they reside (Figures \ref{fig:nc_dfrac_plot} \& \ref{fig:spatial_correls}).
The Fil08 and Fil10 cores have higher mean deuterium fractions than the Fil06 cores in [HDCO]/[H$_2$CO], [pD$_2$CO]/[H$_2$CO], and [CH$_2$DOH/CH$_3$OH] by factors of 2.0, 2.3, and 2.6 respectively.
In contrast, there is a lack of spatial correlation between the central density of the cores and the filaments in which they reside.
This is consistent with the lack of correlation between the organic deuteration ratios and density of the cores.
For example, the lowest central density cores in Fil08 and Fil06 (Seo15 and Seo14 respectively) are similar to each other in density, but [HDCO/H$_2$CO] and [pD$_2$CO/H$_2$CO] are enhanced by factors of 1.9 and 3.2 respectively in Seo15 relative to Seo14.
As another example, the highest central density cores in Fil10 and Fil06 (Seo08 and Seo12) are also similar to each other in density, but [HDCO/H$_2$CO] and [pD$_2$CO/H$_2$CO] are enhanced by factors of 2.0 and 2.7 respectively in Seo08 relative to Seo12.
One notable exception to the spatial correlation patterns is seen for core Seo06 in Fil10 which has a [CH$_2$DOH/CH$_3$OH] ratio close to the average for the Fil06 cores and is an outlier compared to the other cores in Fil10.

There are numerous potential causes for these differences in organic deuteration ratios between the three filaments.
The observed differences could be driven by different filament-scale evolutionary histories or by physical differences between the filaments such as differences in cosmic ray ionization rate or the average density of the filaments.
We shall explore each of these possibilities in turn.

Deuteration is sensitive to the dynamical history of the filaments and the cores.
For example, if two starless cores with identical initial chemical conditions (i.e. same cosmic ray ionziation rate, same initial abundances, same temperature, etc.) are observed to have the same central density but one core has a high molecular D/H ratio and the other core has a low molecular D/H ratio, then the core with the higher ratio has spent longer dynamically evolving to its present density \citep{Galloway-Sprietsma2022}.
Such differentiation due to differences in dynamical history was seen in monotonically increasing density simulations in \cite{Kong2015} where cores with more rapid evolution rates had lower [N$_2$D$^+$]/[N$_2$H$^+$] than the more slowly evolving cores. 

Making detailed, chemo-magnetohydrodynamical models to model the deuterated organics ratios of each core, including gas phase and grain surface chemistry and which explore a range of initial physical and chemical conditions, is beyond the scope of this current paper.
However, we can make qualitative assessments of the relative evolution rates between filaments.
The higher deuterium fractions observed in Fil08 and Fil10 and lower deuterium fractions in Fil06 seems to indicate that Fil08 and Fil10 have evolved more slowly to their present densities, as a whole, than Fil06. 
This implies that the cores in Fil08 and Fil10 have spent more time near the core's present densities to build up deuterium fractions to their present values compared to the cores in  Fil06.
The core Seo12 in Fil06 is a good example of the argument for different filament-scale dynamical histories as it has a high central density of $\sim 10^6$ cm$^{-3}$ but a relatively low deuterium fraction.
At its high density, the timescale to reach a high deuterium fraction is shorter than for lower density cores in Fil08 and Fil10 (see Figure 14 in \citealt{Galloway-Sprietsma2022} for an example of this for [oNH$_2$D]/[pNH$_3$]).
Therefore, to have a low deuterium fraction compared to the other cores in Fil08 and Fil10, it must have spent a relatively shorter time at its present density. 

Within Fil10, the relatively lower values of [HDCO]/[H$_2$CO] and [pD$_2$CO]/[H$_2$CO] in Seo06, Sci7-1, and Sci7-2 compared to Seo08 and Seo09 could mean that these cores are younger and are less dynamically evolved within Fil10 which would indicate that different portions of Fil10 are evolving at different rates.
Another possibility is that these lower density cores, which are $1/3$ to $1/6$ the central density of Seo09, take longer at their lower densities to reach the deuterium fractions observed toward Seo08 and Seo09 and could therefore still be consistent with a more slowly evolving Fil10 relative to Fil06.

Numerous magneto-hydrodynamical simulations of core evolution within molecular clouds indicate that starless cores can have very diverse dynamical histories.
One recent example is the STARFORGE simulations by \cite{Offner2025}, which found that cores evolve at an incredibly broad range of rates with lifetimes spanning 1 to 10 freefall timescales, and that their central density evolution can be non-monotonic.
They additionally found that core evolution is very stochastic, partially due to turbulence in the cloud and partially due to variability caused by the dendrogram algorithm used to define core boundaries, since cores do not have well-defined boundaries.
It is worth noting, however, that the STARFORGE simulations model a much more active star-forming environment with more significant feedback than present in the Taurus Molecular Cloud.
Previous MHD simulations by \cite{Offner2022} and \cite{Smullen2020} found similar qualitative results for core evolutionary histories.
Another recent example are the numerical simulations by \cite{Moon2025,Moon2025b,Moon2025c} analyzing the evolution of individual cores in turbulent, self-gravitating clouds.
Although their simulations omitted magnetic fields, they found cores following a similarly diverse range of evolutionary paths.
They determined that a large factor in this variability is the strong dependence of the characteristic density of collapse on local cloud properties, especially the amount of turbulence.

While the hypothesis of different dynamical histories for the filaments appears to be plausible, 
another possibility for explaining the spatially different deuterium fraction in the filaments is that the filaments have different physical properties that change the deuterium fractionation timescale and equilibrium values.
We can gain insight into the potential impact of different physical factors on deuterium ratios by comparing with published models that have explored variations in physical and chemical parameter space. 
The chemical models of \cite{Kong2015} systematically studied the effect of variations core density, temperature, heavy element depletion, and the cosmic ray ionization rate ($\zeta$) on the equilibrium [N$_2$D$^+$]/[N$_2$H$^+$] ratio.
The cosmic ray ionization rate, in particular, was found to be a sensitive parameter. 
For example, with input parameters held constant at the fiducial value (H$_2$ density of $5 \times 10^4$ cm$^{-3}$ which corresponds to the lowest central densities observed toward our cores), 
\cite{Kong2015} find that the equilibrium [N$_2$D$^+$]/[N$_2$H$^+$] in static density models decreases by a factor of 6 as $\zeta$ increases from $10^{-18}$ to $10^{-15}$ s$^{-1}$.
These models make for useful comparisons for the B10 cores, although the comparison is not perfect.
N$_2$H$^+$ and N$_2$D$^+$ form exclusively via gas-phase chemistry, whereas H$_2$CO and its deuterated isotopologues form via a mixture of gas phase and grain surface routes.

As an illustrative example, we considered the two densest cores in B10, Seo12 in Fil06 and Seo08 in Fil10.
Both have nearly equal central H$_2$ densities slightly below $10^6$ cm$^{-3}$.
The deuterium fraction for which the difference between Seo12 and Seo08 was the smallest is  [HDCO]/[H$_2$CO] with Seo08 exceeding Seo12 by a factor of 2.0.
We consider two scenarios.
First, we assumed that the cosmic ray ionization rate in Seo08 was equal to a ``standard" value of $\zeta_{std} = 2.5 \times10^{-17}$ s$^{-1}$, and calculated what the $\zeta$ must be in Seo12 for the \cite{Kong2015} equilibrium model [N$_2$D$^+$]/[N$_2$H$^+$] to decrease by the same factor of 2.0.
The result was $\zeta=1.9 \times 10^{-16}$, representing an enhancement by a factor of 7.6 of $\zeta$ in Seo12.
If we consider the reverse scenario with Seo12 now at $\zeta_{std}$, then the resulting deuterium fraction variation exceeded those in the \cite{Kong2015} models, meaning $\zeta$ in Seo08 would have to be on the order of at most $10^{-19}$ s$^{-1}$, which is over 25 times lower than in Seo12.
Such variations across a relatively small region of only 0.4 pc between Fil06 and Fil10 in B10 seem unlikely.
For one, as seen in Figure 14 of \cite{Scibelli2023}, there is no evidence that the strength of the interstellar radiation field is higher on one side of B10 than the other.
This conclusion is further supported by a map of $\zeta$ in a section of NGC 1333 of a similar size (in parsecs) to B10 by \cite{Pineda2024}.
NGC 1333 is significantly different compared to B10, as it contains several protostars and is thus generally warmer and less quiescent than B10.
In regions near protostars, there can be rapid spatial variations in $\zeta$.
In regions relatively far from protostars and the edge of the cloud, however, $\zeta$ is relatively constant.
This means it's unlikely that the variations in deuterium fractions seen in B10 are the solely the result of variations in $\zeta$ between the filaments, although we caution that the exact required variation in $\zeta$ is likely different for our deuterated organic ratios than required for [N$_2$D$^+$]/[N$_2$H$^+$].

\begin{figure*}
    \centering
    \includegraphics[width=0.49\linewidth]{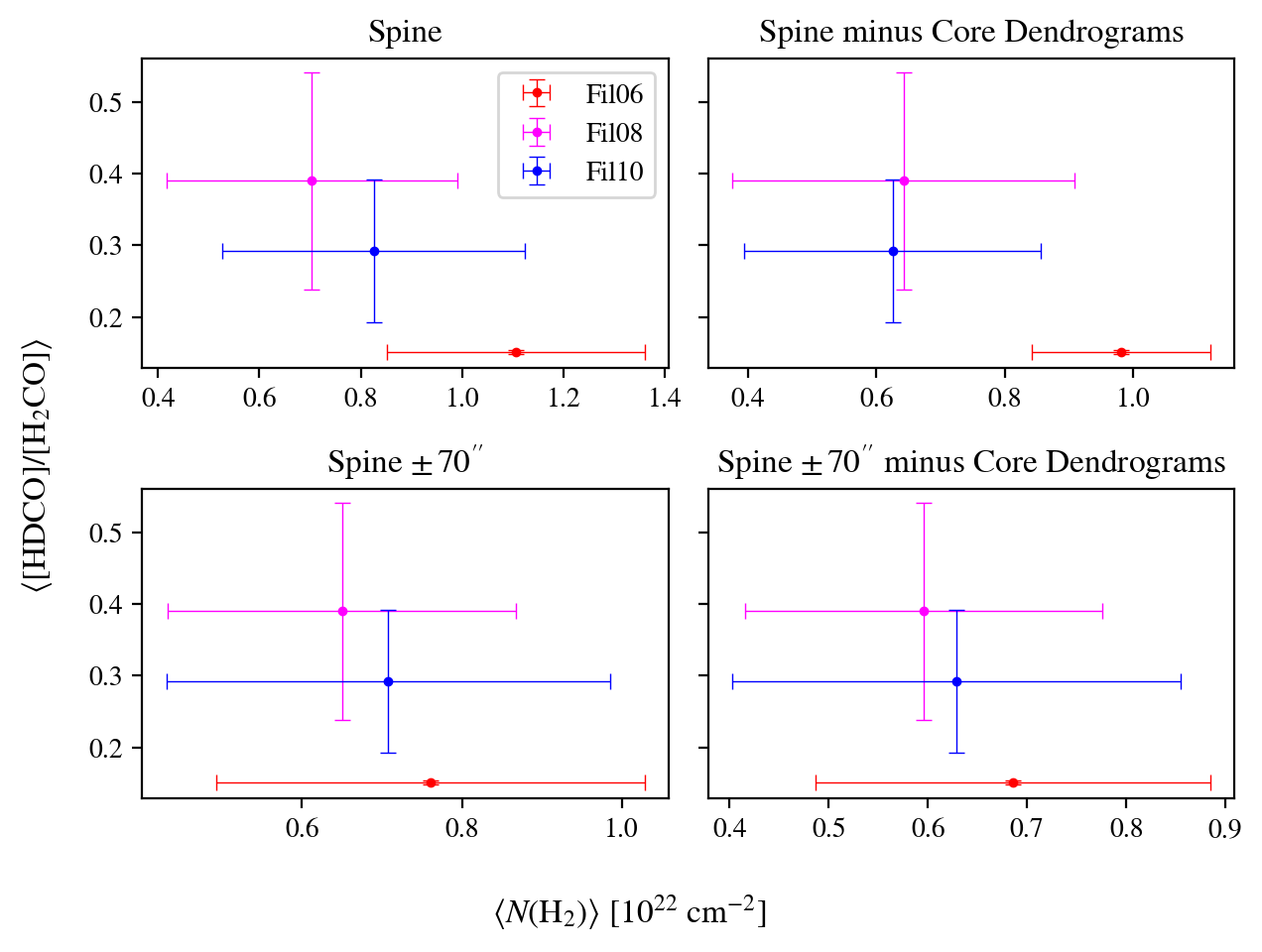}
    \includegraphics[width=0.49\linewidth]{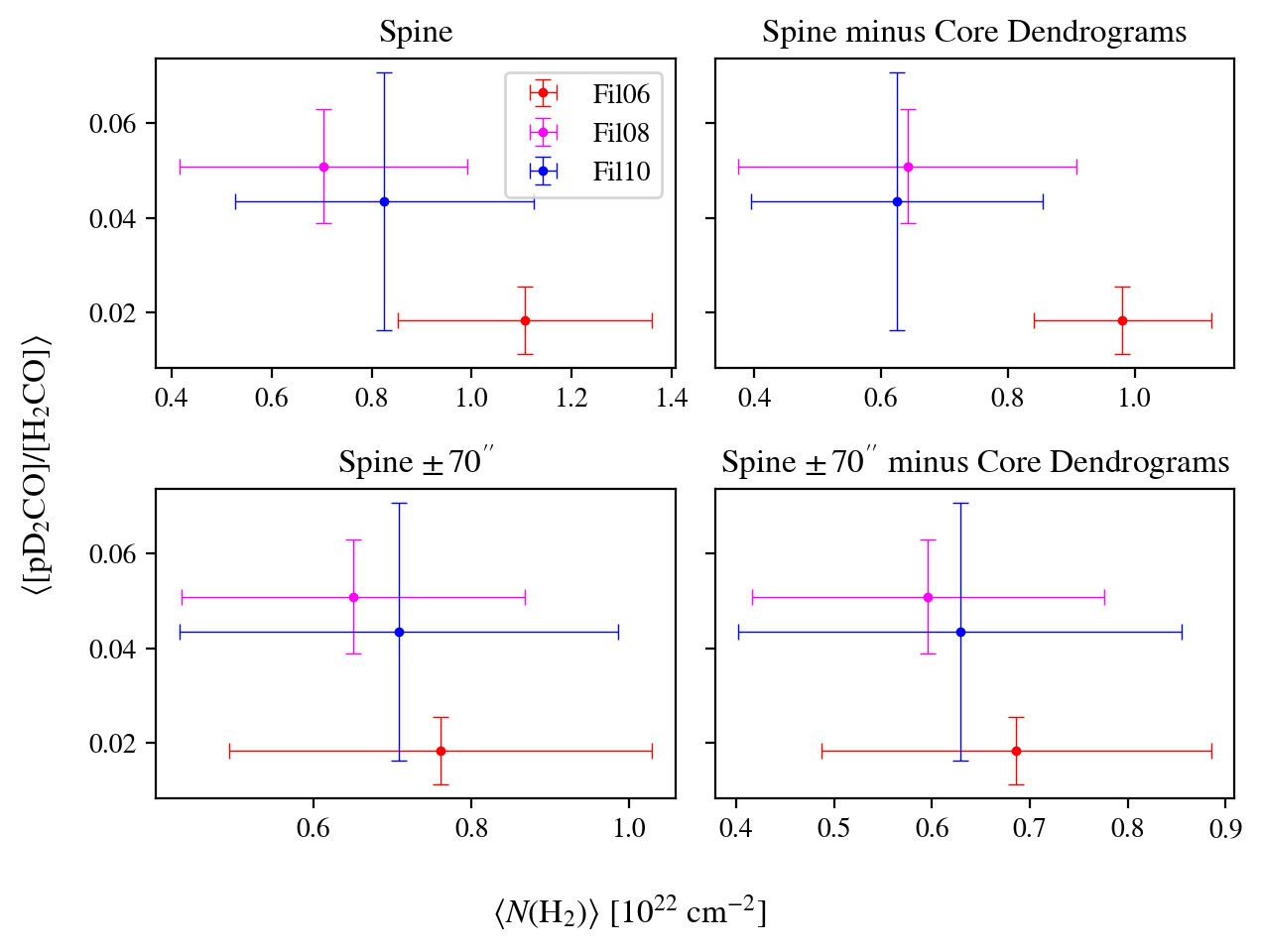}
    \caption{The average deuterium fraction of each B10 filament, labelled by color, plotted against its average H$_2$ column density ($N(\mathrm{H_2})$) in each of the three B10 filaments. 
    The 4 panels on the left use the $\rm[HDCO]/[H_2CO]$ deuterium fraction, and the 4 panels on the right use the $\rm[pD_2CO]/[H_2CO]$ deuterium fraction.
    Each set of 4 panels includes the 4 methods of calculating the average $N(\mathrm{H_2})$ in the filament: averaging $ N(\text{H}_2)$ on pixels along the filament spines and averaging $N(\mathrm{H_2})$ only on pixels within 5 pixels ($\pm0.05$ pc) of the spine and within 3 pixels of the core dendrograms, both including and excluding pixels within the core dendrograms. 
    The errorbars on $\langle N(\mathrm{H_2}) \rangle$ represent the standard deviation of $N(\text{H}_2)$ across the filament.}
    \label{fig:fil_avgNH2}
\end{figure*}

Additional factors to consider in searching for physical differences between the filaments are the average temperature and the average density of the filaments.
We used the average H$_2$ column density calculated by \cite{Singh2022} as a proxy for the average H$_2$ density, since we have no reliable determination of $\langle n_{\text{H}_2}\rangle$ in the filaments outside of the cores.
We calculated $\langle N(\text{H}_2) \rangle$ in the filaments in four different ways, using two basic methods:
one averaging only $N(\mathrm{H_2})$ on pixels along the filament spines, and one averaging $N(\mathrm{H_2})$ only on pixels within 5 pixels ($\pm 0.05$ pc or $\pm 70^{\prime\prime}$) of the spine and within 3.5 pixels of the core dendrograms defined in \cite{Scibelli2023}.
For both methods, we calculated $\langle N(\text{H}_2) \rangle$ both including and excluding the pixels within the core dendrograms.
The results are shown in Figure \ref{fig:fil_avgNH2} plotted against the average $\rm[HDCO]/[H_2CO]$ and $\rm[pD_2CO]/[H_2CO]$ in each filament.
Using the pixels along the spines generally results in a much greater spread of $\langle N(\text{H}_2) \rangle$ between the three filaments. 
This is likely because the spines of Fil08 and Fil10 are much longer than that of Fil06, extending into much more diffuse regions far beyond the locations of the cores.
In both cases, excluding the dendrograms lowered $\langle N(\text{H}_2) \rangle$, although less so in the pixel regions.
Regardless of the method, though, Fil06 actually has the highest $\langle N(\text{H}_2) \rangle$, even if only by a small amount.
This result cannot explain the low average deuterium fractions of Fil06 in comparison deuterium fractions of Fil08 and Fil10.
We then looked at the average temperatures of the filaments.
Using a dust temperature ($T_\mathrm{dust}$) map of B10 from \cite{Singh2022}, we again averaged within $70^{\prime\prime}$ of the filament spines and core dendrograms for each spine.
The filaments are essentially identical in average $T_\mathrm{dust}$, with $13.0\pm0.7$ K for Fil06, $13.5\pm0.6$ K for Fil08, and $13.1\pm0.8$ K for Fil10.
We then repeated this analysis with a maximum likelihood gas kinetic temperature ($T_k$) map of B10 by coauthors B. Svoboda and L. J. Steffes derived from the \texttt{NestFit} re-analysis of pNH$_3$ (1,1) and (2,2) (Steffes et al., in prep.).
The average $T_k$ of the filaments is again effectively identical, with $8.83\pm0.74$ K for Fil06, $8.70\pm0.66$ K for Fil08, and $8.73\pm0.62$ K for Fil10.
In total, in terms of average temperature ($T_\mathrm{dust}$ and $T_k$) the filaments are well within one standard deviation of each other, which is $<1$ K across the board.
Given the spatial proximity of the B10 filaments and the lack of significant physical differences between the filaments, it appears the observed deuterium fractions are tracing the dynamical history; specifically, that Fil06 is dynamically younger and has evolved to its present densities quicker than Fil08 and Fil10.

\subsection{Evolutionary Comparisons}\label{subsec:evo_comp}

\begin{figure*}[ht!]\centering
\includegraphics[width=1\linewidth]{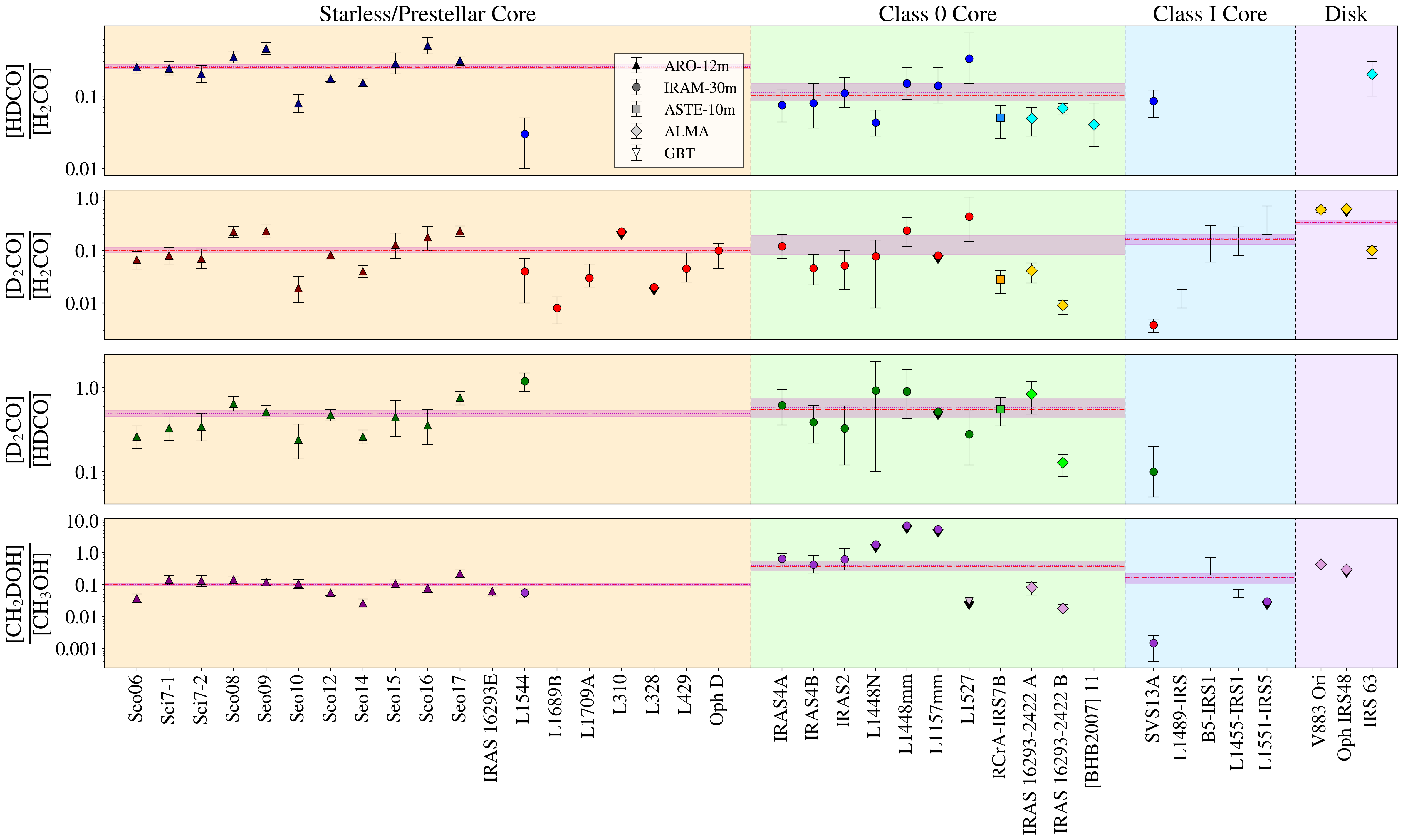}
\caption{Deuterium fractions are plotted for sources grouped into different evolutionary stages: Starless/Prestellar cores (orange shaded region), Class 0 cores (green shaded region), Class I cores (blue shaded region), and interferometric observations in protostellar disks around Class 0/I/II cores (purple shaded region).
The marker shapes and colors correspond to the telescope with which the data was taken; no marker indicates that only a range of values was published rather than a single value with errorbars.
Dashed red lines correspond to the mean deuterium fractions for each evolutionary category with at least two non-upper limit data points.
Dotted magenta lines correspond to median value of the mean deuterium fraction (again excluding upper limits) from a Monte Carlo analysis, with the 1$\sigma$ confidence interval represented by the magenta shaded region.
References for the different types of sources may be found in Section \ref{subsec:evo_comp}.
We note that V883 Ori is transitioning from a Class I to a Class II core and is undergoing an accretion burst \citep{Fadul2025}.
\label{fig:evo_stage_comp}}
\end{figure*}

We compared the deuterium fractions in our core sample to those in other starless and prestellar cores \citep{Scibelli2025, Chacon-Tanarro2019, Bacmann2003}, Class 0 protostars \citep{Parise2006, Watanabe2012, Manigand2020, Persson2018, Evans2023, Sakai2009, Nazari2024}, Class I protostars \citep{Bianchi2017, Mercimek2022}, and Class 0/I/II protoplanetary disks \citep{van_der_Marel2021, Brunken2022, Podio2024, Fadul2025} in the literature.
The resulting plot (Figure \ref{fig:evo_stage_comp}) was adapted from \cite{Podio2024}, with the addition of our B10 core sample and the [CH$_2$DOH]/[CH$_3$OH] deuterium fraction.
Our survey permits a more complete comparison across evolutionary stages, as without the B10 cores, the only starless core with published HDCO data is the well-studied L1544 \citep{Chacon-Tanarro2019}.
We multiplied the [pD$_2$CO]/[H$_2$CO] and [pD$_2$CO]/[HDCO] abundance ratios for the B10 cores by 3 to account for the statistical ortho-to-para ratio of D$_2$CO, which is 6/3 = 2.
The red horizontal dashed lines indicate the mean deuterium fraction across each evolutionary stage excluding upper limits.
We labeled each datapoint by shape according to the telescope used to conduct the observations to check if differences in beam size resulted in systematic differences in the measured deuterium fractions.
We see no such systematic variation across the different evolutionary categories.

To account for the uncertainty in the mean due to the uncertainties in the deuterium fractions of each source, we also repeated the Monte Carlo analysis detailed in Section \ref{subsec:correl} for the mean deuterium fraction of each evolutionary stage, again excluding upper limits (Table \ref{tab:MC_evocomp}).
For cores with only a range of deuterium fractions provided, we used a uniform distribution covering the given range as the PDF to draw values from.
We then took the median and upper and lower 1$\sigma$ confidence intervals of the resulting distribution of mean deuterium fractions, which we plotted on Figure \ref{fig:evo_stage_comp} as a magenta dotted line with a corresponding region shaded in magenta.

\begin{deluxetable*}{llCCR}
\tablewidth{0pt} 
\tablecaption{Monte Carlo Statistics for Mean Deuterium Fractions across Different Evolutionary Stages\label{tab:MC_evocomp}}
\tablehead{
\colhead{Deuterium Fraction $RD$} & \colhead{Evolutionary Stage} &
\colhead{$\langle RD \rangle_\mathrm{obs}$} & 
\colhead{med($\langle RD \rangle$)$_\mathrm{MC}$\tablenotemark{a}} &
\colhead{$\gamma_{1,\langle RD \rangle}$}
}
\startdata
$\mathrm{[HDCO]/[H_2CO]}$ & Starless/Prestellar & 0.25 & 0.26^{+0.02}_{-0.02} & 0.13 \\
  & Class 0 & 0.10 & 0.11^{+0.04}_{-0.03} & 0.66 \\
$\mathrm{[D_2CO]/[H_2CO]}$ & Starless/Prestellar & 0.10 & 0.10^{+0.01}_{-0.01} & 0.12 \\
  & Class 0 & 0.12 & 0.13^{+0.07}_{-0.04} & 0.70 \\
  & Class I & 0.17 & 0.17^{+0.04}_{-0.04} & -0.01 \\
  & Disk & 0.34 & 0.34^{+0.04}_{-0.04} & 0.07 \\
$\mathrm{[D_2CO]/[HDCO]}$ & Starless/Prestellar & 0.49 & 0.50^{+0.04}_{-0.04} & 0.02 \\
  & Class 0 & 0.55 & 0.59^{+0.16}_{-0.14} & 0.20 \\
$\mathrm{[CH_2DOH]/[CH_3OH]}$ & Starless/Prestellar & 0.10 & 0.10^{+0.01}_{-0.01} & 0.07 \\
  & Class 0 & 0.36 & 0.40^{+0.15}_{-0.12} & 0.51 \\
  & Class I & 0.17 & 0.17^{+0.06}_{-0.06} & 0.02 \\
\enddata
\tablecomments{(a) Uncertainties are the 1$\sigma$ confidence intervals of the $\langle RD \rangle$ distribution.}
\end{deluxetable*}

Across the board, the Monte Carlo means are well within 1$\sigma$ of the true mean.
Thus, to quantify the agreement between means across different evolutionary stages, we calculated the number of $\sigma$, $n$, such that $n$ $1\sigma$ intervals of the Monte Carlo means overlapped.
The mean values of [D$_2$CO]/[H$_2$CO] show the most similarity across the most evolutionary stages, with the 
starless and prestellar core mean and the Class I mean within 1$\sigma$ of the Class 0 mean and 1.3$\sigma$ away from each other.
The disk mean is the most discrepant at 5.1$\sigma$ from the starless/prestellar mean, but the disk mean is derived from only two sources, IRS 63 and V883 Ori (Oph IRS48 has only an upper limit).
[D$_2$CO]/[HDCO] only has enough sources for a mean across the starless/prestellar core and Class 0 phases, but those means are well within 1$\sigma$ of each other.
The other deuterium fractions are more discrepant.
For [HDCO]/[H$_2$CO], the starless/prestellar core and Class 0 means are 2.6$\sigma$ apart.
For [CH$_2$DOH]/[CH$_3$OH], Class 0 mean is 2.4$\sigma$ from the starless/prestellar mean and 1.3$\sigma$ from the Class I mean, although the starless/prestellar and Class I means are only 1.0$\sigma$ apart.
However, these deuterium fractions are all less complete than the [D$_2$CO]/[H$_2$CO] sample, particularly in the starless/prestellar stage, where the sample is composed of just the B10 cores, L1544 \citep{Chacon-Tanarro2019}, and for [CH$_2$DOH]/[CH$_3$OH], IRAS 16293E \citep{Scibelli2025}.
The starless/prestellar and Class 0 samples for these deuterium fractions are also the most out of agreement, meaning the incompleteness of the sample may be affecting the means, as the two stages are in very good agreement for [D$_2$CO]/[H$_2$CO].
The completeness of the Class I and especially protoplanetary disk phases are also lacking across the board, with Oph IRS48 having only upper limits for [D$_2$CO]/[H$_2$CO] and  [CH$_2$DOH]/[CH$_3$OH] \citep{Brunken2022, van_der_Marel2021}.

This points towards the inheritance of D$_2$CO molecules in protostellar cores and protostellar disks from the starless and prestellar phase.
In this scenario, these molecules form during the starless core phase, then freeze out towards the central regions of the core as the core grows denser.
After the formation of a protostar, the nearby surrounding environment is heated resulting in the desorption of the deuterated molecules back into the gas phase.
Considering that this agreement is most evident in samples with more sources, the agreement of [HDCO]/[H$_2$CO] and [CH$_2$DOH]/[CH$_3$OH], as well as [D$_2$CO]/[H$_2$CO] in disks, may improve as more surveys of organic deuteration are completed across the phases of low-mass star formation. 
Previous studies have also pointed towards this inheritance of deuteration from this early starless phase.
In particular, protoplanetary disk simulations by \cite{Cleeves2014} showed that reprocessing in disks can not reproduce observed high water deuterium fractions in later stages of star and planet formation, meaning starless cores are the main engines of deuterium fractionation.
ALMA observations of HDO and D$_2$O in Class 0 envelopes are also consistent with inheritance from the prestellar core phase \citep{Jensen2021,Tobin2023}.
High D/H ratios in water and organics in samples of solar system meteorites \citep{Alexander2010, Alexander2012, Busemann2006}, interplanetary dust particles from comets \citep{Messenger2003}, and comets \citep{Cordiner2026,Roth2026,Salazar_Manzano2026} also indicate inheritance from these early stages of star formation.


\section{Conclusion}\label{sec:concl}

We have performed a systematic survey of the starless core population of the Barnard 10 region of the Taurus Molecular cloud in the deuterated isotopologues of formaldehyde, HDCO and pD$_2$CO, as well as the dense gas tracer N$_2$H$^+$.
We also complemented existing observations of CH$_2$DOH and CH$_3$OH by detecting prior non-detections and observing new positions in cores that fragmented with higher resolution continuum imaging.
The main conclusions from this study are:
\begin{enumerate}
    \item Within the B10 cores, the column densities of HDCO and pD$_2$CO do not correlate with that of CH$_2$DOH, and similarly, neither [HDCO]/[H$_2$CO] nor [pD$_2$CO]/[H$_2$CO] correlate with [CH$_2$DOH]/[CH$_3$OH].
    Since CH$_2$DOH deuterates predominately on icy dust grain surfaces, this could indicate that formaldehyde deuteration does not occur primarily on icy grain surfaces. 
    Rather, this shows that formaldehyde deuteration has a significant gas phase route.

    \item There are multiple molecules that correlate well with each other in column density.
    HDCO and pD$_2$CO both correlate well with each other, as well as the beam-averaged H$_2$. 
    N$_2$H$^+$ and pNH$_3$ also correlate well with each other, which is expected as they both trace dense gas.
    They also correlate well with pD$_2$CO, but less so with HDCO, which could indicate that multiply-deuterated isotopologues are better tracers of dense gas than singly-deuterated isotopologues or that D$_2$CO is more dominated by gas-phase formation than HDCO.
    However, spatial maps of these molecules are required to confirm this.

    \item Comparing abundance ratios, including deuterium fractions, of molecules whose abundances tend to peak later (deuterated organics, N$_2$H$^+$, pNH$_3$) vs. earlier (H$_2$CO, CH$_3$OH) in a core's lifetime reveals that 72\% of said ratios correlate well.
    This indicates that, across the board, molecular ratios of late/early tracers behave similarly with respect to each other in the B10 cores and thus supports the notion that deuterium fractions are useful measures of chemical maturity.
    
    \item None of [HDCO]/[H$_2$CO], [pD$_2$CO]/[H$_2$CO], nor [CH$_2$DOH]/[CH$_3$OH] correlate well with any physical or evolutionary parameters in the B10 cores. 
    In particular, the lack of correlation between the deuterium fractions and the core central density $n_c$ means that the B10 cores have a variety of different evolutionary histories in a relatively small ($\sim$ 0.4 pc) region of the Taurus Molecular Cloud.

    \item Comparing [HDCO]/[H$_2$CO], [pD$_2$CO]/[H$_2$CO], nor [CH$_2$DOH]/[CH$_3$OH] spatially in the three filaments in B10 shows significant spatial correlations.
    Specifically, the cores in Fil06 consistently have lower deuterium fractions than cores from other filaments at similar densities.
    It is unlikely that these spatial correlations can be caused by differences in the cosmic ray ionization rate since the filaments are quiescent (no protostars) and are separated by only 0.4 pc in the plane of the sky.
    Additionally, these spatial correlations can not be driven by filament-scale differences in dust temperature, gas kinetic temperature, or H$_2$ column density.
    On average, the filaments are all within 1 K of each other in both temperatures and are also extremely close in average H$_2$ column density, albeit with Fil06 consistently being the densest.
    This thus indicates that whole filaments and spatially-connected sections of filaments may be evolving at different rates from each other, with Fil06 in particular having a shorter evolutionary history at its present range of densities than the other two filaments.
    
    \item Comparing [HDCO]/[H$_2$CO], [D$_2$CO]/[H$_2$CO], [D$_2$CO]/[HDCO], and [CH$_2$DOH]/[CH$_3$OH] in the B10 starless cores to values in starless/prestellar cores, Class 0 and I protostars, and Class 0 and I protoplanetary disks from the literature reveals that mean [D$_2$CO]/[H$_2$CO] does not vary much between evolutionary stages.
    [HDCO]/[H$_2$CO] and [CH$_2$DOH]/[CH$_3$OH] vary more across evolutionary phases, but these deuterium fractions also have significantly less complete samples of sources than [D$_2$CO]/[H$_2$CO].
    This points towards at least D$_2$CO molecules being mostly inherited from the starless and prestellar core phase, rather than being reprocessed during the protostellar phase.
    More surveys of HDCO and CH$_2$DOH across the stages of low-mass star formation may reveal inheritance of these molecules as well.
\end{enumerate}

These results point towards several avenues of future work.
The B10 cores in the present sample are low-mass cores that, if they become unstable in the future and collapse to form protostars, would form M dwarf main sequence stars.
Further systematic surveys, that study the complete starless core population, of organic deuteration toward both low-mass and high-mass star-forming regions are needed to explore relative starless core evolution rates across a wider range of masses and environments.
A larger sample of deuterium fractions in starless and prestellar cores, as well as in later evolutionary stages, would also allow for a more robust investigation of the inheritance of deuterated organics in star and planet formation.
Spatially resolved maps of organic molecules and their deuterated isotopologues across the B10 filaments, along with dense gas tracers like N$_2$H$^+$, would help disentangle some of the correlations (or lack thereof) observed in our sample.
Additionally, a map of the cosmic ray ionization rate in the B10 region similar to the one of NGC 1333 by \cite{Pineda2024} would allow for a more precise comparison of physical conditions across the filaments and measure possible variations in the cosmic ray ionization rate.
Chemo-magnetohydrodynamical simulations of deuteration that follow a variety of starless and prestellar core density evolution paths with different initial conditions are needed for a more quantitative understanding of the relationship between deuterium fractionation and core evolution rates.
This will show exactly how deuterium fractions vary with evolution rate, as well as the effect of non-monotonic density evolution on the observed deuterium fractions. \\

\vspace{0.5cm}
{\large{\textit{Acknowledgments:}}}
We sincerely thank the anonymous referee for careful and thoughtful suggestions that improved this paper.
This work comprised the Senior Honors Thesis for HAL at The University of Arizona.
Support for HAL, YS, LS, and HG was provided by National Science Foundation Astronomy and Astrophysics Grant (AAG) AST-2205474.
HAL was additionally supported by an Arizona NASA Space Grant Internship.
SS acknowledges the National Radio Astronomy Observatory which is a facility of the National Science Foundation operated under cooperative agreement by Associated Universities, Inc.
LS is supported by a National Science Foundation Graduate Research Fellowship Program under Grant No DGC-2137419. 
Any opinions, findings, and conclusions or recommendations expressed in this material are those of the authors and do not necessarily reflect the views of the National Science Foundation.
We are thankful that we have the opportunity to conduct astronomical research on Iolkam Du’ag (Kitt Peak) in Arizona and we recognize and acknowledge the very significant cultural role and reverence that these sites have to the Tohono O’odham Nation. 
We sincerely thank the operators of the Arizona Radio Observatory for their assistance with the observations. 
The 12m Telescope is operated by the Arizona Radio Observatory (ARO), Steward Observatory, University of Arizona, with funding from the State of Arizona, NSF MRI Grant AST-1531366 (PI Ziurys), NSF MSIP grant SV5- 85009/AST- 1440254 (PI Marrone), NSF CAREER grant AST- 1653228 (PI Marrone), and a PIRE grant OISE-1743747 (PI Psaltis).
This research has made use of spectroscopic and collisional data from the EMAA database (https://emaa.osug.fr and https://dx.doi.org/10.17178/EMAA). 
EMAA is supported by the Observatoire des Sciences de l’Univers de Grenoble (OSUG).
This research made use of Photutils, an Astropy package for
detection and photometry of astronomical sources \citep{photutils}.

\vspace{5mm}
\facilities{ARO-12m}

\software{
    asepy \citep{Barlow2026},
    Astropy \citep{astropy2013,astropy2018},
    GILDAS CLASS 
    \citep{Pety2005} (http://www.iram.fr/IRAMFR/GILDAS),
    NetworkX \citep{networkx},
    NumPy \citep{numpy},
    Matplotlib \citep{matplotlib},
    pandas \citep{pandas,pandasv2.1.0},
    Photutils \citep{photutils},
    PySpecKit \citep{Ginsburg2011},
    RADEX \citep{vanderTak2007},
    SciPy \citep{scipy,scipyv1.9.3}
          }


\bibliography{biblio}{}
\bibliographystyle{aasjournal}


\appendix

\restartappendixnumbering

\section{References for Spectroscopic Data}\label{ap:spec_data_refs}

We list here the references for the spectroscopic data of each observed molecule.
The spectroscopic data for CH$_2$DOH was taken from the Jet Propulsion Laboratory Molecular Spectroscopy catalogue (\citealt{JPL}; \url{https://spec.jpl.nasa.gov/}).
This data is contained in the file located at \url{https://spec.jpl.nasa.gov/ftp/pub/catalog/c033004.cat} and was derived from \cite{Quade1980_CH2DOH}, \cite{Su1989_CH2DOH}, \cite{Liu1991_CH2DOH}, \cite{Jacq1993_CH2DOH}, \cite{Mukhopadhyay1997_CH2DOH}, \cite{El_Hilali2011_CH2DOH}, and \cite{Pearson2012_CH2DOH}.
The spectroscopic data for the remaining molecules was taken from the Cologne Database for Molecular Spectroscopy (\citealt{CDMS}; \url{https://cdms.astro.uni-koeln.de/}).
The data for both o/pH$_2$CO can be found in the catalog file at \url{https://cdms.astro.uni-koeln.de/classic/entries/c030501.cat} and was derived from \cite{Takami1968_H2CO}, \cite{Tucker1971_H2CO_HDCO}, \cite{Johnson1972_H2CO}, \cite{Tucker1972_H2CO}, \cite{Chardon1973_H2CO}, \cite{Chardon1977_H2CO}, \cite{Fabricant1977_H2CO_D2CO}, \cite{Dangoisse1978_H2CO_HDCO_D2CO}, \cite{Cornet1980_H2CO}, \cite{Chardon1981_H2CO}, \cite{Bocquet1996_H2CO}, \cite{Muller2000_H2CO}, \cite{Brunken2003_H2CO}, \cite{Eliet2012_H2CO}, and \cite{Muller2017_H2CO}.
The HDCO data is located in the catalog file at \url{https://cdms.astro.uni-koeln.de/classic/entries/c031501.cat} and was derived from \cite{Tucker1971_H2CO_HDCO}, \cite{Johns1977_HDCO}, \cite{Winnewisser1977_HDCO}, \cite{Dangoisse1978_H2CO_HDCO_D2CO}, \cite{Glorieux1978_HDCO}, \cite{Bocquet1999_HDCO_D2CO}, and \cite{Zakharenko2015_HDCO_D2CO}.
The pD$_2$CO data can be found at \url{https://cdms.astro.uni-koeln.de/classic/entries/c032502.cat} and was derived from \cite{Tucker1973_D2CO}, \cite{Chardon1974_D2CO}, \cite{Fabricant1977_H2CO_D2CO}, \cite{Dangoisse1978_H2CO_HDCO_D2CO}, \cite{Baskakov1988_D2CO}, \cite{Bocquet1999_HDCO_D2CO}, \cite{Lohilahti2004_D2CO}, and \cite{Zakharenko2015_HDCO_D2CO}.
All data for A/E-CH$_3$OH is at \url{https://cdms.astro.uni-koeln.de/classic/entries/c032504.cat} and was derived from \cite{Ivash1952_CH3OH}, \cite{Lees1968_CH3OH}, \cite{Pickett1981_CH3OH}, \cite{Herbst1984_CH3OH}, \cite{Sastry1984_CH3OH}, \cite{Anderson1990_CH3OH}, \cite{Matsushima1994_CH3OH}, \cite{Belov1995_CH3OH}, \cite{Odashima1995_CH3OH}, \cite{Tsunekawa1995_CH3OH}, \cite{Xu1995a_CH3OH}, \cite{Xu1995b_CH3OH}, \cite{Muller2004_CH3OH}, \cite{Voronkov2006_CH3OH}, \cite{Xu2008_CH3OH}, \cite{Voronkov2011_CH3OH}, \cite{Voronkov2014_CH3OH}, and \cite{Coudert2015_CH3OH}.
The data for N$_2$H$^+$ is located at \url{https://cdms.astro.uni-koeln.de/classic/entries/c029506.cat} and was derived from \cite{Havenith1990_N2H+}, \cite{Verhoeve1990_N2H+}, \cite{Caselli1995_N2H+}, \cite{Amano2005_N2H+}, and \cite{Cazzoli2012_N2H+}.
We used the hyperfine frequencies listed in \cite{PaganiN2HP2009}.

\section{Details of Column Density Calculations}\label{ap:columndensityappexdix}

In this appendix, we give the details for the calculation of the column densities for A/E-CH$_3$OH, CH$_2$DOH, pD$_2$CO, N$_2$H$^+$ and pNH$_3$.
The calculation of the column densities of HDCO and o/pH$_2$CO are described in Section \ref{subsec:coldens_calcs}.

\subsection{Calculating CH$_3$OH Column Densities with RADEX}\label{subsubsec:meth_coldens}

Although \cite{Scibelli2020} had already calculated CH$_3$OH column densities  for most of our core sample using the non-LTE radiative transfer code \texttt{RADEX} \citep{vanderTak2007}, their analysis was performed prior to the discovery of Sci7-1 and Sci7-2 as distinct cores.
Additionally, \cite{Dagdigian2024} has recently published new estimates of the A/E-CH$_3$OH-pH$_2$ collision rates on the EMAA database (\citealt{EMAA}; \url{https://emaa.osug.fr}).
We note that \texttt{RADEX} calculations work well for the 96.7 GHz A/E-CH$_3$OH transitions because, as was demonstrated in \cite{Scibelli2020}, the column density is a weak function of the core density input into the calculation, varying by less than $10$\% for cores in B10 with $\log n_{\rm{H}_2} = 4.5 - 6.0$ cm$^{-3}$.

We recalculated $N$(CH$_3$OH) for all cores in our sample using the same \texttt{RADEX} methodology as \cite{Scibelli2020}.
For each core, we fixed the gas kinetic temperature $T_k$ to equal the average and standard deviation of $T_k$ within the CH$_3$OH FWHM beamsize of the ARO 12m Table \ref{tab:temps_tau_thetas}.
We also calculated the average H$_2$ volume density within our average methanol beam on the 12m using a spherical average,
\begin{equation}\label{eq:nH2}
    n_{H_2} = \frac{3}{2}\frac{\langle N(\mathrm{H}_2)\rangle}{D\theta_\mathrm{beam}}
\end{equation}
where $\langle N(\mathrm{H}_2)\rangle$ is the average H$_2$ column density within the average CH$_3$OH beam, $D$ is the average distance from Earth to the Taurus Molecular Cloud (135 pc, from \citealt{Schlafly2014}), and $\theta_\mathrm{beam}$ is the average FWHM size of the beams for the relevant A-CH$_3$OH and E-CH$_3$OH transitions on the 12m\footnote{Note that if $D$ is converted to cm and $\theta_{\mathrm{beam}}$ is converted to radians, then $n_{H_2}$ has units of cm$^{-3}$.}.
We then allowed the input methanol column density to vary until the model integrated intensity calculated by \texttt{RADEX} matched the observed integrated intensity.
We also repeated this process with the appropriate upper and lower limits of $T_k$, $n_{H_2}$, $\Delta v$, and $I_\mathrm{obs}$ to get upper and lower limits on the column densities of A-CH$_3$OH and E-CH$_3$OH. 
Finally, we estimated the total methanol column density of a core to be $N$(CH$_3$OH) = $N$(A-CH$_3$OH) + $N$(E-CH$_3$OH).

\cite{Scibelli2020} had used  A/E-CH$_3$OH-pH$_2$ collision rates from \cite{Rabli2010}, so as a check, we calculated methanol column densities using both the \cite{Dagdigian2024} (D24) and \cite{Rabli2010} (RF10) rates.
Our new calculations with the RF10 rates were very close to the ones published in \cite{Scibelli2020} with the ratios of our values to \cite{Scibelli2020} of 1.013 for $N$(A-CH$_3$OH) and 1.012 for $N$(E-CH$_3$OH).  
The ratios of excitation temperatures was also very similar at 1.042 and 0.990 respectively for A-CH$_3$OH and  E-CH$_3$OH.
These small differences are likely due to differences in the volume density of H$_2$ in our calculations and those of \cite{Scibelli2020} derived from an improved H$_2$ column density map \citep{Singh2022}.
Our new calculations with the D24 rates agree well for A-CH$_3$OH with D24-derived column densities 1.004 times the RF10-derived column densities on average.
The D24 $T_{ex}$ values are only 1.006 times the RF10 ones.
However, the results for E-CH$_3$OH differed significantly, with the D24/RF10 column density ratio averaging 1.219 and the excitation temperature ratio averaging 0.980.
Additionally, unlike A-CH$_3$OH, for E-CH$_3$OH the D24 column densities were always higher than they were with the RF10 rates, and the D24 excitation temperatures always below the RF10 ones.
A comparison by \cite{Dagdigian2024} between their rates and the \cite{Rabli2010} rates indeed shows greater disagreement for E-CH$_3$OH.
However, it is not clear which set of collision rates is more accurate. 
We report the excitation temperatures, column densities, and resulting methanol deuterium fractions and [D]/[H] ratios from the D24 rates, since the RF10 results differ very little from \cite{Scibelli2020}.
For $N$(CH$_2$DOH) we only use the D24 A-CH$_3$OH excitation temperatures Table \ref{tab:temps_tau_thetas} since they differed very little from the RF10 ones.
Which set of CH$_3$OH column densities is used ultimately does not affect the results of our correlation analysis (which is described in Section \ref{subsec:correl}), though, as the ranks of the column densities remain identical.

\subsection{CTEX Column Density Analysis for pD$_2$CO, CH$_2$DOH, and N$_2$H$^+$} \label{subsubsec:CTEX}

Like for HDCO, the observed transitions of pD$_2$CO and CH$_2$DOH are optically thin in our sources.
However, we only observed one transition for each molecule and cannot simultaenously constrain $N_{tot}$ and $T_{ex}$.
For pD$_2$CO, we assumed that $T_{ex}$ was the same as for HDCO (Table \ref{tab:temps_tau_thetas}) to find pD$_2$CO column density with uncertainties (Table \ref{tab:coldens}). 

We determined column densities for CH$_2$DOH by assuming the same excitation temperatures from an analysis of the A-CH$_3$OH 96.7 GHz transition, which had a similar $E_u/k$ as our CH$_2$DOH transition (see Section \ref{subsubsec:meth_coldens}).
Just prior to the submission of this paper, new calculations became available for CH$_2$DOH on the Lille Spectroscopic Database (\citealt{LSD}; \url{https://lsd.univ-lille.fr/table/42}) based upon laboratory data from \cite{Oyama2023} and \cite{Coudert2014}.
The partition function and $E_u/k$ are identical to the JPL catalog entry, but the frequency of the transition is 90.7 kHz higher and $A_{ul} = 2.33 \times 10^{-6}$ s$^{-1}$ which is $1.15$ times higher than the JPL Catalog value of $A_{ul} = 2.02 \times 10^{-6}$ s$^{-1}$.
Since the column densities of CH$_2$DOH are calculated with Case 1 of Equation \ref{eqn:CTEX_Ntot}, then using the Lille Database entry instead of the JPL catalog entry would result in our reported column densities and errors being scaled by a factor of $0.87$.
We do not apply this correction to our column densities so that we may make fair comparisons to CH$_2$DOH column densities calculated in the literature which used the JPL catalog.

While \cite{Tafalla2015} had already mapped B10 in N$_2$H$^+$ $1 \rightarrow 0$ at higher spatial resolution than our observations using the IRAM 30m radio telescope, the published map has gaps that make convolving it to the beam size of the 12m for comparison with other molecules in our survey challenging.
Also, the signal-to-noise made hyperfine fitting, and calculating an accurate optical depth, a challenge for some parts of the map.
Therefore, we re-observed N$_2$H$^+$ $1 \rightarrow 0$ with the ARO 12m telescope but needed to consider the effect of a finite source size and filling fraction within the ARO beam.
For a Gaussian source intensity distribution with FWHM of $\theta_s$ on the sky the filling fraction is 
\begin{equation}\label{eq:fillfrac}
    f = \frac{\theta_s^2}{\theta^2_s+\theta_\mathrm{beam}^2}
\end{equation}
for a Gaussian beam FWHM $\theta_\mathrm{beam} = \theta_{12} = 64.71^{\prime\prime}$ for the ARO 12m and $\theta_\mathrm{beam} = \theta_{30} = 27.83^{\prime\prime}$ for the IRAM 30m at the N$_2$H$^+$ $1 \rightarrow 0$ frequency.
N$_2$H$^+$ $1 \rightarrow 0$ is an excellent dense gas tracer that has intensity distributions that typically peak coincident with the dust continuum peak, therefore a centrally-peaked Gaussian intensity distribution is a reasonable assumption \citep{Caselli2002n2hp,Tafalla2015,Spezzano2017}. 
We calculate $\theta_s$ from the ratio of observed intensities, $R_{30/12} = I_{30}/I_{12}$ using 
\begin{equation}
    \theta_s = \sqrt{\frac{\theta_{12}^2 - R_{30/12}\theta_{30}^2}{R_{30/12}-1}} \ \ \ .
\end{equation}
Thanks to high signal-to-noise spectra observed with the ARO 12m telescope, the hyperfine splitting of N$_2$H$^+$ $1 \rightarrow 0$ was fit using \texttt{GILDAS-CLASS} to derive the total optical depth, $\tau_{tot}$ Table \ref{tab:temps_tau_thetas}.
Using our calculated source sizes Table \ref{tab:temps_tau_thetas}, we then corrected our observed N$_2$H$^+$ $1 \rightarrow 0$ spectra by dividing the spectra by the appropriate beam filling fraction, $f$. 
We fit filling fraction corrected line profiles using the built-in N$_2$H$^+$ $1 \rightarrow 0$ hyperfine profile fitter in \texttt{PySpecKit} \citep{Ginsburg2011} to get the excitation temperature $T_{ex}$ for each source.
We then calculated the N$_2$H$^+$ column density from Case 2 of Equation \ref{eqn:CTEX_Ntot}.

\subsection{Calculating pNH$_3$ Column Densities}

We calculate pNH$_3$ column densities at the core peak locations within the $33^{\prime\prime}$ FWHM main beam of the 100m Green Bank Telescope from the (1,1) and (2,2) observations in \cite{Seo2015}.
Since the pNH$_3$ K = 1 and K = 2 rotational ladders are not connected via electric dipole transitions, the traditional single-temperature CTEX approximation does not work.
Ammonia column densities, with well determined optical depths, are determined from a hybrid version of the Case 2 CTEX approximation that uses $T_{ex}$ for the (1,1) transition and a rotation temperature, $T_{rot}$, to connect the different K ladders. 
We first calculated the optical depth of the central hyperfine peak\footnote{The complete hyperfine pattern of the pNH$_3$ (1,1) transition consists of 18 hyperfine transitions (see \citealt{Kukolich1967,Mangum2015}).  Due to the stronger coupling with the $^{14}$N nucleus ($\vec{F_1} = \vec{J} + \vec{I}(^{14}$N)), the dominant hyperfine pattern results in 5 peaks composed of blended hyperfine lines due to the weaker coupling with the $^{1}$H  nuclei ($\vec{F} = \vec{F_1} + \vec{I}_{tot}(^{1}$H$_3$) which are in an $I_{tot} = 1/2$ spin state for pNH$_3$).
The central peak contains 8 blended hyperfine transitions from $F_1 = 1 \rightarrow 1$ and $F_1 = 2 \rightarrow 2$ with an optical depth that is $1/2$ of the total optical depth summed over all 18 hyperfine components.} ($\tau_{(1,1,m)} = \tau_{\rm{tot}}/2$), FWHM linewidth ($\Delta v$), and $T_{ex}$\footnote{The ratio of the first and fourth parameters in a \texttt{GILDAS-CLASS} hyperfine fit are equal to $J_{\nu}(T_{ex}) - J_{\nu}(T_{cmb})$.} at each of the core peak locations by fitting the hyperfine pattern using \texttt{GILDAS-CLASS}.
The combined column density of (1,1), summed over the upper and lower energy levels of the transition, is determined from (see \citealt{Friesen2009})
\begin{equation}
    \begin{split}
        N(1,1) &= N(1,1)^s + N(1,1)^a \\
        &= \frac{4 \pi^{3/2} \nu_{(1,1)}^3}{\sqrt{\ln(2)} A_{(1,1)}c^3} \frac{1 + e^{-h\nu_{(1,1)}/kT_{ex}}}{1 - e^{-h\nu_{(1,1)}/kT_{ex}}} \,2\tau_{(1,1,m)} \Delta v_{(1,1)} \; ,
    \end{split}
\end{equation}
where $N(1,1)^s$ and $N(1,1)^a$ indicate the column densities of the symmetric and antisymmetric inversion levels respectively.
The rotation temperature is defined to relate the level populations of the $(1,1)$ and $(2,2)$ energy levels (see Equation E9 in  \citealt{Mangum2015}), 
\begin{equation}
    T_{rot} = \frac{-(E_{(2,2)} - E_{(1,1)})/k}{ \ln \left\{ \frac{-243 \Delta v_{(2,2)}}{860 \tau_{(1,1,m)}\Delta v_{(1,1)}}  L \right\}} \; ,
\end{equation}
where $(E_{(2,2)} - E_{(1,1)})/k \approx 41.5$ K and 
\begin{equation}
    L = \ln \left[ 1 - \frac{T_A^*(2,2,m)}{T_A^*(1,1,m)} \left(1 - e^{-\tau_{(1,1,m)}} \right) \right] \: .
\end{equation}
This rotation temperature is then used to calculate the pNH$_3$ column density with a partition function summed over only the pNH$_3$ (1,1) and (2,2) energy levels,
\begin{equation}
    N(\mathrm{pNH_3}) = N(1,1) \left[ 1 + \frac{5}{3}e^{-\frac{(E_{(2,2)} - E_{(1,1)})}{kT_{rot}}} \right] \;.
\end{equation}
At the low $T_{rot}$ and T$_{ex}$ observed in the B10 cores, the level populations in the $(2,1)^{s}$, $(2,1)^{a}$, and higher energy levels may be ignored.

\section{Details of the Correlation Analysis} \label{ap:correl_details}

In this appendix, we give further details for the Monte Carlo analysis of the Spearman rank correlation coefficients of our comparisons between quantities.

\subsection{Spearman Rank Correlation Coefficient Distributions from Monte Carlo Methods}\label{ap:MC_PDFS}
\begin{figure}[ht!]
    \centering
    \includegraphics[width=\linewidth]{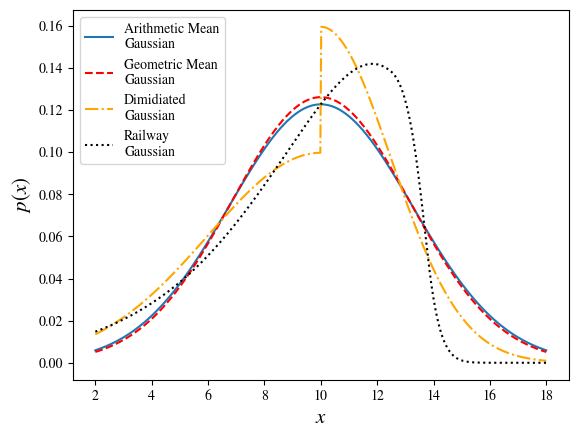}
    \caption{Examples of the probability distribution functions we used for data with asymmetric errorbars, here shown for a hypothetical datapoint $x_o=10.0^{+2.5}_{-4.0}$. }
    \label{fig:pdf_ex}
\end{figure}

We performed Monte Carlo sampling to assess the effect of errors on the calculation of $r_s$ using several different types of PDF with the \texttt{asepy} Python package\footnote{https://github.com/igvgit/AsymmetricErrorsPy} \citep{Barlow2026}.
For data with even errorbars, we always used the usual normal Gaussian distibution
\begin{equation}\label{eq:gaussian}
    p(x) = \frac{1}{\sqrt{2\pi}\sigma}e^{-\frac{(x-\mu)^2}{2\sigma^2}}
\end{equation}
with $p$ the probability density, $x$ the value of the datapoint, $\mu$ the mean (the observed data), and $\sigma$ the standard deviation/width (the uncertainty in the data).
However, our data frequently had uneven errorbars. We tried four different types of PDF to account for this unevenness.
The first two were Gaussians where we defined the width to be either the arithemtic mean of the upper and lower errorbars ($\sigma^+$ and $\sigma^-$ respectively),
\begin{equation}\label{eq:amean}
    \sigma = \frac{\sigma^+ + \sigma^-}{2} \;\;,
\end{equation}
or the geometric mean of the errorbars,
\begin{equation}\label{eq:gmean}
    \sigma = \sqrt{\sigma^+\sigma^-} \ \ .
\end{equation}
Another type of distribution we tried was the dimidiated Gaussian \citep{Barlow2026}, as described in Equation \ref{eq:dimid_gauss}.
This PDF essentially uses the left half of a Gaussian with $\sigma=\sigma^-$ and the right half of a Gaussian with $\sigma=\sigma^+$.
The dimidiated Gaussian has the benefit of being a simple representation of asymmetric errors, but also has an unphysical discontinuity at the median $M$. 
The final PDF we used, the railway Gaussian, circumvents the issue of the discontinuity by creating a smooth, but distorted Gaussian. 
The method for calculating the expression for a railway Gaussian is discussed in depth in Appendix A.3 of \cite{Barlow2026}.
Examples of each type of Gaussian PDF we used for data with asymmetric uncertainties, are shown in Figure \ref{fig:pdf_ex}.
All four of these types of PDFs are equivalent to a normal Gaussian when $\sigma^+ = \sigma^-$.

\begin{figure*}[ht!]\centering
\includegraphics[width=1\linewidth]{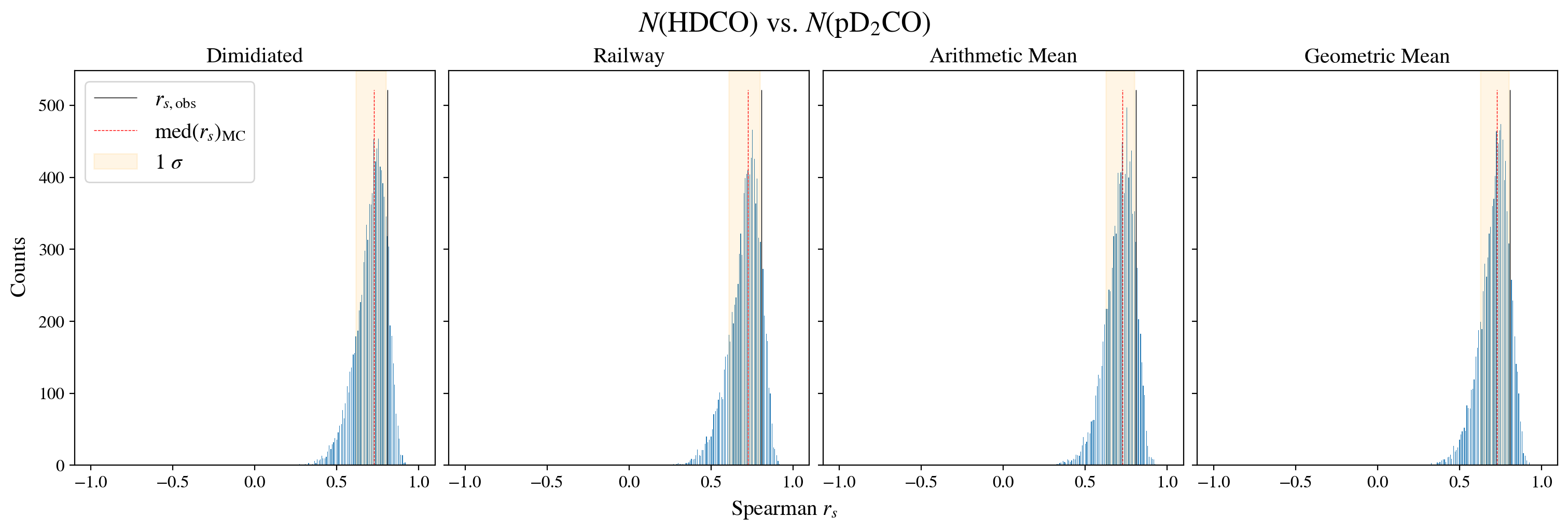}
\caption{Histograms of the distributions of $r_s$ resulting from the different PDFs for $N$(HDCO) vs. $N$(pD$_2$CO).
The median (red dashed line), 1 $\sigma$ confidence interval (orange shaded region) and $r_{s,\mathrm{obs}}$ (black line) are shown on each.
\label{fig:pdf_agreement}}
\end{figure*}

In general, the distributions derived from the four different PDFs were very similar. 
An example for the correlation of $N$(HDCO) vs. $N$(pD$_2$CO) is shown in Figure \ref{fig:pdf_agreement}.
The median with $1\sigma$ uncertainties and the mean with the standard deviation are $0.73^{+0.07}_{-0.11}$ and $0.71\pm0.09$ for the dimidiated Gaussian, $0.73^{+0.07}_{-0.12}$ and $0.71\pm0.10$ for the railway Gaussian, $0.73^{+0.07}_{-0.10}$ and $0.71\pm0.09$ for the arithmetic mean Gaussian, and $0.73^{+0.07}_{-0.10}$ and $0.72\pm0.09$ for the geometric mean Gaussian.
Given these very small differences between the PDFs, we chose to use the distributions derived from the dimidiated Gaussian PDF, since it accounts for uneven uncertainties, unlike arithmetic and geometric mean Gaussian PDFs, while being easier to implement than the railway Gaussian PDF.

\subsection{Spearman Rank Correlation Coefficient Distribution Statistics}\label{ap:spearman_MC_stats}

We include here tables of the Monte Carlo distribution and jackknife resampling statistics for the column density (Table \ref{tab:MC_jk_N}), deuterium fraction (Table \ref{tab:MC_jk_RD}), and molecular abundance ratio (Table \ref{tab:MC_jk_molratios}) comparisons.

\begin{deluxetable*}{llRRRR}
\tablewidth{0pt} 
\tablecaption{Monte Carlo \& Jackknife Resampling Statistics for Column Density Comparisons \label{tab:MC_jk_N}}
\tablehead{
\colhead{Quantity 1} & \colhead{Quantity 2} &
\colhead{$r_{s,\mathrm{{obs}}}$} & 
\colhead{med($r_s$)$_\mathrm{MC}$\tablenotemark{a}} &
\colhead{$\gamma_{1,r_s}$} &
\colhead{$\langle r_s \rangle_\mathrm{jk}$\tablenotemark{b}}
} 
\startdata
$N(\mathrm{CH_2DOH})$ & $N(\mathrm{CH_3OH})$ & -0.25 & -0.23^{+0.17}_{-0.17} & 0.17 & -0.24\pm0.12 \\
  & $N(\mathrm{HDCO})$ & 0.20 & 0.17^{+0.17}_{-0.17} & -0.09 & 0.20\pm0.12 \\
  & $\langle N(\mathrm{H_2}) \rangle_\mathrm{beam}$ & 0.06 & 0.05^{+0.19}_{-0.20} & 0.01 & 0.06\pm0.12 \\
  & $N(\mathrm{H_2CO})$ & -0.06 & -0.01^{+0.19}_{-0.19} & -0.11 & -0.06\pm0.12 \\
  & $N(\mathrm{N_2H^+})$ & 0.28 & 0.27^{+0.18}_{-0.18} & 0.03 & 0.28\pm0.12 \\
  & $N(\mathrm{pD_2CO})$ & 0.39 & 0.35^{+0.18}_{-0.19} & -0.14 & 0.39\pm0.12 \\
  & $N(\mathrm{pNH_3})$ & 0.32 & 0.36^{+0.22}_{-0.22} & -0.16 & 0.32\pm0.12 \\
$N(\mathrm{CH_3OH})$ & $N(\mathrm{HDCO})$ & 0.50 & 0.44^{+0.09}_{-0.09} & -0.21 & 0.50\pm0.07 \\
  & $\langle N(\mathrm{H_2}) \rangle_\mathrm{beam}$ & 0.31 & 0.39^{+0.14}_{-0.11} & 0.26 & 0.31\pm0.11 \\
  & $N(\mathrm{H_2CO})$ & 0.56 & 0.54^{+0.09}_{-0.08} & 0.16 & 0.56\pm0.09 \\
  & $N(\mathrm{N_2H^+})$ & 0.09 & 0.05^{+0.06}_{-0.09} & -0.28 & 0.09\pm0.13 \\
  & $N(\mathrm{pD_2CO})$ & -0.02 & -0.08^{+0.14}_{-0.15} & 0.02 & -0.02\pm0.10 \\
  & $N(\mathrm{pNH_3})$ & -0.13 & -0.09^{+0.14}_{-0.13} & 0.17 & -0.13\pm0.13 \\
$N(\mathrm{HDCO})$ & $\langle N(\mathrm{H_2}) \rangle_\mathrm{beam}$ & 0.73 & 0.71^{+0.07}_{-0.09} & -0.39 & 0.72\pm0.05 \\
  & $N(\mathrm{H_2CO})$ & 0.53 & 0.48^{+0.09}_{-0.09} & -0.23 & 0.52\pm0.10 \\
  & $N(\mathrm{N_2H^+})$ & 0.56 & 0.55^{+0.06}_{-0.08} & -0.01 & 0.56\pm0.08 \\
  & $N(\mathrm{pD_2CO})$ & 0.81 & 0.73^{+0.07}_{-0.11} & -0.81 & 0.80\pm0.04 \\
  & $N(\mathrm{pNH_3})$ & 0.38 & 0.39^{+0.14}_{-0.13} & 0.07 & 0.38\pm0.10 \\
$N(\mathrm{H_2CO})$ & $\langle N(\mathrm{H_2}) \rangle_\mathrm{beam}$ & 0.12 & 0.10^{+0.12}_{-0.13} & 0.04 & 0.12\pm0.12 \\
  & $N(\mathrm{N_2H^+})$ & -0.20 & -0.20^{+0.11}_{-0.10} & 0.38 & -0.20\pm0.13 \\
  & $N(\mathrm{pD_2CO})$ & 0.08 & 0.05^{+0.13}_{-0.15} & -0.12 & 0.08\pm0.12 \\
  & $N(\mathrm{pNH_3})$ & -0.42 & -0.29^{+0.17}_{-0.15} & 0.57 & -0.41\pm0.10 \\
$N(\mathrm{N_2H^+})$ & $N(\mathrm{H_2})_\mathrm{peak}$ & 0.75 & 0.76^{+0.05}_{-0.05} & -0.39 & 0.75\pm0.06 \\
  & $N(\mathrm{pD_2CO})$ & 0.78 & 0.75^{+0.07}_{-0.07} & -0.13 & 0.78\pm0.04 \\
  & $N(\mathrm{pNH_3})$ & 0.93 & 0.87^{+0.06}_{-0.11} & -1.10 & 0.92\pm0.02 \\
$N(\mathrm{pD_2CO})$ & $\langle N(\mathrm{H_2}) \rangle_\mathrm{beam}$ & 0.76 & 0.62^{+0.11}_{-0.13} & -0.31 & 0.76\pm0.06 \\
  & $N(\mathrm{pNH_3})$ & 0.73 & 0.68^{+0.10}_{-0.11} & -0.18 & 0.72\pm0.05 \\
$N(\mathrm{pNH_3})$ & $N(\mathrm{H_2})_\mathrm{peak}$ & 0.63 & 0.64^{+0.10}_{-0.14} & -0.43 & 0.62\pm0.08 \\
\enddata
\tablecomments{(a) Uncertainties are the 1$\sigma$ confidence intervals of the $r_s$ distribution. (b) Uncertainties are the standard deviation of the $r_s$ jackknife replicates.}
\end{deluxetable*}

\begin{deluxetable*}{llRRRR}
\tablewidth{0pt} 
\tablecaption{Monte Carlo \& Jackknife Resampling Statistics for Deuterium Fraction Comparisons \label{tab:MC_jk_RD}}
\tablehead{
\colhead{Quantity 1} & \colhead{Quantity 2} &
\colhead{$r_{s,\mathrm{{obs}}}$} & 
\colhead{med($r_s$)$_\mathrm{MC}$\tablenotemark{a}} &
\colhead{$\gamma_{1,r_s}$} &
\colhead{$\langle r_s \rangle_\mathrm{jk}$\tablenotemark{b}}
} 
\startdata
$\mathrm{[CH_2DOH]/[CH_3OH]}$ & $\langle n_\mathrm{H_2} \rangle_\mathrm{dendro}$ & 0.08 & -0.03^{+0.24}_{-0.23} & 0.01 & 0.08\pm0.13 \\
  & $\mathrm{[HDCO]/[H_2CO]}$ & 0.39 & 0.28^{+0.20}_{-0.20} & -0.11 & 0.39\pm0.09 \\
  & $N(\mathrm{H_2})_\mathrm{peak}$ & -0.41 & -0.34^{+0.17}_{-0.16} & 0.22 & -0.40\pm0.08 \\
  & $M(R_C)$ & -0.09 & -0.10^{+0.25}_{-0.23} & 0.16 & -0.09\pm0.12 \\
  & $N(\mathrm{N_2H^+})$ & 0.20 & 0.14^{+0.17}_{-0.17} & -0.08 & 0.20\pm0.11 \\
  & $n_c$ & -0.07 & 0.01^{+0.22}_{-0.21} & 0.06 & -0.07\pm0.15 \\
  & $\mathrm{[pD_2CO]/[H_2CO]}$ & 0.57 & 0.42^{+0.20}_{-0.20} & -0.25 & 0.57\pm0.07 \\
  & $\mathrm{[pD_2CO]/[HDCO]}$ & 0.61 & 0.45^{+0.19}_{-0.22} & -0.36 & 0.60\pm0.08 \\
  & $N(\mathrm{pNH_3})$ & 0.38 & 0.27^{+0.19}_{-0.21} & -0.18 & 0.38\pm0.10 \\
  & $r_\mathrm{plat}$ & -0.15 & -0.12^{+0.22}_{-0.18} & 0.31 & -0.15\pm0.15 \\
  & ISRF & -0.63 & -0.38^{+0.29}_{-0.25} & 0.36 & -0.63\pm0.05 \\
  & $\alpha_\mathrm{K,G,P}$ & 0.04 & 0.12^{+0.23}_{-0.21} & 0.11 & 0.04\pm0.14 \\
$\mathrm{[HDCO]/[H_2CO]}$ & $\langle n_\mathrm{H_2} \rangle_\mathrm{dendro}$ & 0.20 & 0.13^{+0.17}_{-0.18} & 0.03 & 0.20\pm0.14 \\
  & $N(\mathrm{H_2})_\mathrm{peak}$ & 0.15 & 0.17^{+0.15}_{-0.15} & -0.01 & 0.15\pm0.12 \\
  & $M(R_C)$ & 0.43 & 0.36^{+0.16}_{-0.16} & -0.23 & 0.43\pm0.11 \\
  & $N(\mathrm{N_2H^+})$ & 0.46 & 0.45^{+0.12}_{-0.13} & -0.27 & 0.46\pm0.12 \\
  & $n_c$ & 0.19 & 0.21^{+0.13}_{-0.15} & -0.30 & 0.18\pm0.13 \\
  & $\mathrm{[pD_2CO]/[H_2CO]}$ & 0.85 & 0.75^{+0.11}_{-0.17} & -1.05 & 0.85\pm0.03 \\
  & $\mathrm{[pD_2CO]/[HDCO]}$ & 0.67 & 0.42^{+0.21}_{-0.24} & -0.33 & 0.67\pm0.07 \\
  & $N(\mathrm{pNH_3})$ & 0.65 & 0.51^{+0.15}_{-0.19} & -0.65 & 0.64\pm0.08 \\
  & $r_\mathrm{plat}$ & 0.21 & 0.18^{+0.13}_{-0.16} & -0.16 & 0.21\pm0.12 \\
  & ISRF & 0.07 & 0.07^{+0.27}_{-0.29} & -0.19 & 0.07\pm0.14 \\
  & $\alpha_\mathrm{K,G,P}$ & 0.12 & 0.09^{+0.21}_{-0.18} & 0.17 & 0.11\pm0.15 \\
$\mathrm{[pD_2CO]/[H_2CO]}$ & $\langle n_\mathrm{H_2} \rangle_\mathrm{dendro}$ & 0.35 & 0.26^{+0.18}_{-0.21} & -0.23 & 0.34\pm0.12 \\
  & $N(\mathrm{H_2})_\mathrm{peak}$ & 0.26 & 0.21^{+0.16}_{-0.15} & 0.06 & 0.26\pm0.11 \\
  & $M(R_C)$ & 0.55 & 0.42^{+0.16}_{-0.19} & -0.37 & 0.55\pm0.09 \\
  & $N(\mathrm{N_2H^+})$ & 0.69 & 0.59^{+0.13}_{-0.13} & -0.32 & 0.69\pm0.07 \\
  & $n_c$ & 0.48 & 0.41^{+0.16}_{-0.18} & -0.44 & 0.47\pm0.06 \\
  & $\mathrm{[pD_2CO]/[HDCO]}$ & 0.93 & 0.62^{+0.16}_{-0.21} & -0.60 & 0.92\pm0.01 \\
  & $N(\mathrm{pNH_3})$ & 0.85 & 0.65^{+0.13}_{-0.17} & -0.66 & 0.85\pm0.03 \\
  & $r_\mathrm{plat}$ & 0.02 & -0.01^{+0.17}_{-0.18} & -0.05 & 0.02\pm0.15 \\
  & ISRF & 0.01 & -0.03^{+0.30}_{-0.30} & -0.02 & 0.01\pm0.12 \\
  & $\alpha_\mathrm{K,G,P}$ & 0.08 & 0.13^{+0.22}_{-0.22} & -0.17 & 0.08\pm0.14 \\
$\mathrm{[pD_2CO]/[HDCO]}$ & $\langle n_\mathrm{H_2} \rangle_\mathrm{dendro}$ & 0.49 & 0.35^{+0.23}_{-0.29} & -0.45 & 0.48\pm0.11 \\
  & $N(\mathrm{H_2})_\mathrm{peak}$ & 0.24 & 0.18^{+0.20}_{-0.22} & -0.14 & 0.23\pm0.11 \\
  & $M(R_C)$ & 0.50 & 0.33^{+0.24}_{-0.27} & -0.23 & 0.50\pm0.08 \\
  & $N(\mathrm{N_2H^+})$ & 0.74 & 0.56^{+0.15}_{-0.18} & -0.46 & 0.73\pm0.04 \\
  & $n_c$ & 0.58 & 0.47^{+0.21}_{-0.22} & -0.35 & 0.57\pm0.08 \\
  & $N(\mathrm{pNH_3})$ & 0.85 & 0.60^{+0.15}_{-0.19} & -0.52 & 0.84\pm0.04 \\
  & $r_\mathrm{plat}$ & -0.21 & -0.20^{+0.22}_{-0.21} & 0.13 & -0.21\pm0.15 \\
  & ISRF & -0.07 & -0.14^{+0.34}_{-0.29} & 0.20 & -0.07\pm0.13 \\
  & $\alpha_\mathrm{K,G,P}$ & 0.05 & 0.10^{+0.28}_{-0.29} & -0.02 & 0.05\pm0.14 \\
\enddata
\tablecomments{(a) Uncertainties are the 1$\sigma$ confidence intervals of the $r_s$ distribution. (b) Uncertainties are the standard deviation of the $r_s$ jackknife replicates.}
\end{deluxetable*}

\begin{deluxetable*}{llRRRR}
\tablewidth{0pt} 
\tablecaption{Monte Carlo \& Jackknife Resampling Statistics for Molecular Abundance Ratio Comparisons \label{tab:MC_jk_molratios}}
\tablehead{
\colhead{Quantity 1} & \colhead{Quantity 2} &
\colhead{$r_{s,\mathrm{{obs}}}$} & 
\colhead{med($r_s$)$_\mathrm{MC}$\tablenotemark{a}} &
\colhead{$\gamma_{1,r_s}$} &
\colhead{$\langle r_s \rangle_\mathrm{jk}$\tablenotemark{b}}
} 
\startdata
$\mathrm{[CH_2DOH]/[CH_3OH]}$ & $\mathrm{[CH_3OH]/[H_2CO]}$ & -0.43 & -0.35^{+0.23}_{-0.19} & 0.42 & -0.42\pm0.05 \\
  & $\mathrm{[N_2H^+]/[CH_3OH]}$ & 0.87 & 0.66^{+0.14}_{-0.16} & -0.58 & 0.87\pm0.03 \\
  & $\mathrm{[N_2H^+]/[H_2CO]}$ & 0.35 & 0.26^{+0.20}_{-0.16} & 0.06 & 0.34\pm0.10 \\
  & $\mathrm{[N_2H^+]/[pNH_3]}$ & -0.50 & -0.28^{+0.29}_{-0.26} & 0.29 & -0.50\pm0.10 \\
  & $\mathrm{[pNH_3]/[CH_3OH]}$ & 0.73 & 0.60^{+0.15}_{-0.18} & -0.47 & 0.72\pm0.07 \\
  & $\mathrm{[pNH_3]/[H_2CO]}$ & 0.30 & 0.32^{+0.18}_{-0.19} & -0.06 & 0.30\pm0.10 \\
$\mathrm{[CH_3OH]/[H_2CO]}$ & $\mathrm{[N_2H^+]/[CH_3OH]}$ & -0.15 & -0.09^{+0.19}_{-0.20} & -0.00 & -0.15\pm0.11 \\
  & $\mathrm{[N_2H^+]/[H_2CO]}$ & 0.63 & 0.57^{+0.13}_{-0.15} & -0.42 & 0.62\pm0.08 \\
  & $\mathrm{[N_2H^+]/[pNH_3]}$ & 0.00 & -0.04^{+0.27}_{-0.28} & 0.04 & -0.00\pm0.12 \\
  & $\mathrm{[pNH_3]/[CH_3OH]}$ & 0.10 & 0.01^{+0.20}_{-0.21} & -0.14 & 0.10\pm0.10 \\
  & $\mathrm{[pNH_3]/[H_2CO]}$ & 0.63 & 0.48^{+0.15}_{-0.16} & -0.38 & 0.62\pm0.08 \\
$\mathrm{[HDCO]/[H_2CO]}$ & $\mathrm{[CH_3OH]/[H_2CO]}$ & 0.59 & 0.51^{+0.16}_{-0.21} & -0.52 & 0.59\pm0.07 \\
  & $\mathrm{[N_2H^+]/[CH_3OH]}$ & 0.58 & 0.58^{+0.14}_{-0.15} & -0.21 & 0.58\pm0.09 \\
  & $\mathrm{[N_2H^+]/[H_2CO]}$ & 0.92 & 0.82^{+0.08}_{-0.13} & -1.07 & 0.92\pm0.03 \\
  & $\mathrm{[N_2H^+]/[pNH_3]}$ & -0.46 & -0.29^{+0.32}_{-0.35} & 0.20 & -0.46\pm0.08 \\
  & $\mathrm{[pNH_3]/[CH_3OH]}$ & 0.78 & 0.62^{+0.13}_{-0.16} & -0.70 & 0.77\pm0.03 \\
  & $\mathrm{[pNH_3]/[H_2CO]}$ & 0.93 & 0.79^{+0.09}_{-0.14} & -1.05 & 0.92\pm0.02 \\
$\mathrm{[N_2H^+]/[CH_3OH]}$ & $\mathrm{[N_2H^+]/[H_2CO]}$ & 0.53 & 0.58^{+0.14}_{-0.12} & 0.06 & 0.52\pm0.10 \\
  & $\mathrm{[N_2H^+]/[pNH_3]}$ & -0.49 & -0.35^{+0.31}_{-0.33} & 0.23 & -0.49\pm0.11 \\
  & $\mathrm{[pNH_3]/[CH_3OH]}$ & 0.86 & 0.84^{+0.08}_{-0.10} & -0.95 & 0.86\pm0.04 \\
  & $\mathrm{[pNH_3]/[H_2CO]}$ & 0.53 & 0.61^{+0.14}_{-0.14} & -0.16 & 0.52\pm0.10 \\
$\mathrm{[N_2H^+]/[H_2CO]}$ & $\mathrm{[N_2H^+]/[pNH_3]}$ & -0.44 & -0.32^{+0.33}_{-0.35} & 0.19 & -0.43\pm0.09 \\
  & $\mathrm{[pNH_3]/[CH_3OH]}$ & 0.75 & 0.65^{+0.13}_{-0.16} & -0.68 & 0.74\pm0.05 \\
  & $\mathrm{[pNH_3]/[H_2CO]}$ & 0.98 & 0.91^{+0.05}_{-0.08} & -2.16 & 0.98\pm0.01 \\
$\mathrm{[N_2H^+]/[pNH_3]}$ & $\mathrm{[pNH_3]/[CH_3OH]}$ & -0.74 & -0.45^{+0.30}_{-0.29} & 0.35 & -0.73\pm0.07 \\
  & $\mathrm{[pNH_3]/[H_2CO]}$ & -0.50 & -0.40^{+0.30}_{-0.32} & 0.29 & -0.50\pm0.09 \\
$\mathrm{[pD_2CO]/[H_2CO]}$ & $\mathrm{[CH_3OH]/[H_2CO]}$ & 0.33 & 0.29^{+0.20}_{-0.23} & -0.36 & 0.33\pm0.10 \\
  & $\mathrm{[N_2H^+]/[CH_3OH]}$ & 0.81 & 0.74^{+0.11}_{-0.14} & -0.53 & 0.80\pm0.05 \\
  & $\mathrm{[N_2H^+]/[H_2CO]}$ & 0.82 & 0.78^{+0.09}_{-0.16} & -1.30 & 0.81\pm0.03 \\
  & $\mathrm{[N_2H^+]/[pNH_3]}$ & -0.67 & -0.38^{+0.30}_{-0.34} & 0.25 & -0.67\pm0.09 \\
  & $\mathrm{[pNH_3]/[CH_3OH]}$ & 0.93 & 0.76^{+0.11}_{-0.15} & -0.84 & 0.92\pm0.01 \\
  & $\mathrm{[pNH_3]/[H_2CO]}$ & 0.86 & 0.79^{+0.09}_{-0.15} & -1.28 & 0.86\pm0.03 \\
$\mathrm{[pD_2CO]/[HDCO]}$ & $\mathrm{[CH_3OH]/[H_2CO]}$ & 0.15 & -0.02^{+0.27}_{-0.27} & -0.06 & 0.14\pm0.10 \\
  & $\mathrm{[N_2H^+]/[CH_3OH]}$ & 0.85 & 0.67^{+0.14}_{-0.15} & -0.57 & 0.84\pm0.05 \\
  & $\mathrm{[N_2H^+]/[H_2CO]}$ & 0.68 & 0.49^{+0.22}_{-0.24} & -0.41 & 0.68\pm0.07 \\
  & $\mathrm{[N_2H^+]/[pNH_3]}$ & -0.62 & -0.36^{+0.29}_{-0.28} & 0.32 & -0.61\pm0.09 \\
  & $\mathrm{[pNH_3]/[CH_3OH]}$ & 0.89 & 0.68^{+0.14}_{-0.18} & -0.58 & 0.89\pm0.03 \\
  & $\mathrm{[pNH_3]/[H_2CO]}$ & 0.74 & 0.55^{+0.19}_{-0.24} & -0.43 & 0.73\pm0.06 \\
$\mathrm{[pNH_3]/[CH_3OH]}$ & $\mathrm{[pNH_3]/[H_2CO]}$ & 0.77 & 0.70^{+0.12}_{-0.15} & -0.70 & 0.77\pm0.04 \\
\enddata
\tablecomments{(a) Uncertainties are the 1$\sigma$ confidence intervals of the $r_s$ distribution. (b) Uncertainties are the standard deviation of the $r_s$ jackknife replicates.}
\end{deluxetable*}

\section{Raw Spectra} \label{ap:spectra}

\begin{figure*}[ht!]\centering
\includegraphics[width=1\linewidth]{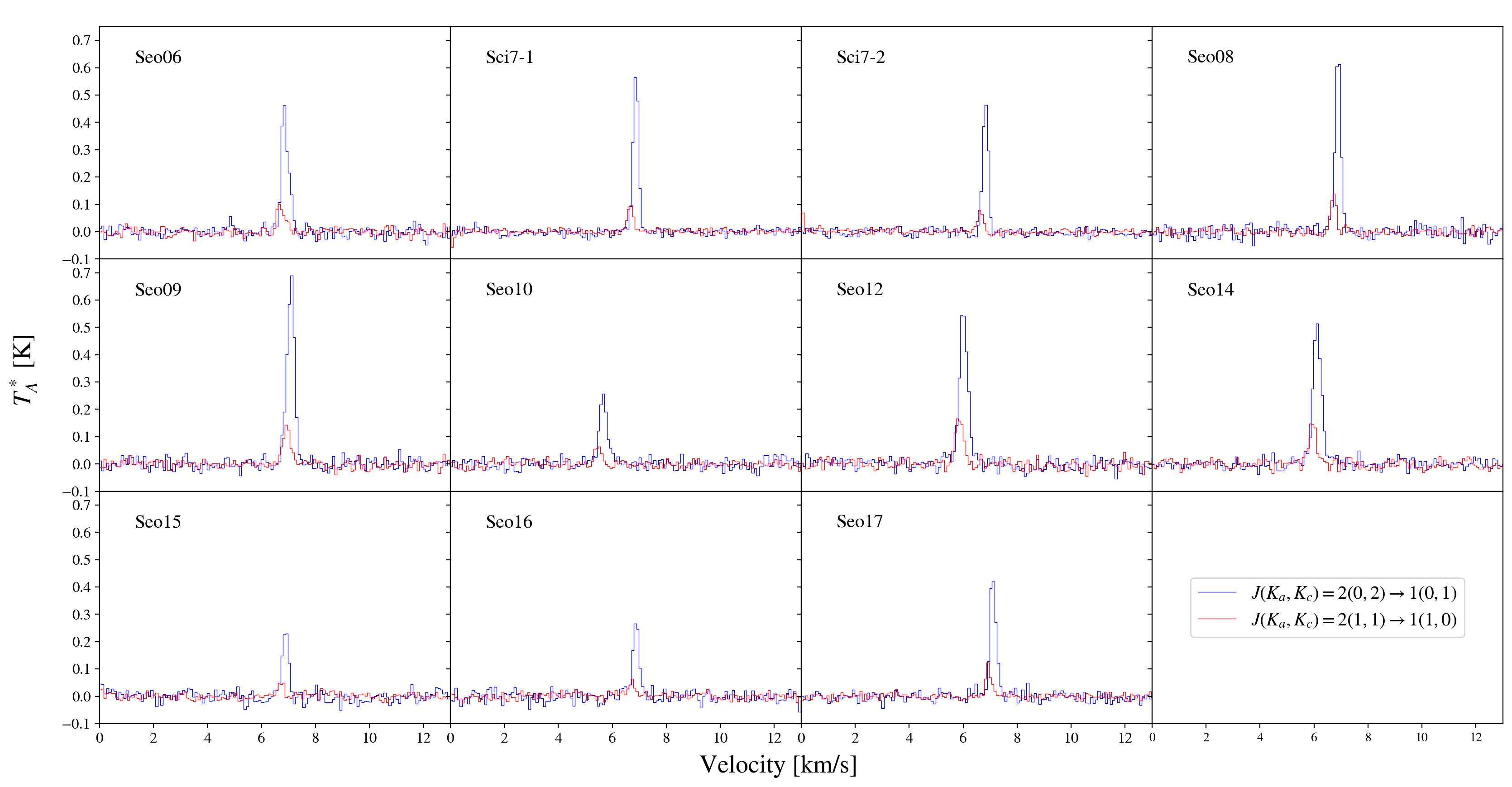}
\caption{
Spectra for all observed sources of the HDCO $J(K_a,K_c) = 2(0,2) \rightarrow 1(0,1)$ (blue) and $J(K_a,K_c) = 2(1,1) \rightarrow 1(1,0)$ (red) transitions on the $T_{A}^*$ scale.
The tuning frequency and beam efficiency $\eta_{mb}$ are listed in Table \ref{tab:obs_transitions}.
\label{fig:HDCO_spectra}}
\end{figure*}

\begin{figure*}[ht!]\centering
\includegraphics[width=1\linewidth]{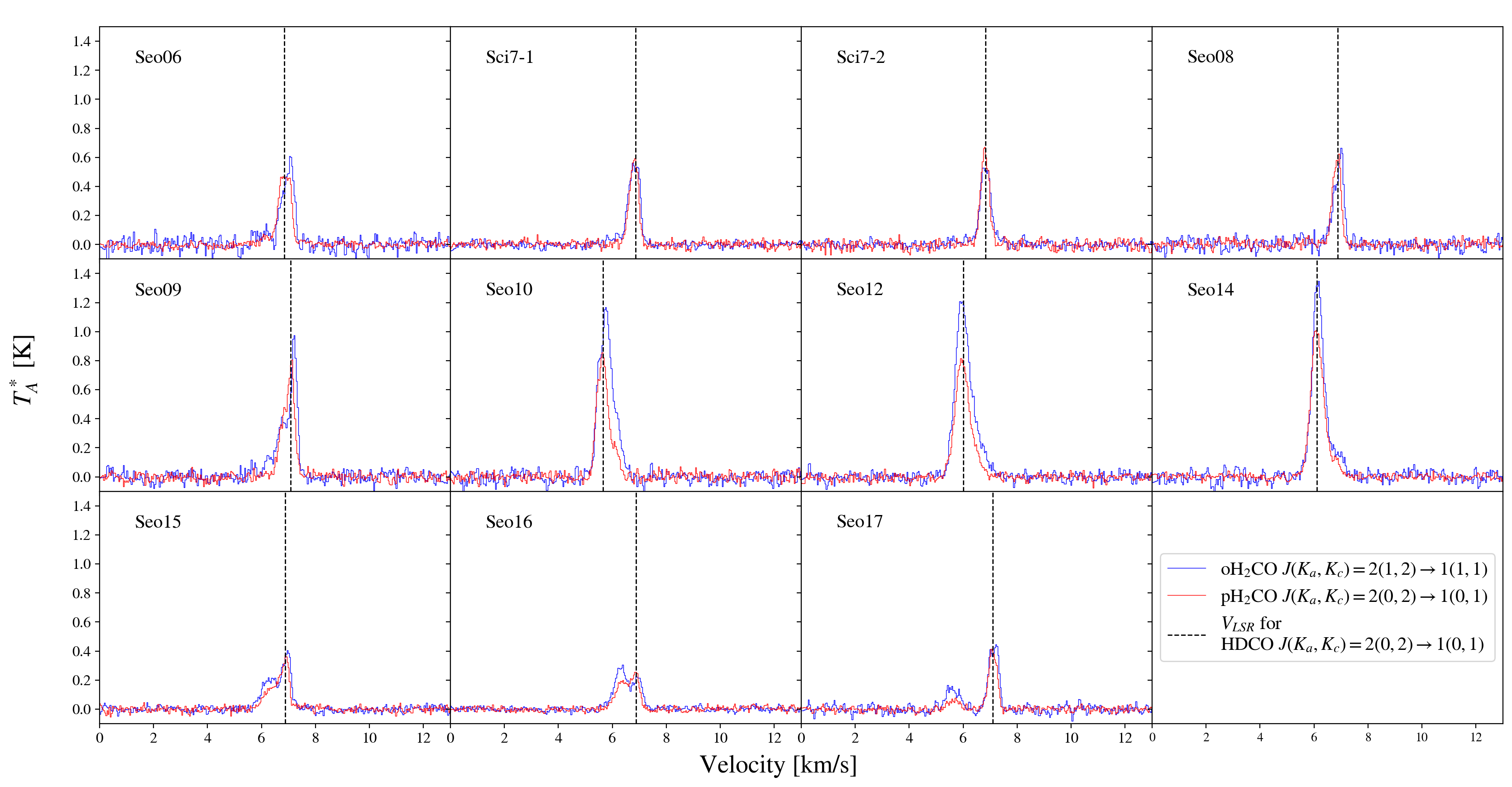}
\caption{
Spectra for all observed sources of the oH$_2$CO $J(K_a,K_c) = 2(1,2) \rightarrow 1(1,1)$ (blue) and pH$_2$CO $J(K_a,K_c) = 2(0,2) \rightarrow 1(0,1)$ (red) transitions on the $T_{A}^*$ scale.
The dashed black lines indicate the $v_{LSR}$ of the HDCO $J(K_a,K_c) = 2(0,2) \rightarrow 1(0,1)$ transition to clarify which velocity component of the H$_2$CO emission was fit.
The tuning frequencies and beam efficiencies $\eta_{mb}$ are listed in Table \ref{tab:obs_transitions}.
\label{fig:H2CO_spectra}}
\end{figure*}

\begin{figure*}[ht!]\centering
\includegraphics[width=1\linewidth]{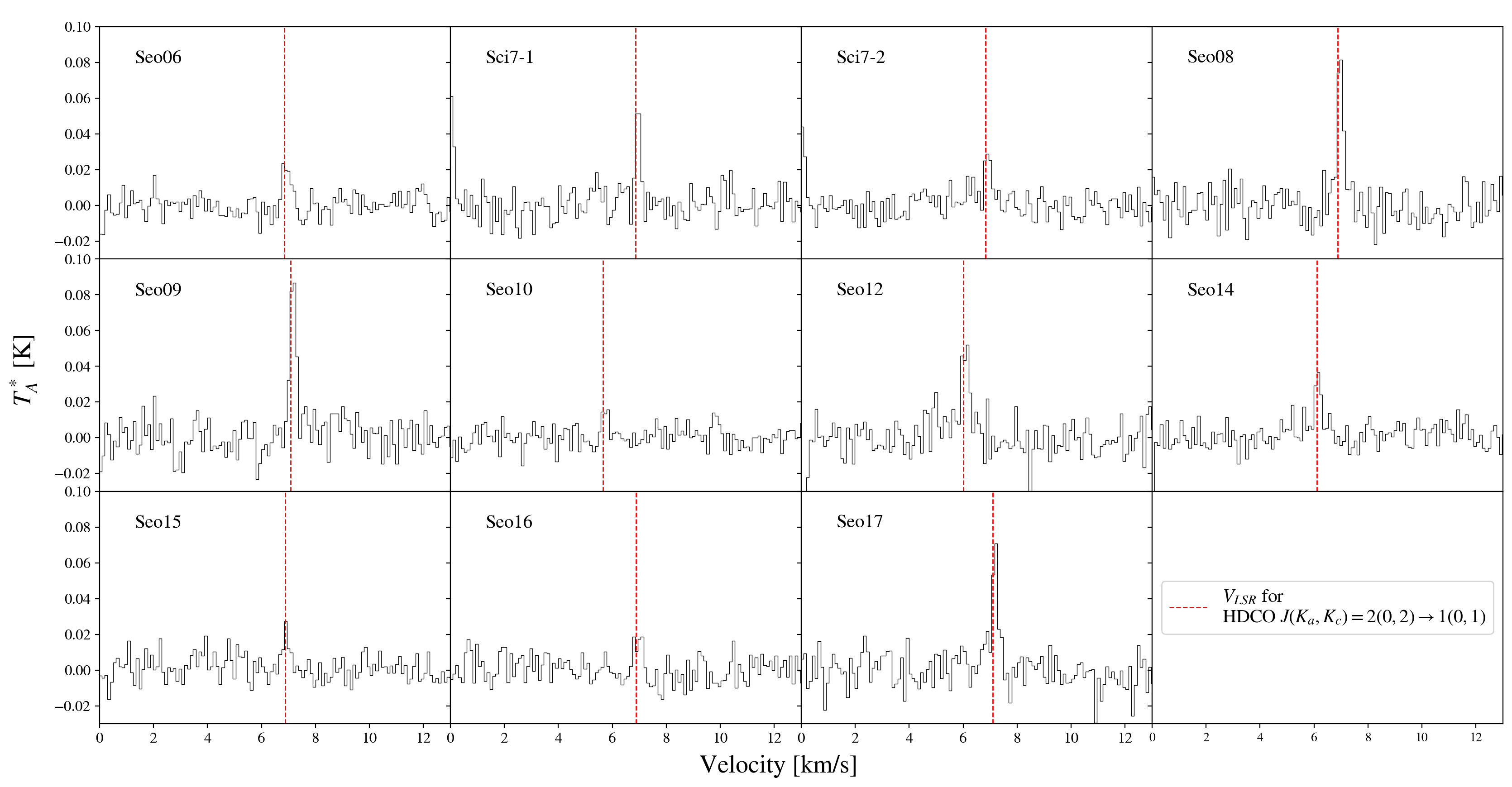}
\caption{
Spectra for all observed sources of the pD$_2$CO $J(K_a,K_c) = 2(1,2) \rightarrow 1(1,1)$ transition on the $T_{A}^*$ scale. 
$v_{LSR}$ for HDCO $2(0,2) \rightarrow 1(0,1)$} (black dashed line) is provided for reference.
The tuning frequency and beam efficiency $\eta_{mb}$ are listed in Table \ref{tab:obs_transitions}.
\label{fig:pD2CO_spectra}
\end{figure*}

\begin{figure*}[ht!]\centering
\includegraphics[width=1\linewidth]{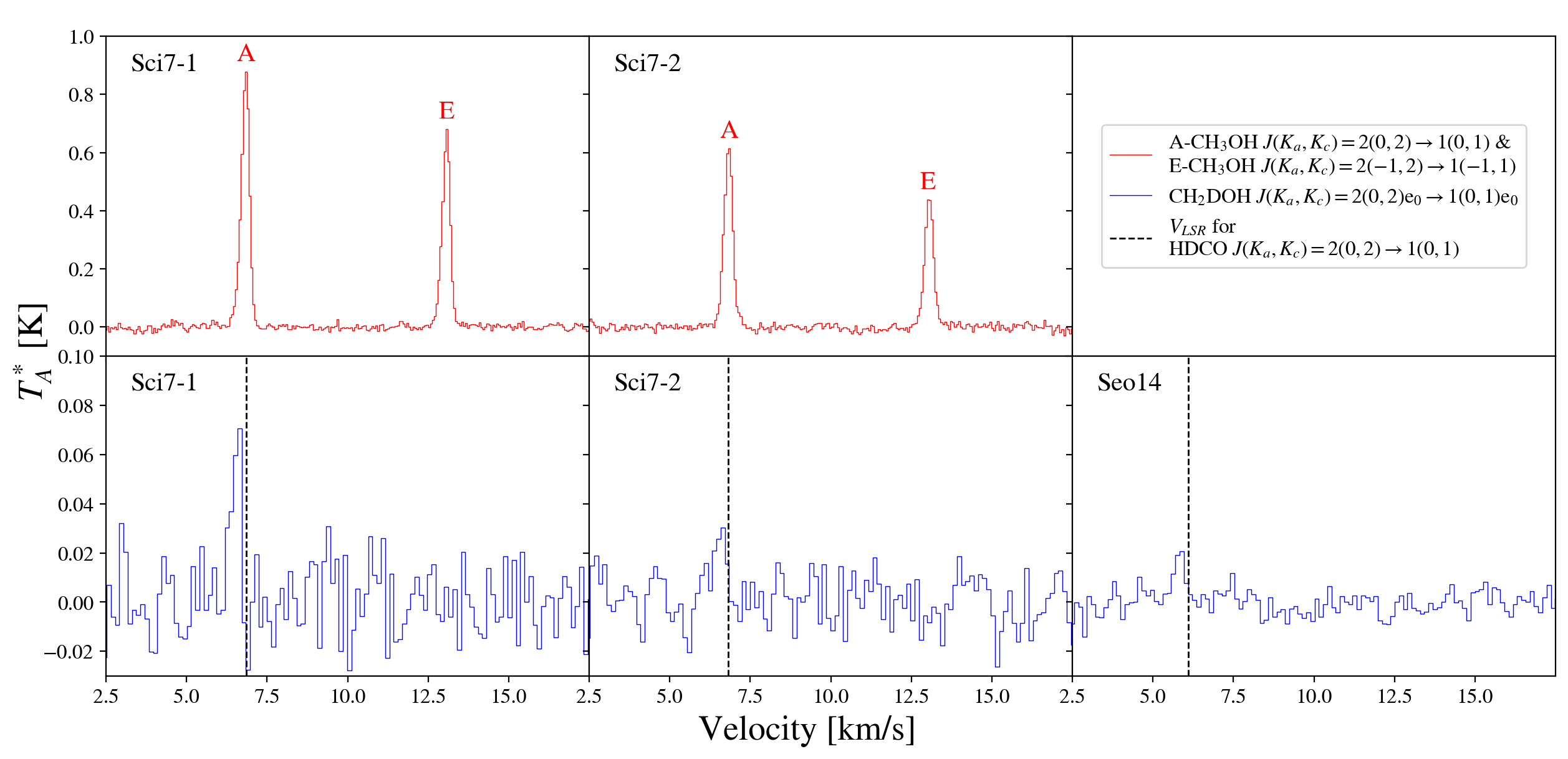}
\caption{
Spectra of the A-CH$_3$OH $J(K_a,K_c) = 2(0,2) \rightarrow 1(0,1)$ and E-CH$_3$OH $J(K_a,K_c) = 2(-1,2) \rightarrow 1(-1,1)$ (top row, red), as well as the CH$_2$DOH $J(K_a,K_c) = 2(0,2)\mathrm{e}_0 \rightarrow 1(0,1)\mathrm{e}_0$} (bottom row, blue) transitions on the $T_{A}^*$ scale. 
These spectra are only for sources not previously observed (Sci7-1 and Sci7-2) or detected (Seo14) in these transitions. 
For the remaining sources, the CH$_3$OH spectra may be found in \cite{Scibelli2020} and the CH$_2$DOH spectra in \cite{Ambrose2021}.
$v_{LSR}$ for HDCO $2(0,2) \rightarrow 1(0,1)$ (black dashed line) is provided for reference for the CH$_2$DOH spectra.
The tuning frequencies and beam efficiencies $\eta_{mb}$ are listed in Table \ref{tab:obs_transitions}. 
The CH$_3$OH observations were tuned to the frequency of A-CH$_3$OH $J(K_a,K_c) = 2(0,2) \rightarrow 1(0,1)$.
\label{fig:methanol_spectra}
\end{figure*}

\begin{figure*}[ht!]\centering
\includegraphics[width=1\linewidth]{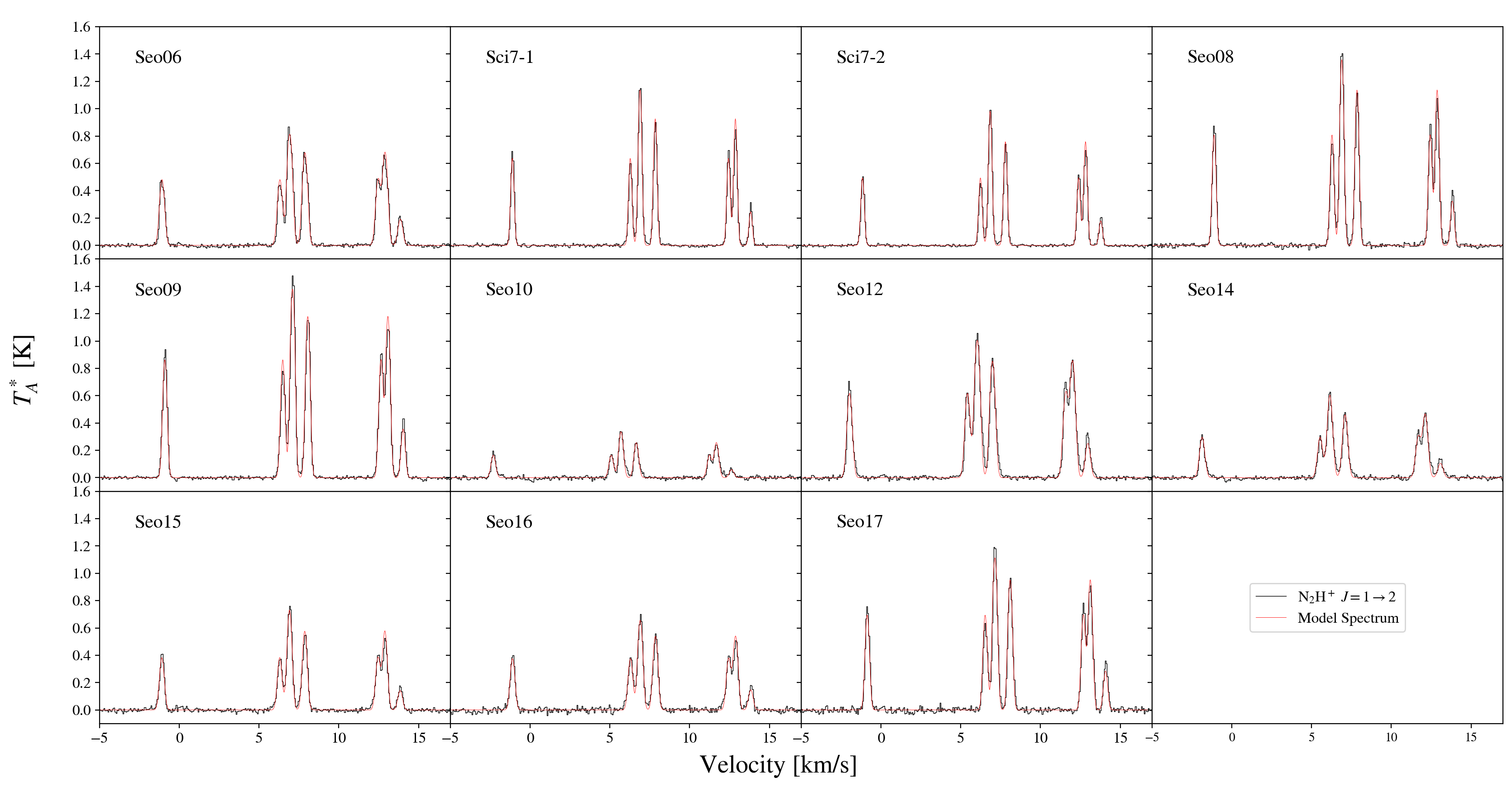}
\caption{
Spectra for all observed sources of the N$_2$H$^+$ $J = 1 \rightarrow 0$ transition on the $T_{A}^*$ scale with model spectra derived from line parameters from the \texttt{CLASS} and \texttt{PySpecKit} fits overlaid in red. 
The tuning frequency and beam efficiency $\eta_{mb}$ are listed in Table \ref{tab:obs_transitions}.
\label{fig:N2H+_spectra}}
\end{figure*}

We show all of the baselined and averaged spectra for all spectral lines observed with the ARO 12m telescope plotted on the $T_{A}^*$ scale in Figures \ref{fig:HDCO_spectra}, \ref{fig:H2CO_spectra}, \ref{fig:pD2CO_spectra}, \ref{fig:methanol_spectra}, and  \ref{fig:N2H+_spectra}.

\section{Miscellaneous Tables}

We include here tables of the linewidths (Table \ref{tab:delV}) and velocities at the local standard of rest (Table \ref{tab:vLSR}) for observed transitions across the B10 core sample.
We also include the beam-averaged H$_2$ column densities for each core within each relevant beam size (Table \ref{tab:beam_avg_H2}). 
The beam sizes are in Table \ref{tab:obs_transitions} for each transition and the process of calculating the column densities is detailed in Section \ref{subsec:beam-avg_NH2}.

\begin{longrotatetable}
\movetabledown=2cm
\begin{deluxetable*}{lCCCCCCCCC}
\tablewidth{0pt} 
\tablecaption{Linewidths\label{tab:delV}}
\tablehead{
\colhead{} & \multicolumn{1}{c}{oH$_2$CO\tablenotemark{a}} & \multicolumn{1}{c}{pH$_2$CO\tablenotemark{a}} & \multicolumn{2}{c}{HDCO} & \multicolumn{1}{c}{pD$_2$CO} & \multicolumn{1}{c}{A-CH$_3$OH} & \multicolumn{1}{c}{E-CH$_3$OH} & \multicolumn{1}{c}{CH$_2$DOH} & \multicolumn{1}{c}{N$_2$H$^+$} \\
\colhead{} & \multicolumn{1}{c}{$2(1,2) \rightarrow 1(1,1)$} & \multicolumn{1}{c}{$2(0,2) \rightarrow 1(0,1)$} & \multicolumn{1}{c}{$2(0,2) \rightarrow 1(0,1)$} & \multicolumn{1}{c}{$2(1,1) \rightarrow 1(1,0)$} & \multicolumn{1}{c}{$2(1,2) \rightarrow 1(1,1)$} & \multicolumn{1}{c}{$2(0,2) \rightarrow 1(0,1)$} & \multicolumn{1}{c}{$2(-1,2) \rightarrow 1(-1,1)$} & \multicolumn{1}{c}{$2(0,2) \mathrm{e}_0 \rightarrow 1(0,1) \mathrm{e}_0$} & \multicolumn{1}{c}{$1 \rightarrow 0$} \\
\colhead{Core} & \colhead{$\Delta v$} & \colhead{$\Delta v$} & \colhead{$\Delta v$} & \colhead{$\Delta v$} & \colhead{$\Delta v$} & \colhead{$\Delta v$} & \colhead{$\Delta v$} & \colhead{$\Delta v$} & \colhead{$\Delta v$} \\
\colhead{} & \colhead{(km s$^{-1}$)} & \colhead{(km s$^{-1}$)} & \colhead{(km s$^{-1}$)} & \colhead{(km s$^{-1}$)} & \colhead{(km s$^{-1}$)} &  \colhead{(km s$^{-1}$)} & \colhead{(km s$^{-1}$)} & \colhead{(km s$^{-1}$)} & \colhead{(km s$^{-1}$)} \\
}
\startdata 
Seo06 & 0.485\pm0.042 & 0.520\pm0.040 & 0.325\pm0.012 & 0.35\pm0.04 & 0.35\pm0.07 &   &   &   & 0.352\pm0.001 \\
Sci7-1 & 0.485\pm0.008 & 0.381\pm0.007 & 0.242\pm0.004 & 0.26\pm0.02 & 0.21\pm0.03 & 0.297\pm0.003 & 0.292\pm0.003 & 0.32\pm0.06 & 0.241\pm0.001 \\
Sci7-2 & 0.451\pm0.010 & 0.368\pm0.008 & 0.245\pm0.005 & 0.25\pm0.03 & 0.30\pm0.07 & 0.319\pm0.004 & 0.320\pm0.006 & 0.50\pm0.21 & 0.253\pm0.000 \\
Seo08 & 0.296\pm0.042 & 0.365\pm0.007 & 0.251\pm0.007 & 0.24\pm0.02 & 0.26\pm0.03 &   &   &   & 0.266\pm0.001 \\
Seo09 & 0.244\pm0.042 & 0.304\pm0.040 & 0.336\pm0.008 & 0.35\pm0.03 & 0.27\pm0.03 &   &   &   & 0.300\pm0.002 \\
Seo10 & 0.457\pm0.042 & 0.467\pm0.040 & 0.339\pm0.022 & 0.29\pm0.03 & 0.28\pm0.06 &   &   &   & 0.329\pm0.005 \\
Seo12 & 0.716\pm0.042 & 0.564\pm0.040 & 0.369\pm0.010 & 0.36\pm0.02 & 0.39\pm0.09 &   &   &   & 0.363\pm0.001 \\
Seo14 & 0.469\pm0.042 & 0.507\pm0.040 & 0.372\pm0.009 & 0.32\pm0.03 & 0.28\pm0.05 &   &   & 0.41\pm0.03\tablenotemark{b} & 0.381\pm0.002 \\
Seo15 & 0.341\pm0.042 & 0.254\pm0.040 & 0.280\pm0.019 & 0.26\pm0.04 & 0.29\pm0.09 &   &   &   & 0.337\pm0.002 \\
Seo16 & 0.368\pm0.042 & 0.318\pm0.040 & 0.273\pm0.018 & 0.29\pm0.07 & 0.36\pm0.07 &   &   &   & 0.330\pm0.003 \\
Seo17 & 0.423\pm0.042 & 0.365\pm0.009 & 0.270\pm0.009 & 0.23\pm0.02 & 0.24\pm0.05 &   &   &   & 0.298\pm0.002 \\
\enddata
\tablecomments{(a) For sources with multiple velocity components, the given $\Delta v$ is that of the component that most closely matched the $v_{LSR}$ of HDCO $2(0,1) \rightarrow 1(0,1)$ in a 2-component fit. Additionally, the linewidth uncertainties given by multi-component fitting in \texttt{RADEX} were unreliable. Thus the linewidth uncertainties for sources with multiple velocity components were replaced by the channel width in km/s (0.042 km/s for oH$_2$CO $2(1,2)\rightarrow1(1,1)$ and 0.040 km/s for pH$_2$CO $2(0,2)\rightarrow1(0,1)$). (b) Combined observations from this work and 
\cite{Ambrose2021}.}
\end{deluxetable*}
\end{longrotatetable}

\begin{longrotatetable}
\movetabledown=2cm
\begin{deluxetable*}{lCCCCCCCCCCC}
\tablewidth{0pt} 
\tablecaption{Velocities at the Local Standard of Rest\label{tab:vLSR}}
\tablehead{
\colhead{} & \multicolumn{1}{c}{oH$_2$CO\tablenotemark{a}} & \multicolumn{1}{c}{pH$_2$CO\tablenotemark{a}} & \multicolumn{2}{c}{HDCO} & \multicolumn{1}{c}{pD$_2$CO} & \multicolumn{1}{c}{A-CH$_3$OH} & \multicolumn{1}{c}{E-CH$_3$OH} & \multicolumn{1}{c}{CH$_2$DOH} & \multicolumn{1}{c}{N$_2$H$^+$} \\
\colhead{} & \multicolumn{1}{c}{$2(1,2) \rightarrow 1(1,1)$} & \multicolumn{1}{c}{$2(0,2) \rightarrow 1(0,1)$} & \multicolumn{1}{c}{$2(0,2) \rightarrow 1(0,1)$} & \multicolumn{1}{c}{$2(1,1) \rightarrow 1(1,0)$} & \multicolumn{1}{c}{$2(1,2) \rightarrow 1(1,1)$} & \multicolumn{1}{c}{$2(0,2) \rightarrow 1(0,1)$} & \multicolumn{1}{c}{$2(-1,2) \rightarrow 1(-1,1)$} & \multicolumn{1}{c}{$2(0,2) \mathrm{e}_0 \rightarrow 1(0,1) \mathrm{e}_0$} & \multicolumn{1}{c}{$1 \rightarrow 0$} \\
\colhead{Core} & \colhead{$v_{LSR}$} & \colhead{$v_{LSR}$} & 
\colhead{$v_{LSR}$} & \colhead{$v_{LSR}$} & \colhead{$v_{LSR}$} & \colhead{$v_{LSR}$} & \colhead{$v_{LSR}$} & \colhead{$v_{LSR}$} & \colhead{$v_{LSR}$} \\
\colhead{} & \colhead{(km s$^{-1}$)} & \colhead{(km s$^{-1}$)} & \colhead{(km s$^{-1}$)} & \colhead{(km s$^{-1}$)} & \colhead{(km s$^{-1}$)} &  \colhead{(km s$^{-1}$)} & \colhead{(km s$^{-1}$)} & \colhead{(km s$^{-1}$)} & \colhead{(km s$^{-1}$)} \\
}
\startdata 
Seo06 & 6.957\pm0.025 & 6.859\pm0.040 & 6.852\pm0.005 & 6.701\pm0.017 & 6.92\pm0.04 &   &   &   & 6.9202\pm0.0006 \\
Sci7-1 & 6.798\pm0.022 & 6.793\pm0.003 & 6.861\pm0.002 & 6.663\pm0.009 & 6.95\pm0.01 & 6.819\pm0.001 & 6.825\pm0.001 & 6.53\pm0.03 & 6.8877\pm0.0003 \\
Sci7-2 & 6.797\pm0.017 & 6.795\pm0.003 & 6.821\pm0.002 & 6.631\pm0.012 & 6.91\pm0.03 & 6.813\pm0.002 & 6.816\pm0.002 & 6.54\pm0.08 & 6.8438\pm0.0003 \\
Seo08 & 6.907\pm0.014 & 6.835\pm0.003 & 6.885\pm0.003 & 6.715\pm0.007 & 6.97\pm0.01 &   &   &   & 6.8917\pm0.0004 \\
Seo09 & 7.083\pm0.023 & 7.104\pm0.040 & 7.076\pm0.003 & 6.923\pm0.011 & 7.17\pm0.01 &   &   &   & 7.0975\pm0.0020 \\
Seo10 & 5.716\pm0.014 & 5.619\pm0.040 & 5.651\pm0.008 & 5.464\pm0.017 & 5.73\pm0.03 &   &   &   & 5.6970\pm0.0018 \\
Seo12 & 5.960\pm0.012 & 5.925\pm0.040 & 6.002\pm0.004 & 5.845\pm0.010 & 6.07\pm0.03 &   &   &   & 6.0350\pm0.0006 \\
Seo14 & 6.096\pm0.011 & 6.085\pm0.040 & 6.106\pm0.004 & 5.935\pm0.010 & 6.15\pm0.02 &   &   & 5.81\pm0.01\tablenotemark{b} & 6.1450\pm0.0008 \\
Seo15 & 6.674\pm0.075 & 6.897\pm0.040 & 6.875\pm0.009 & 6.716\pm0.019 & 6.90\pm0.03 &   &   &   & 6.9090\pm0.0009 \\
Seo16 & 6.678\pm0.067 & 6.867\pm0.040 & 6.869\pm0.007 & 6.714\pm0.023 & 6.96\pm0.05 &   &   &   & 6.8985\pm0.0013 \\
Seo17 & 7.137\pm0.020 & 7.087\pm0.004 & 7.096\pm0.004 & 6.927\pm0.008 & 7.18\pm0.01 &   &   &   & 7.1411\pm0.0007 \\
\enddata
\tablecomments{(a) For sources with multiple velocity components, the given $v_{LSR}$ is that of the component that most closely matched the $v_{LSR}$ of HDCO $2(0,1) \rightarrow 1(0,1)$ in a 2-component fit. (b) Combined observations from this work and \cite{Ambrose2021}.}
\end{deluxetable*}
\end{longrotatetable}

\begin{longrotatetable}
\begin{deluxetable*}{lCCCCCCC}
\tablewidth{0pt} 
\tablecaption{Beam-Averaged H$_2$ Column Densities\label{tab:beam_avg_H2}}
\tablehead{
\colhead{Core} & \colhead{$\langle N(\mathrm{H}_2) \rangle_{\langle\theta(\mathrm{H}_2\mathrm{CO})\rangle}$} & \colhead{$\langle N(\mathrm{H}_2) \rangle_{\langle\theta(\mathrm{HDCO})\rangle}$} & \colhead{$\langle N(\mathrm{H}_2) \rangle_{\theta(\mathrm{pD}_2\mathrm{CO})}$} & \colhead{$\langle N(\mathrm{H}_2) \rangle_{\langle\theta(\mathrm{CH}_3\mathrm{OH})\rangle}$} &  \colhead{$\langle N(\mathrm{H}_2) \rangle_{\theta(\mathrm{CH}_2\mathrm{DOH})}$} & \colhead{$N(\mathrm{H}_2)_\mathrm{peak}$} \\
\colhead{} & \colhead{($10^{22}$ cm$^{-2}$)} & \colhead{($10^{22}$ cm$^{-2}$)} & \colhead{($10^{22}$ cm$^{-2}$)} & \colhead{($10^{22}$ cm$^{-2}$)} & \colhead{($10^{22}$ cm$^{-2}$)} & \colhead{($10^{22}$ cm$^{-2}$)}
} 
\startdata 
Seo06 & 1.25\pm0.03 & 1.24\pm0.03 & 1.22\pm0.03 & 1.19\pm0.03 & 1.18\pm0.03 & 1.27\pm0.04 \\
Sci7-1 & 1.00\pm0.03 & 0.99\pm0.03 & 0.96\pm0.03 & 0.93\pm0.03 & 0.91\pm0.03 & 1.08\pm0.03 \\
Sci7-2 & 0.73\pm0.02 & 0.72\pm0.02 & 0.71\pm0.02 & 0.70\pm0.02 & 0.69\pm0.02 & 0.76\pm0.02 \\
Seo08 & 1.02\pm0.03 & 1.01\pm0.03 & 0.99\pm0.03 & 0.97\pm0.03 & 0.96\pm0.03 & 1.05\pm0.03 \\
Seo09 & 1.41\pm0.04 & 1.39\pm0.04 & 1.33\pm0.04 & 1.28\pm0.04 & 1.25\pm0.04 & 1.59\pm0.04 \\
Seo10 & 0.71\pm0.02 & 0.70\pm0.02 & 0.68\pm0.02 & 0.66\pm0.02 & 0.64\pm0.02 & 0.80\pm0.02 \\
Seo12 & 1.45\pm0.05 & 1.43\pm0.05 & 1.36\pm0.05 & 1.30\pm0.05 & 1.26\pm0.05 & 1.69\pm0.05 \\
Seo14 & 1.04\pm0.04 & 1.02\pm0.04 & 0.97\pm0.03 & 0.93\pm0.03 & 0.90\pm0.03 & 1.20\pm0.04 \\
Seo15 & 0.85\pm0.02 & 0.85\pm0.02 & 0.84\pm0.02 & 0.83\pm0.02 & 0.82\pm0.02 & 0.86\pm0.02 \\
Seo16 & 0.96\pm0.03 & 0.96\pm0.03 & 0.95\pm0.03 & 0.94\pm0.03 & 0.92\pm0.03 & 1.09\pm0.04 \\
Seo17 & 1.05\pm0.04 & 1.03\pm0.04 & 0.99\pm0.04 & 0.95\pm0.04 & 0.92\pm0.04 & 1.19\pm0.04 \\
\enddata
\end{deluxetable*}
\end{longrotatetable}

\end{document}